\documentclass[a4paper,12pt]{article}

\pdfoutput=1
\usepackage{amsmath}
\usepackage{amssymb}
\usepackage{amsfonts}
\usepackage{mathrsfs}
\usepackage{bbm}
\usepackage{graphicx,subfigure}

\usepackage{bookmark}

\usepackage{cite}

\usepackage{calc}
\usepackage{hyperref}
\usepackage{array}

\usepackage{ulem}

\usepackage{multirow}
\usepackage{color}
\usepackage{xcolor,colortbl}

\usepackage{psfrag}
\usepackage{pstricks}

\numberwithin{equation}{section}
\allowdisplaybreaks

\newlength{\dinwidth}
\newlength{\dinmargin}
\definecolor{nicered}{rgb}{1.0,0.0,0.2}

\definecolor{color1}{rgb}{0.9,.4,.2}

\definecolor{color2}{rgb}{0.3,.6,.7}

\definecolor{color3}{rgb}{0.7,.2,.7}

\usepackage{color}

\begin{document}

\title{
\vspace*{-0.5cm}
\bf \Large
$\mathbf{\boldsymbol{B_{s,d}-\bar{B}_{s,d}}}$ mixings and $\mathbf{\boldsymbol{B_{s,d}\to\ell^+\ell^-}}$ decays\\[-2pt] within the Manohar-Wise model}

\author{Xiao-Dong Cheng$^{1}$\footnote{chengxd@mails.ccnu.edu.cn}, Xin-Qiang Li$^{1,2}$\footnote{xqli@mail.ccnu.edu.cn}, Ya-Dong Yang$^{1}$\footnote{yangyd@mail.ccnu.edu.cn} and Xin Zhang$^{1}$\footnote{zhangxin027@mails.ccnu.edu.cn}\\
{$^1$\small Institute of Particle Physics and Key Laboratory of Quark and Lepton Physics~(MOE)}\\[-0.2cm]
{    \small Central China Normal University, Wuhan, Hubei 430079, P.~R.~China}\\[-0.1cm]
{$^2$\small State Key Laboratory of Theoretical Physics, Institute of Theoretical Physics,}\\[-0.2cm]
{    \small Chinese Academy of Sciences, Beijing 100190, P.~R.~China}}

\date{}
\maketitle
\bigskip\bigskip
\maketitle
\vspace{-1.2cm}

\begin{abstract}
{\noindent}In this paper, we perform a complete one-loop computation of the short-distance Wilson coefficients for $B_{s,d}-\bar{B}_{s,d}$ mixings and $B_{s,d}\to\ell^+\ell^-$ decays in the Manohar-Wise model, which extends the SM scalar sector by a colour-octet and weak-doublet scalar. Based on these calculations, combined constraints on the model parameters are derived from the current flavour data, including the $B_{s,d}-{\bar{B}}_{s,d}$ mixings, $B_{s,d}\to\mu^+\mu^-$, $B\to X_s\gamma$, $B\to K^{\ast}\gamma$, $B\to\rho\gamma$, and $Z\to b \bar{b}$ decays. The future sensitivity to the model is also explored in the observables achievable with $50~{\rm fb}^{-1}$ of LHCb and $50~{\rm ab}^{-1}$ of Belle-II data. We find that the Manohar-Wise model could explain the current data, especially when the couplings of the charged colour-octet scalars to quarks are complex, with the resulting $\chi^2_{\rm min}$ being significantly smaller than that of the SM. Finally, we investigate correlations among the two isospin asymmetries $\Delta(K^*\gamma)$ and $\Delta(\rho\gamma)$, as well as the averaged time-integrated branching ratios $\overline{{\mathcal B}}(B_{s,d}\to \mu^+\mu^-)$; some of them are found to be quite strong and could provide, therefore, further insights into the model, once more precise experimental measurements and theoretical predictions for these observables are available in the future.
\end{abstract}

\newpage

\section{Introduction}
\label{sec:intro}

The discovery of a new boson, with a mass close to $125~{\rm GeV}$, by the ATLAS~\cite{ATLAS:1} and CMS~\cite{CMS:1} collaborations stands as a remarkable success of the Standard Model~(SM) of particle physics. The measured properties of this boson are so far in agreement with those of the SM Higgs~\cite{ATLAS:coupling,CMS:coupling,Aaltonen:2013kxa}, suggesting that the electroweak symmetry breaking~(EWSB) is probably realized in the most elegant and simple way, \textit{i.e.}, via the Higgs mechanism implemented through one scalar doublet. It is, however, noted that none of the fundamental principles of the SM forbids the possibility of an enlarged scalar sector associated with the EWSB. A natural question we are now facing is then whether the discovered state corresponds to the unique Higgs boson predicted by the SM, or it is just the first signal of a much richer scenario of EWSB.

Among the many possible scenarios for new physics~(NP) beyond the SM, the two-Higgs-doublet model~(2HDM)~\cite{Lee:1973iz} provides a minimal extension of the SM, by adding a second scalar doublet to the SM field content. It can easily accommodate the electroweak~(EW) precision tests and give rise, at the same time, to a very rich phenomenology~\cite{Branco:2011iw}. As a potential theory of nature, the 2HDM is very interesting on its own, since it allows for CP violation beyond what is provided by the SM. It is also helpful to gain further insights into the scalar sector of supersymmetry and other models that contain similar scalar contents. The direct search for the scalar spectrum of the model at high-energy collisions or through indirect constraints via precision flavour experiments are, therefore, an important task for the next years.

Within the SM, the flavour-changing neutral currents~(FCNCs) are forbidden at tree level and, due to the Glashow-Iliopoulos-Maiani~(GIM) mechanism~\cite{GIM}, are highly suppressed at loop level. In the most general version of 2HDM, however, unwanted FCNCs appear even at tree level, which represents a major shortcoming of the model. The hypothesis of natural flavour conservation~(NFC) is the usual way out to this issue. By limiting the number of scalar doublets coupling to a given type of right-handed fermion to be at most one, the absence of dangerous FCNCs is guaranteed~\cite{Glashow:1976nt}. This can be explicitly implemented via a discrete $\mathcal{Z}_2$ symmetry acting differently on the two scalar doublets, leading to four types of 2HDM~(usually named as type-I, II, X and Y models), which has been studied extensively for many years~\cite{Branco:2011iw}.

Another efficient way to guarantee the smallness of FCNCs is to impose the principle of minimal flavour violation~(MFV)~\cite{MFV:1,MFV:2,Buras:2010mh}, a concept that can be traced back to Refs.~\cite{Chivukula:1987py,Hall:1990ac} and has become very popular during the past decade. It amounts to assuming that all the flavour-violating interactions, including those mediated by the electrically neutral scalars, are controlled by the Cabibbo-Kobayashi-Maskawa~(CKM) matrix~\cite{Cabibbo:1963yz,Kobayashi:1973fv}, as happens in the SM. This can be implemented by requiring all the scalar Yukawa couplings be composed of the SM ones $Y^{U}$ and $Y^{D}$. It has been shown in Refs.~\cite{Manohar:2006ga,Arnold:2009ay} that, under the MFV hypothesis, the allowed $SU(3)_C \otimes SU(2)_L \otimes U(1)_Y$ representations of the second scalar that couples to quarks via Yukawa interactions are fixed to be either $(\mathbf{1}, \mathbf{2})_{1/2}$ or $(\mathbf{8}, \mathbf{2})_{1/2}$; namely, the second scalar can be either colour-singlet or colour-octet. Examples of the former include the aligned 2HDM~(A2HDM)~\cite{Pich:2009sp} and the four types of 2HDM reviewed in Ref.~\cite{Branco:2011iw}. In the following, we shall refer to the 2HDM with the second scalar being colour-octet as the Manohar-Wise~(MW) model~\cite{Manohar:2006ga}. A characteristic feature of the MW model is that its scalar spectrum contains, besides a CP-even and colour-singlet Higgs boson~(the usual SM one), four colour-octet particles~(one CP-even, one CP-odd and two electrically charged), giving rise to many interesting phenomena~\cite{Manohar:2006ga,He:2013tla,Li:2013vlx,Degrassi:2010ne,Heo:2008sr,Gresham:2007ri,
Carpenter:2011yj}.

In this paper, motivated by the latest experimental data on $B_{s,d}-\bar{B}_{s,d}$ mixings and $B_{s,d}\to\mu^+\mu^-$ decays~\cite{Agashe:2014kda,Amhis:2014hma}, we shall perform a detailed study of these processes within the MW model. While the $B_{s,d}-\bar{B}_{s,d}$ mixings have already been addressed within the model~\cite{Manohar:2006ga}, it should be noted that the Wilson coefficient~(Eq.~(26) in \cite{Manohar:2006ga}) is obtained only in the limit of heavy charged colour-octet scalars and with zero external momenta. In a more general case for the model parameters, contributions from the other operators become non-negligible and, in order to get a gauge-independent result, the box diagrams should be calculated by keeping the external momenta up to the second order. The same observation is also applied to the rare leptonic $B_{s,d}\to\ell^+\ell^-$ decays, as demonstrated in Ref.~\cite{Li:2014fea}. The current experimental data on the averaged time-integrated branching ratios of $B_{s,d}\to\mu^+\mu^-$ decays, averaged over the CMS~\cite{Chatrchyan:2013bka} and LHCb~\cite{Aaij:2013aka} measurements, read~\cite{CMS:2014xfa}
\begin{align}\label{Bsd2mumu:data}
\overline{ {\mathcal B}}({{{B}}_{s}}\to \mu^+\mu^-)=(2.8_{-0.6}^{+0.7})\times 10^{-9}\,, \qquad \qquad \overline{ {\mathcal B}}({{{B}}_{d}}\to \mu^+\mu^-)=(3.9^{+1.6}_{-1.4})\times 10^{-10}\,,
\end{align}
which are in remarkable agreement with the latest updated predictions within the SM~\cite{Bobeth:2013uxa}
\begin{equation} \label{eq:BqmmSM}
 \overline{\mathcal{B}}({{{B}}_{s}}\to \mu^+\mu^-) = (3.65 \pm 0.23) \times 10^{-9}\,, \qquad \overline{\mathcal{B}}({{{B}}_{d}}\to \mu^+\mu^-) = (1.06 \pm 0.09) \times 10^{-10}\,.
\end{equation}
All the experimental and theoretical progresses will lead to more stringent constraints on physics beyond the SM. Combining these processes with the interesting $B\to X_s\gamma$, $B\to K^{\ast}\gamma$, $B\to\rho\gamma$, and $Z\to b \bar{b}$ decays~\cite{Li:2013vlx,Degrassi:2010ne}, we shall investigate the combined constraints on the model parameters from the current flavour data. In addition, the future sensitivity to the model is also explored in the observables achievable with $50~{\rm fb}^{-1}$ of LHCb~\cite{Bediaga:2012py} and $50~{\rm ab}^{-1}$ of Belle-II data~\cite{Aushev:2010bq}~(the so-called ``stage II" projection introduced in Ref.~\cite{Charles:2013aka}).

The outline of this paper is as follows. In Sec.~\ref{sec:MWmodel} we give a brief review of the MW model. In Sec.~\ref{sec:framework} we summarize the theoretical framework used to calculate the flavour observables both within the SM and in the MW model. A complete one-loop computation of the relevant Wilson coefficients for $B_{s,d}-\bar{B}_{s,d}$ mixings and $B_{s,d}\to\ell^+\ell^-$ decays in the MW model is also described in this section. Detailed numerical results and discussion are then presented in Sec.~\ref{sec:result}. Our conclusions are finally made in Sec.~\ref{sec:conclusion}. Explicit analytical results for the Wilson coefficients relevant to $B_{s,d}-\bar{B}_{s,d}$ mixings are given in Appendix~\ref{appendix:1}.

\section{The Manohar-Wise model}
\label{sec:MWmodel}

In the MW model~\cite{Manohar:2006ga}, the scalar sector of the SM is supplemented with a colour-octet and weak-doublet scalar. The two scalar fields in the model can be parametrized by
\begin{align} \label{eq:scalars}
H=\begin{pmatrix}
  \omega^+\\
  \frac1{\sqrt2}(v+h-iz^0)
  \end{pmatrix}, \qquad\quad
S^A=\begin{pmatrix}
    {S^{A}_{+}}\\
    \frac1{\sqrt2}(S^A_R+i S^A_I)
    \end{pmatrix},
\end{align}
with $v=(\sqrt{2} G_F )^{-1/2} \simeq 246~{\rm GeV}$. Here $H$ is identified as the SM scalar doublet, with $\omega^{\pm}$ and $z^0$ being the three would-be Goldstone bosons, whereas $h$ is the SM Higgs. $S^A$ denotes the colour-octet and weak-doublet scalar, with $A$ being an adjoint colour index; $S^A_{\pm}$, $S^A_R$, and $S^A_I$ denote the electrically charged, neutral CP-even, and CP-odd colour-octet scalars, respectively.

\subsection{Yukawa couplings to fermions}

The Yukawa interactions of the two scalar fields with the SM fermions are given by~\cite{Manohar:2006ga,Degrassi:2010ne}
\begin{align} \label{eq:yukawa1}
- \mathcal{L}_Y = \bar Q_L^0(Y^d H + \bar{Y}^d S^A T^A) d_R^0
                 +\bar Q_L^0(Y^u \tilde{H} + \bar{Y}^u \tilde{S}^A T^A) u_R^0
                 +\bar{L}_L^0 Y^\ell H e_R + \rm{h.c.},
\end{align}
where $\tilde{H}=i\sigma_2 H^{\ast}$ and $\tilde{S}^A=i\sigma_2 S^{A\ast}$, with $\sigma_2$ being the Pauli matrix. $Q_L^0$ and $L_L^0$ denote the left-handed quark and lepton doublets, and $u_R^0$, $d_R^0$ and $e_R^0$ the right-handed up-type quark, down-type quark and lepton singlets, respectively, in the weak interaction basis. The fundamental representation in colour space of the $SU(3)_C$ generators $T^A~(A=1,\cdots,8)$ acts on the quark fields, and determines the colour nature of the second scalar doublet.

All fermionic fields in Eq.~(\ref{eq:yukawa1}) are written as 3-dimensional flavour vectors and the Yukawa couplings $Y^f$ and $\bar{Y}^f$~($f=u,d,\ell$) are, therefore, general $3\times3$ complex matrices in flavour space. According to the MFV hypothesis, the matrices $\bar{Y}^{u,d}$ should be composed of the combination of $Y^{u}$ and $Y^{d}$, and transform under the $SU(3)_{Q_L}\otimes SU(3)_{U_R} \otimes SU(3)_{D_R}$ flavour symmetry in the same way as $Y^{u,d}$ themselves~\cite{Manohar:2006ga}. This can be achieved by requiring the alignment in flavour space of the Yukawa matrices~\cite{Pich:2009sp}
\begin{align} \label{eq:alignedcondition}
\bar{Y}^d=\eta_d\, Y^d, \qquad \bar{Y}^u=\eta_u^\ast\, Y^u,
\end{align}
where the two parameters $\eta_f$~($f=u,d$) are generally arbitrary complex numbers. The alignment condition also guarantees automatically the absence of tree-level FCNCs~\cite{Pich:2009sp,Bai:2012ex,Altmannshofer:2012ar}. Applying the SM unitary transformations to rotate the fermionic fields from the interaction to the mass-eigenstate basis, one can finally obtain the Yukawa interaction of physical colour-octet scalars with quarks in the mass-eigenstate basis~\cite{Manohar:2006ga}
\begin{align} \label{wiseyukawaphy}
\mathcal{L}_Y = & -\sqrt{2}{{\eta}_{u}}\bar{u}_{R}^{i}\frac{m_{u}^{i}}{v}{{T}^{A}}u_{L}^{i}S_{0}^{A} - \sqrt{2}{{\eta }_{d}}\bar{d}_{L}^{i}{{T}^{A}}\frac{m_{d}^{i}}{v}d_{R}^{i}{S_{0}^{A}} \nonumber\\[0.2cm]
& + \sqrt{2}{{\eta}_{u}}\bar{u}_{R}^{i}\frac{m_{u}^{i}}{v}{{T}^{A}}{{V}_{ij}}d_{L}^{j}S_{+}^{A} - \sqrt{2}{{\eta }_{d}}\bar{u}_{L}^{i}\frac{m_{d}^{j}}{v}{{T}^{A}}{{V}_{ij}}d_{R}^{j}S_{+}^{A} + {\rm h.c.},
\end{align}
where $S_{0}^{A}=\frac{S^A_R+i S^A_I}{\sqrt2}$ and ${V}_{ij}$~($i=u,c,t$ and $j=d,s,b$) is the CKM matrix element.

\subsection{Scalar self-couplings and couplings to gauge bosons}

The most general and renormalizable scalar potential of the MW model is given by~\cite{Manohar:2006ga}
\begin{align} \label{pot}
V = & \frac{\lambda}{4}\left(H^{\dagger i} H_i-\frac {v^2}{2}\right)^2 + 2 m_S^2 {\rm Tr} S^{\dagger i} S_i + \lambda_1 H^{\dagger i} H_i {\rm Tr} S^{\dagger j} S_j + \lambda_2 H^{\dagger i} H_j {\rm Tr} S^{\dagger j} S_i \nonumber\\[0.2cm]
& + \Bigl[ \lambda_3  H^{\dagger i} H^{\dagger j} {\rm Tr} S_i S_j  + \lambda_4 H^{\dagger i} {\rm Tr} S^{\dagger j} S_j  S_i + \lambda_5 H^{\dagger i} {\rm Tr} S^{\dagger j} S_i  S_j + \text{h.c.}\Bigr] \nonumber\\[0.2cm]
& + \lambda_6 {\rm Tr} S^{\dagger i} S_i S^{\dagger j} S_j + \lambda_7 {\rm Tr} S^{\dagger i} S_j S^{\dagger j} S_i + \lambda_8 {\rm Tr} S^{\dagger i} S_i {\rm Tr} S^{\dagger j} S_j + \lambda_9 {\rm Tr} S^{\dagger i} S_j {\rm Tr} S^{\dagger j} S_i \nonumber\\[0.2cm]
& + \lambda_{10} {\rm Tr} S_i   S_j {\rm Tr} S^{\dagger i} S^{\dagger j} + \lambda_{11} {\rm Tr} S_i  S_j  S^{\dagger j} S^{\dagger i},
\end{align}
where the $SU(2)_L$ indices on the scalar doublets have been explicitly displayed. Traces are taken over the colour indices and the notation $S=S^AT^A$ is used. Due to the Hermiticity of the scalar potential, all the parameters are real except for $\lambda_4$ and $\lambda_5$~($\lambda_3$ has been made real by a phase rotation of the $S$ fields)~\cite{Manohar:2006ga}. From this potential, one can easily obtain the trilinear coupling involving two charged colour-octet scalars and the SM Higgs boson $h$
\begin{align}
g_{S^A_+ S^B_- h}=-\frac{v}{2}\lambda_1 \delta^{AB}.
\end{align}
which is the only cubic coupling relevant in our calculation.

The scalar couplings to the gauge bosons arise from the gauge-kinetic Lagrangian
\begin{align}
\mathcal{L}_{kin} = {\left({D_{\mu}}H\right)}^{\dagger}D^{\mu}H + \left({D_{\mu}}S^A\right)^{\dagger}D^{\mu}S^A,
\end{align}
where $D^{\mu}$ denotes the SM gauge-covariant derivative. From this Lagrangian, one can easily get the trilinear coupling of two charged colour-octet scalars with the $Z$ boson
\begin{align}
g_{S^A_+ S^B_- Z^{\mu}}= g \frac{\cos 2\theta_W}{2\cos \theta_W} \left(p_{S^B_-}^{\mu} -p_{S^A_+}^{\mu}\right) \delta_{AB},
\end{align}
where $g$ is the $SU(2)_L$ gauge coupling constant, and $\theta_W$ the weak mixing angle. $p_{S^A_+}^{\mu}$ and $p_{S^B_-}^{\mu}$ denote, respectively, the incoming momenta of the two charged colour-octet scalars.

As all the other couplings are irrelevant to our calculation, they are not mentioned here. It is, however, noted that all the Feynman rules in the MW model have been cross-checked with the package \texttt{FeynRules}~\cite{Christensen:2008py}.

\section{Theoretical framework for flavour observables}
\label{sec:framework}

\subsection{\texorpdfstring{$\mathbf{\boldsymbol{B_{s,d}-\bar B_{s,d}}}$}{Lg} mixings}
\label{sec:Bmixing}

\begin{figure}[t]
\centering
\includegraphics[width=0.24\textwidth]{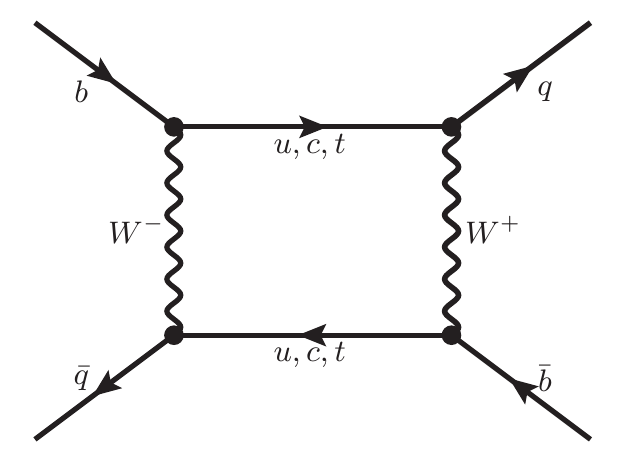}~
\includegraphics[width=0.24\textwidth]{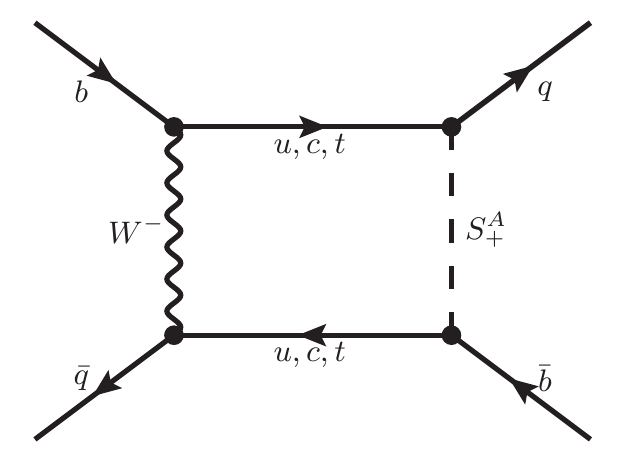}~
\includegraphics[width=0.24\textwidth]{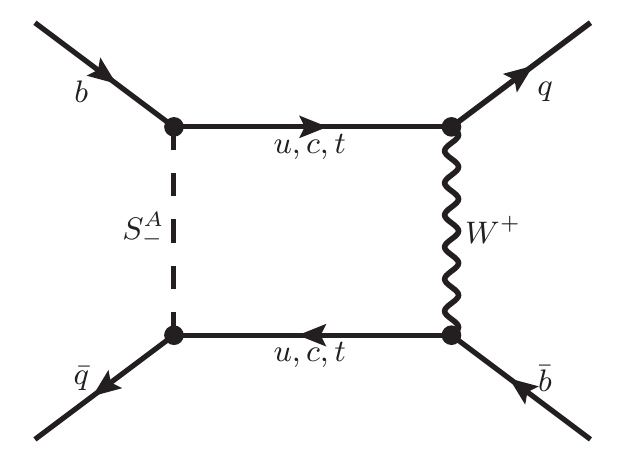}~
\includegraphics[width=0.24\textwidth]{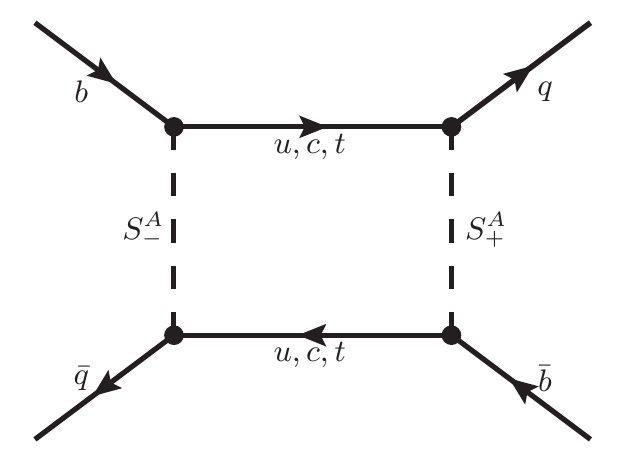}
\caption{\small Box diagrams for the $B_q-\bar B_q$ mixing in the unitary gauge both within the SM (the first one)and in the MW model (the last three). Crossed diagrams have also been taken into account.}
\label{LOboxen}
\end{figure}

The B-meson mixings are governed by the SM box diagrams with exchanges of up-type quarks and $W^{\pm}$ bosons, and the box diagrams mediated by up-type quarks and charged colour-octet scalars in the MW model, as shown in Fig.~\ref{LOboxen}. The resulting effective weak Hamiltonian reads~\cite{Buras:2001ra}
\begin{align}
\mathcal{H}^{|\Delta B|=2}_{\rm eff} = \frac{G_{F}^2}{16 \pi^2}\,m_{W}^{2}\, \sum\limits_{i}\tilde{C}_{i}(\mu)Q_i + \text{h.c.}.
\end{align}
Here $G_F$ is the Fermi coupling constant, and $\tilde{C}_{i}(\mu)=V_{\rm CKM}^{i}C_{i}(\mu)$ with $V_{\rm CKM}^{i}$ being the CKM factor and $C_{i}(\mu)$ the scale-dependent Wilson coefficients. When QCD renormalization group effects are taken into account, there are totally $8$ operators responsible for the $B_{s,d}-\bar B_{s,d}$ mixings, which can be split further into $5$ separate sectors according to the chirality of the quark fields they contain ($q=d$ or $s$)~\cite{Buras:2001ra,Buras:2000if}:
\begin{align}
&Q_1^{VLL} = (\bar{q}^{\alpha} \gamma_{\mu}    P_L b^{\alpha})
                  (\bar{q}^{ \beta} \gamma^{\mu}    P_L b^{ \beta}),
\nonumber \\[0.2cm]
&Q_1^{LR} =  (\bar{q}^{\alpha} \gamma_{\mu}    P_L b^{\alpha})
                  (\bar{q}^{ \beta} \gamma^{\mu}    P_R b^{ \beta}), \hspace{4mm}
 Q_2^{LR} =  (\bar{q}^{\alpha}                 P_L b^{\alpha})
                  (\bar{q}^{ \beta}                 P_R b^{ \beta}),
\nonumber\\[0.2cm]
&Q_1^{SLL} = (\bar{q}^{\alpha}                 P_L b^{\alpha})
                  (\bar{q}^{ \beta}                 P_L b^{ \beta}), \hspace{4mm}
 Q_2^{SLL} = (\bar{q}^{\alpha} \sigma_{\mu\nu} P_L b^{\alpha})
                  (\bar{q}^{ \beta} \sigma^{\mu\nu} P_L b^{ \beta}),
\label{normal}
\end{align}
where $\alpha, \beta$ are the colour indices, $\sigma_{\mu\nu}= \frac{1}{2}[\gamma_{\mu},\gamma_{\nu}]$ and $P_{L,R}=\frac{1}{2}(1\mp \gamma_5)$. The remaining two sectors ($Q_1^{VRR}$ and $Q_i^{SRR}$) are obtained from $Q_1^{VLL}$ and $Q_i^{SLL}$ by interchanging $P_L$ and $P_R$.

Within the SM, only the operator $Q_1^{VLL}$ contributes and the corresponding Wilson coefficient up to next-to-leading order (NLO) is given by~\cite{Buras:1990fn,Urban:1997gw}
\begin{align}
\tilde{C}_{VLL}^{SM}(\mu_W) \;=\; 4\lambda_t^2\,\left[S_{WW}(x_t)+\frac{\alpha_s (\mu_W)}{4\pi}D_{SM}(x_t,x_{\mu_W})\right]\,,
\label{bsbsmixingcvllsm}
\end{align}
with $\lambda_t = V_{tb}V_{tq}^{*}$, $x_t={m_t^2}/{m_W^2}$ and $x_{\mu_W}={\mu_W^2}/{m_W^2}$. The leading order (LO) coefficient $S_{WW}(x_t)$ is the known Inami-Lim function~\cite{Inami:1980fz}, and the NLO coefficient $D_{SM}(x_t,x_{\mu_W})$ reads~\cite{Urban:1997gw}
\begin{align}
D_{SM}(x_t,x_{\mu_W}) \;=\; & C_F \Biggl\{ L^{(1,SM)}(x_t) + \biggl[ 6 \ln (x_{\mu_W}) \, \left( x_t {\frac{\partial}{\partial x_t}} \right) + 3 \biggr] \, S_{WW}(x_t) \Biggr\} \nonumber\\[0.2cm]
& + C_A  \Biggl\{ L^{(8,SM)}(x_t) + \biggl[ 6 \ln (x_{\mu_W})  +5 \biggr] \, S_{WW}(x_t) \Biggr\} \,,
\label{bsbsmixingcvllsmDfunction}
\end{align}
where $C_F=4/3$ and $C_A=1/3$. Explicit expressions for $S_{WW}(x_t)$, $L^{(1,SM)}(x_t)$ and $L^{(8,SM)}(x_t)$ can be found in Refs.~\cite{Buras:1990fn,Urban:1997gw}. Note that the SM box diagrams are evaluated with zero external momenta, which is a good approximation for $B_{s,d}-\bar B_{s,d}$ mixings within the SM~\cite{Urban:1997cm}.

The additional contribution to the effective weak Hamiltonian arises from the SM box diagrams with the $W^{\pm}$ bosons replaced by the charged colour-octet scalars (the last three ones in Fig.~\ref{LOboxen}). These diagrams depend on the two parameters $\eta_u$ and $\eta_d$, which characterise the Yukawa couplings of charged colour-octet scalars to quarks. For most general values of these parameters, especially when $\eta_d/\eta_u \simeq m_t/m_b$, each term of the second line in Eq.~\eqref{wiseyukawaphy} can give a comparable contribution and should be, therefore, taken into account simultaneously. In such a specific case, in order to obtain a gauge-independent result, it is necessary to keep at least the external $b$-quark momenta up to the second order~\cite{Li:2014fea}. This is contrary to the SM case where all the momenta of external quarks can be safely set to zero.

Following the same computational procedure as used in Ref.~\cite{Li:2014fea}, we set the light-quark masses $m_{d,s}$ to zero; while for the $b$-quark mass $m_b$, we keep it up to the second order. As the external momenta are much smaller than the masses of internal propagators, the Feynman integrands are firstly expanded in external momenta before performing the loop integration~\cite{Smirnov:1994tg}
\begin{equation} \label{eq:HME}
 \frac{1}{(k+l)^2-M^2} = \frac{1}{k^2-M^2}\,\left[1 - \frac{l^2+2(k\cdot l)}{k^2-M^2} + \frac{4(k\cdot l)^2}{(k^2-M^2)^2}\right]+\mathcal{O}(l^4/M^4)\,,
\end{equation}
where $M$ denotes a heavy mass, $k$ is the loop momentum and $l$ an arbitrary external momentum. After factorizing out the external momenta in Eq.~\eqref{eq:HME}, we are left with integrals that depend only on the loop momentum $k$ and the heavy masses $m_i$. They can be further reduced, with the help of partial fraction decomposition~\cite{Smirnov:2004ym}
\begin{equation} \label{eq:PFD}
 \frac{1}{(q^2-m_1^2)(q^2-m_2^2)} = \frac{1}{m_1^2-m_2^2}\,\left[\frac{1}{q^2-m_1^2} - \frac{1}{q^2-m_2^2}\right]\,,
\end{equation}
to the ones in which only a single mass occurs in the propagator denominators. Finally, after the tensor reduction~\cite{Passarino:1978jh}, the only non-vanishing one-loop scalar integrals take the form~\cite{Peskin:1995ev}
\begin{equation} \label{eq:scalarint}
 \int \frac{d^D k}{(2\pi)^D}\,\frac{1}{(k^2-m^2)^n} = \frac{(-1)^n i}{(4\pi)^{D/2}}\,\frac{\Gamma(n-D/2)}{\Gamma(n)}\, \left(\frac{1}{m^2}\right)^{n-D/2}\,,
\end{equation}
with an arbitrary integer power $n$ and with $m\neq0$. In addition, the naive dimensional regularization scheme is employed to regularize the divergences appearing in Feynman integrals.

The LO non-vanishing Wilson coefficients in the MW model at the matching scale $\mu_S\sim \mathcal{O}(m_{S^A_{+}})$, where heavy degrees of freedom are integrated out, are given by
\begin{align}
\tilde{C}_{VLL}^{NP}(\mu_S)&=\lambda_t^2 C_{VLL}^{NP,tt}(\mu_S)+2 \lambda_c \lambda_t C_{VLL}^{NP,ct}(\mu_S)+\lambda_c^2 C_{VLL}^{NP,cc}(\mu_S), \label{bsbsmixingcvllsm:1} \\[0.2cm]
\tilde{C}_{SRR,1}^{NP}(\mu_S)&=\lambda_t^2 C_{SRR,1}^{NP,tt}(\mu_S)+2 \lambda_c \lambda_t C_{SRR,1}^{NP,ct}(\mu_S)+\lambda_c^2 C_{SRR,1}^{NP,cc}(\mu_S), \label{bsbsmixingcvllsm:2} \\[0.2cm]
\tilde{C}_{SRR,2}^{NP}(\mu_S)&=\lambda_t^2 C_{SRR,2}^{NP,tt}(\mu_S)+2 \lambda_c \lambda_t C_{SRR,2}^{NP,ct}(\mu_S)+\lambda_c^2 C_{SRR,2}^{NP,cc}(\mu_S),
\label{bsbsmixingcvllsm:3}
\end{align}
where $\lambda_c = V_{cb}V_{cq}^{*}$ and the explicit expressions for $C_{VLL}^{NP,ij}$, $C_{SRR,1}^{NP,ij}$ and $C_{SRR,2}^{NP,ij}$ are listed in Appendix~\ref{appendix:1}. Here we have used the Fierz identities
\begin{align}
& (\bar{q} \gamma_{\mu} P_L T^A b)(\bar{q} \gamma^{\mu} P_L T^A b) = \frac{1}{3} \, Q_1^{VLL} \,, \\[0.2cm]
& (\bar{q} \gamma_{\mu} P_L T^A T^B b)(\bar{q} \gamma^{\mu} P_L T^B T^A b) = \frac{11}{18} \, Q_1^{VLL} \,, \\[0.2cm]
& (\bar{q} P_R T^A b)(\bar{q} P_R T^A b) = -\frac{5}{12} \, Q_1^{SRR} + \frac{1}{16} \, Q_2^{SRR} \,.
\end{align}
It should be noted that the limit $m_{u}\to 0$ and the unitarity relation $\lambda_u + \lambda_c + \lambda_t = 0$ have been implicitly exploited during the calculation. To make sure the gauge independence of our results, we have performed the calculation both in the Feynman and in the unitary gauge.

Together with the analytical formulae for the QCD renormalization group factors given in Ref.~\cite{Buras:2001ra}, one can obtain the corresponding Wilson coefficients at the lower scale $\mu_b\sim \mathcal{O}(m_b)$. The effective weak Hamiltonian for $|\Delta B|=2$ transitions in the MW model reads
\begin{align}
\mathcal{H}^{|\Delta B|=2}_{\rm eff} &= \frac{G_{F}^2}{16 \pi^2}m_{W}^{2} \bigg\{ \left[\tilde{C}_{VLL}^{SM}(\mu_b)+\tilde{C}_{VLL}^{NP}(\mu_b)\right]Q_1^{VLL} + \tilde{C}_{SRR,1}^{NP}(\mu_b)Q_1^{SRR} \nonumber \\[0.2cm]
& \hspace{2.2cm} +\tilde{C}_{SRR,2}^{NP}(\mu_b)Q_2^{SRR} \bigg\} + \text{h.c.}\,,
\label{heffdeltf2}
\end{align}
from which the off-diagonal matrix element for $B_q-\bar B_q$ mixing is derived as
\begin{align}
\langle B_q |\mathcal{H}^{|\Delta B|=2}_{\rm eff} |\bar{B}_q \rangle &= \frac{G_{F}^2}{16 \pi^2}m_{W}^{2} \biggl\{ \left[\tilde{C}_{VLL}^{SM}(\mu_b)+\tilde{C}_{VLL}^{NP}(\mu_b)\right]\, \langle B_q | Q_1^{VLL}(\mu_b)|\bar{B}_q \rangle \nonumber\\[0.2cm]
& \hspace{0.2cm} + \tilde{C}_{SRR,1}^{NP}(\mu_b)\, \langle B_q |Q_1^{SRR}(\mu_b)|\bar{B}_q \rangle +\tilde{C}_{SRR,2}^{NP}(\mu_W)\, \langle B_q |Q_2^{SRR}(\mu_b)|\bar{B}_q \rangle \biggr\}\,,
\label{heffmatrixelement}
\end{align}
with the operator matrix elements $\langle B_q |Q_i(\mu_b)|\bar{B}_q \rangle$ given, respectively, by~\cite{Buras:2001ra}
\begin{align}
\langle B_q | Q_1^{VLL}(\mu_b)|\bar{B}_q \rangle &=\frac{1}{3} \, m_{B_q} \, f_{B_q}^2 \, B_1^{VLL}(\mu_b), \\[0.2cm]
\langle B_q |Q_1^{SRR}(\mu_b)|\bar{B}_q \rangle &=-\frac{5}{24} \left(\frac{m_{B_q}}{m_b(\mu_b)+m_q (\mu_b)}\right)^2 m_{B_q} \, f_{B_q}^2 \, B_1^{SRR}(\mu_b),\\[0.2cm]
\langle B_q |Q_2^{SRR}(\mu_b)|\bar{B}_q \rangle &=-\frac{1}{2} \left(\frac{m_{B_q}}{m_b(\mu_b)+m_q (\mu_b)}\right)^2 m_{B_q} \, f_{B_q}^2 \, B_2^{SRR}(\mu_b).
\label{hadronmatrixelement}
\end{align}
Here $f_{B_q}$ is the $B_q$-meson decay constant, and $B_1^{VLL}$, $B_1^{SRR}$ and $B_2^{SRR}$ are the non-perturbative Bag parameters, whose values can be obtained in lattice QCD~\cite{Carrasco:2013zta}. The mass difference $\Delta M_{B_{q}^{0}}$ is then given by
\begin{align}
\Delta M_{B_q^{0}}=2\left|\langle B_q |\mathcal{H}^{|\Delta B|=2}_{\rm eff}|\bar{B}_q \rangle\right|,
\label{deltmbds}
\end{align}
which can be used to constrain the model parameters, taking into account the current precise experimental measurements~\cite{Agashe:2014kda,Amhis:2014hma}.

\subsection{\texorpdfstring{$\mathbf{\boldsymbol{B_{s,d}\rightarrow \ell^+ \ell^-}}$}{Lg} decays}
\label{sec:bsd2mumu}

The most general effective weak Hamiltonian for the purely leptonic $B_{s,d}\rightarrow \ell^+ \ell^-$ decays can be written as~\cite{Buras:2013uqa}
\begin{align}  \label{eq:Heff}
 {\cal H}_{\rm eff} \; =\; -\frac{G_F^2\,m_W^2}{\pi^2}\left[
		V_{tb}^{\phantom{*}} V_{tq}^* \, \sum_{i}^{10,S,P} \left( C_i\,{\cal O}_i + C'_i\,{\cal O}'_i\right) + \mathrm{h.c.}\right]\,,
\end{align}
with the corresponding operators given by
\begin{align} \label{eq:operators}
	{\cal O}_{10} &= (\bar q\gamma_\mu P_L b)\, (\bar{\ell} \gamma^\mu \gamma_5 \ell)\,, &
	{\cal O}'_{10} &= (\bar q\gamma_\mu P_R b)\, (\bar{\ell} \gamma^\mu \gamma_5 \ell)\,, \notag\\[0.2cm]	
	{\cal O}_{S} &= \frac{m_\ell m_b}{M^2_W}\; (\bar q P_R b)\, (\bar{\ell}\ell)\,, &
	{\cal O}'_{S} &= \frac{m_\ell m_b}{M^2_W}\; (\bar q P_L b)\, (\bar{\ell}\ell)\,, \notag\\[0.2cm]	
	{\cal O}_{P} &= \frac{m_\ell m_b}{M^2_W}\; (\bar q P_R b)\, (\bar{\ell} \gamma_5 \ell)\,, &
	{\cal O}'_{P} &= \frac{m_\ell m_b}{M^2_W}\; (\bar q P_L b)\, (\bar{\ell} \gamma_5 \ell)\,,
\end{align}
where $\ell=e,\mu,\tau$ and $q=d,s$. In this paper, we shall neglect contributions from the operators ${\cal O}'_i$, because they only give contributions proportional to the light-quark mass $m_q$. These rare processes are then governed only by the operators ${\cal O}_{10}$, ${\cal O}_{S}$ and ${\cal O}_{P}$ in our approximation.

\begin{figure}[t]
\centering
\includegraphics[width=0.24\textwidth]{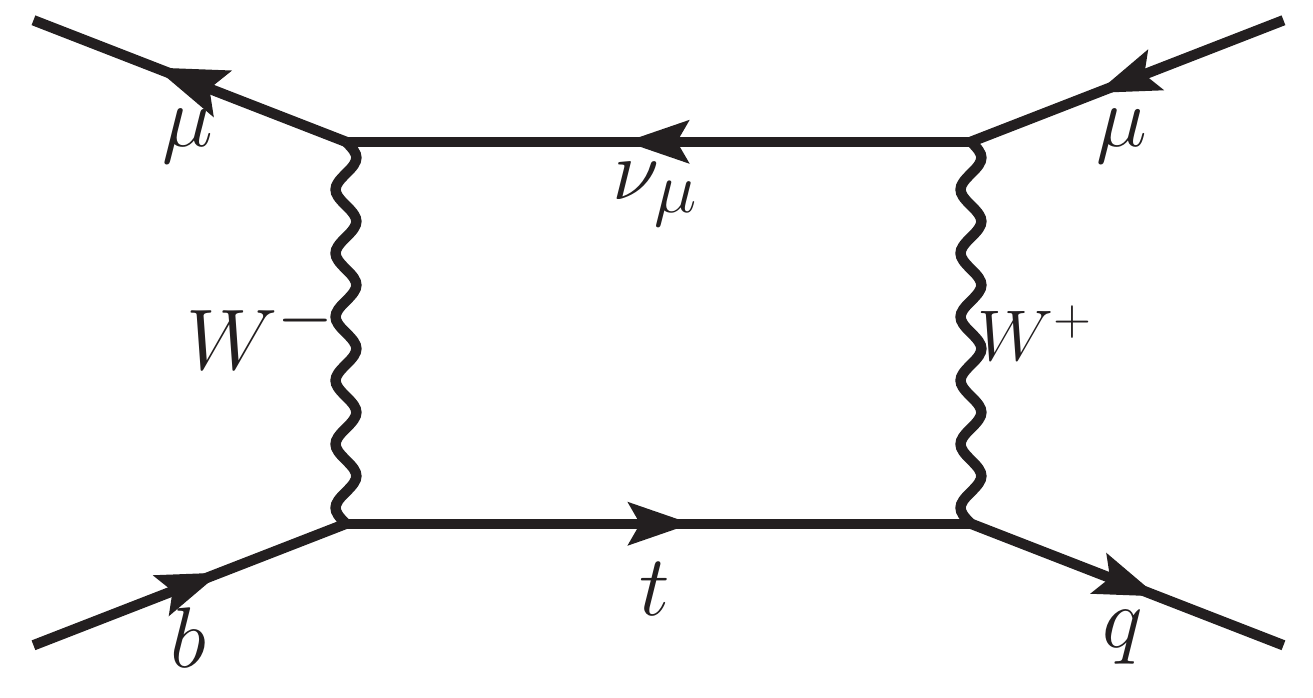}\quad \quad
\includegraphics[width=0.24\textwidth]{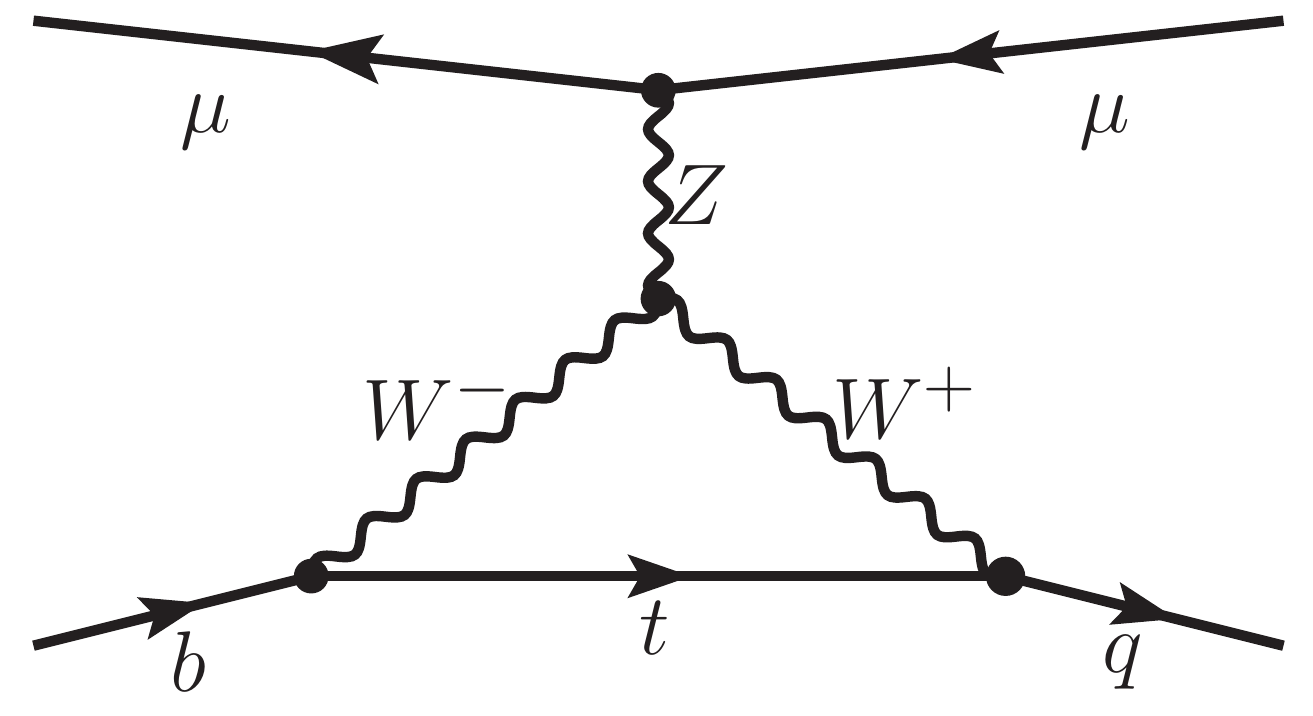}\quad \quad
\includegraphics[width=0.24\textwidth]{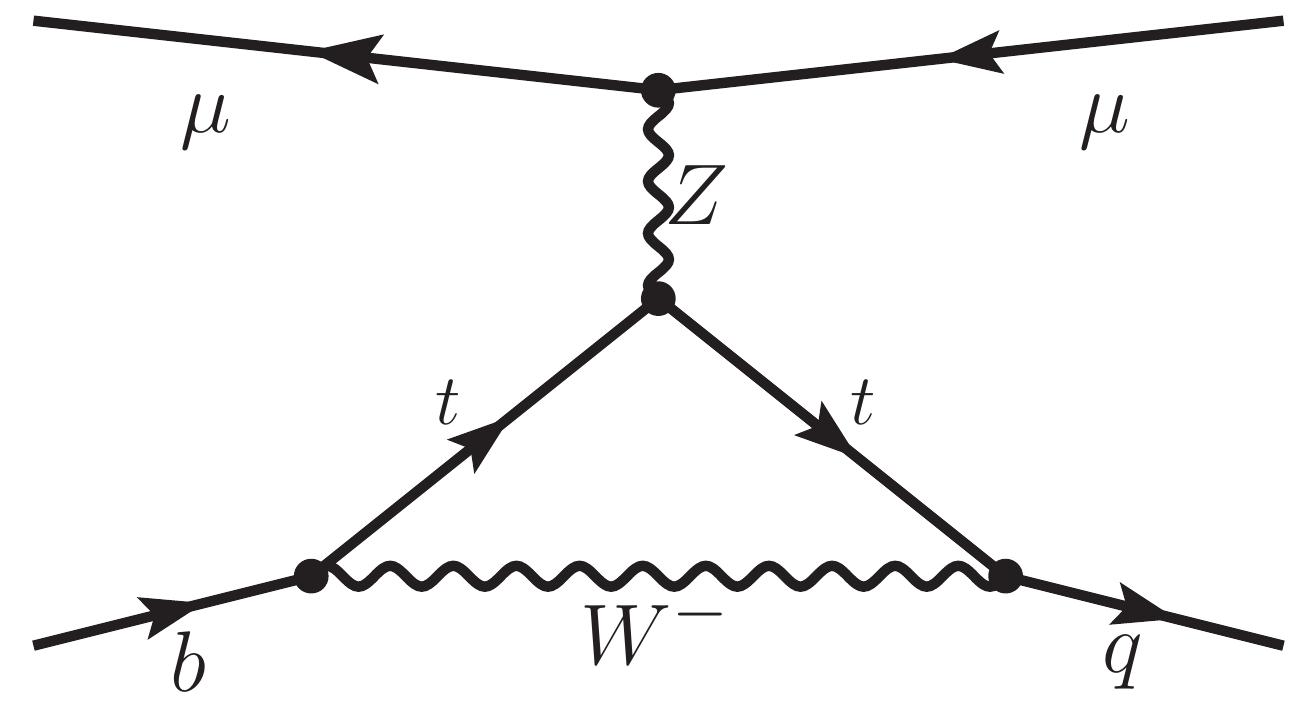}\\
\vskip 0.6cm
\includegraphics[width=0.24\textwidth]{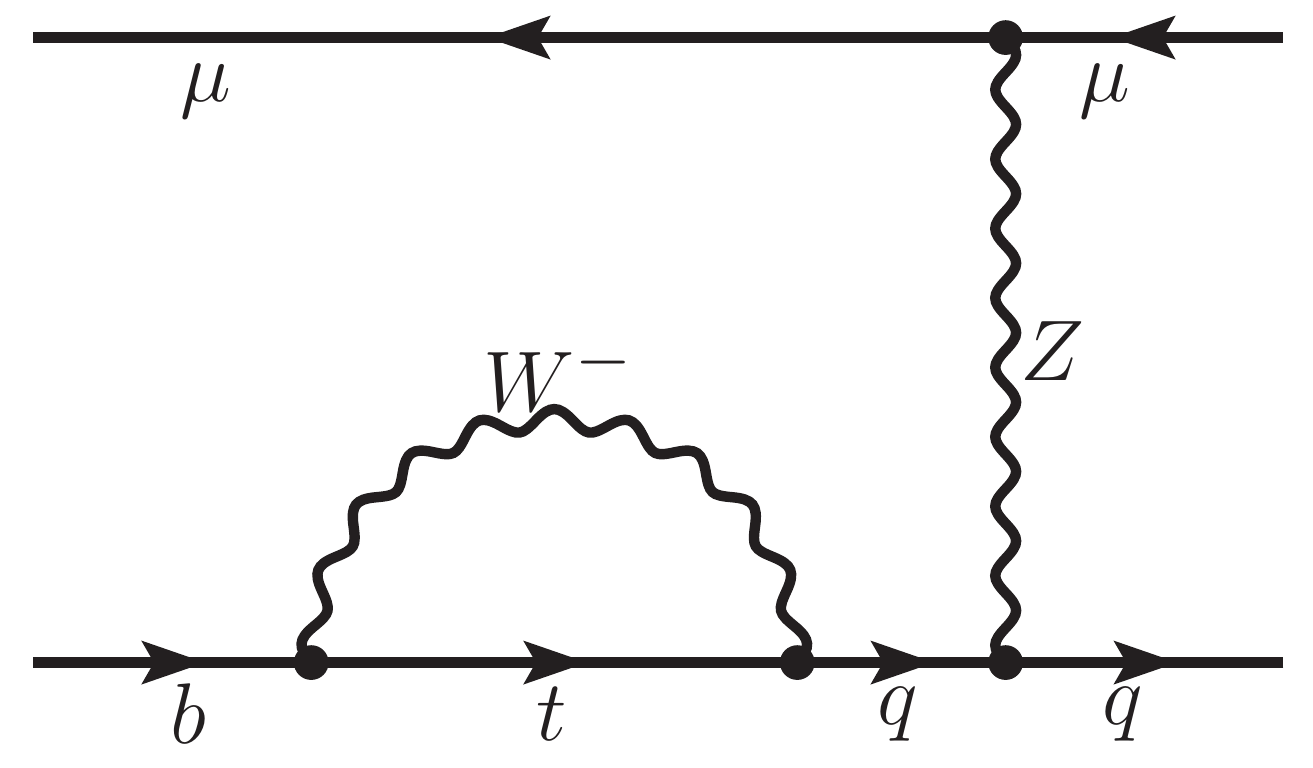}\quad \quad
\includegraphics[width=0.24\textwidth]{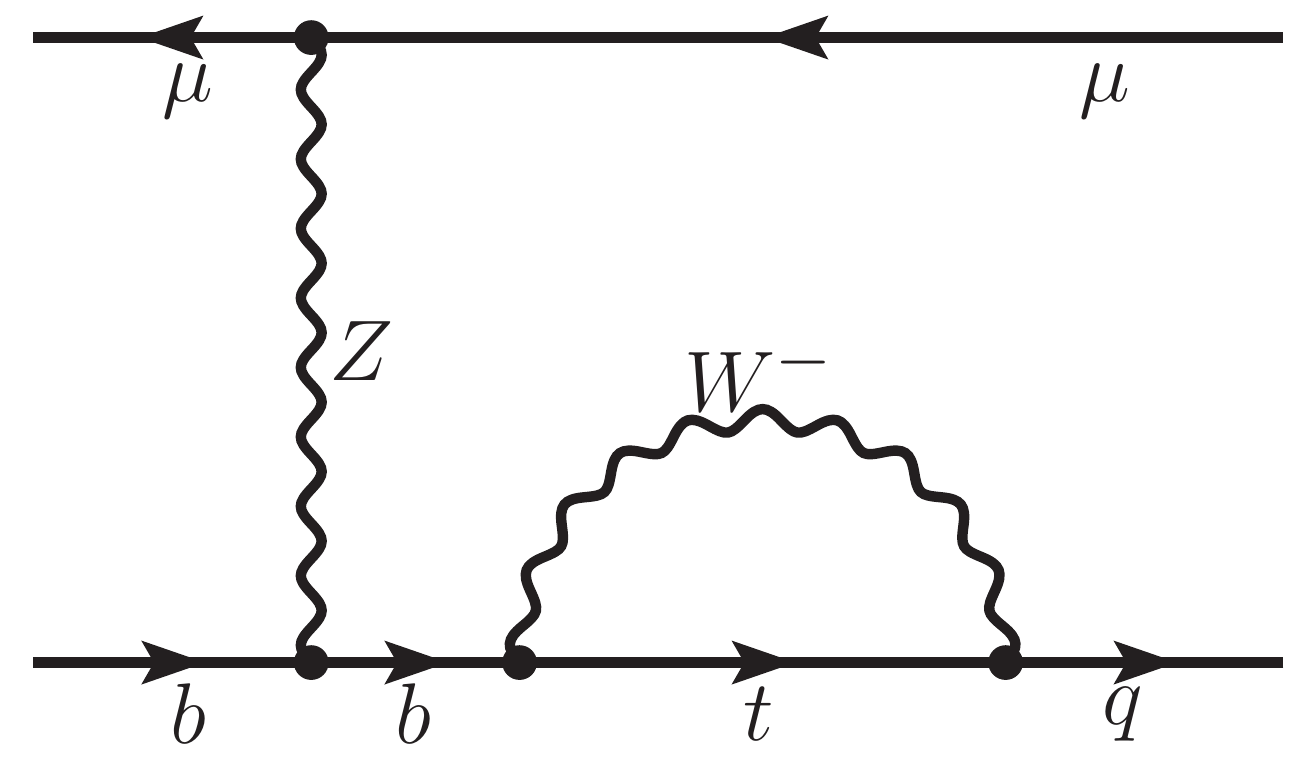}\quad \quad
\includegraphics[width=0.24\textwidth]{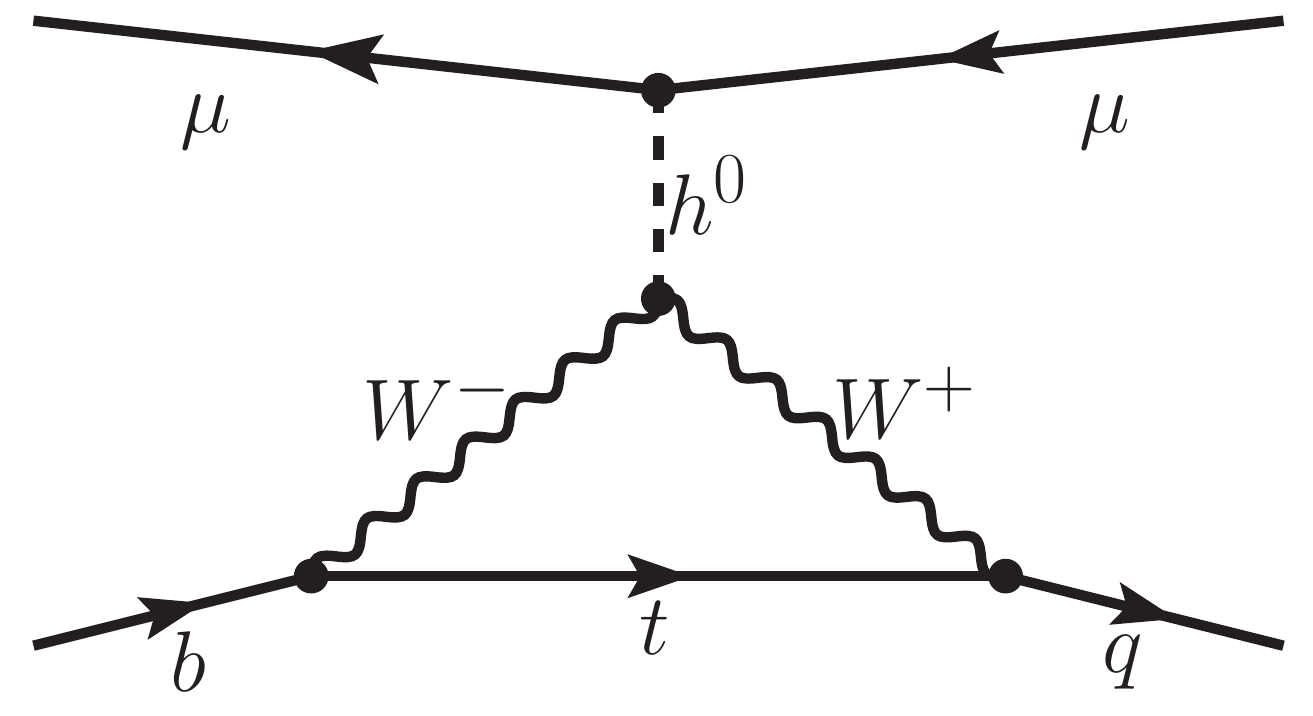}\\
\vskip 0.6cm
\includegraphics[width=0.24\textwidth]{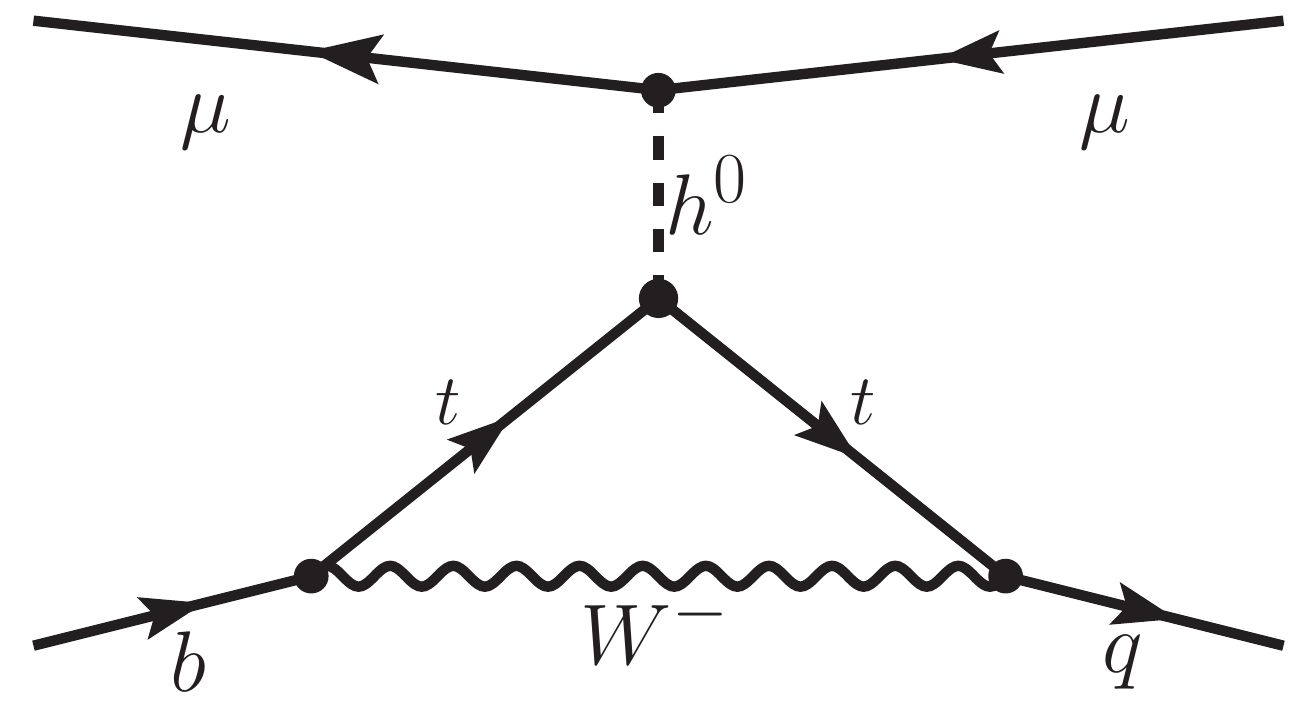}\quad \quad
\includegraphics[width=0.24\textwidth]{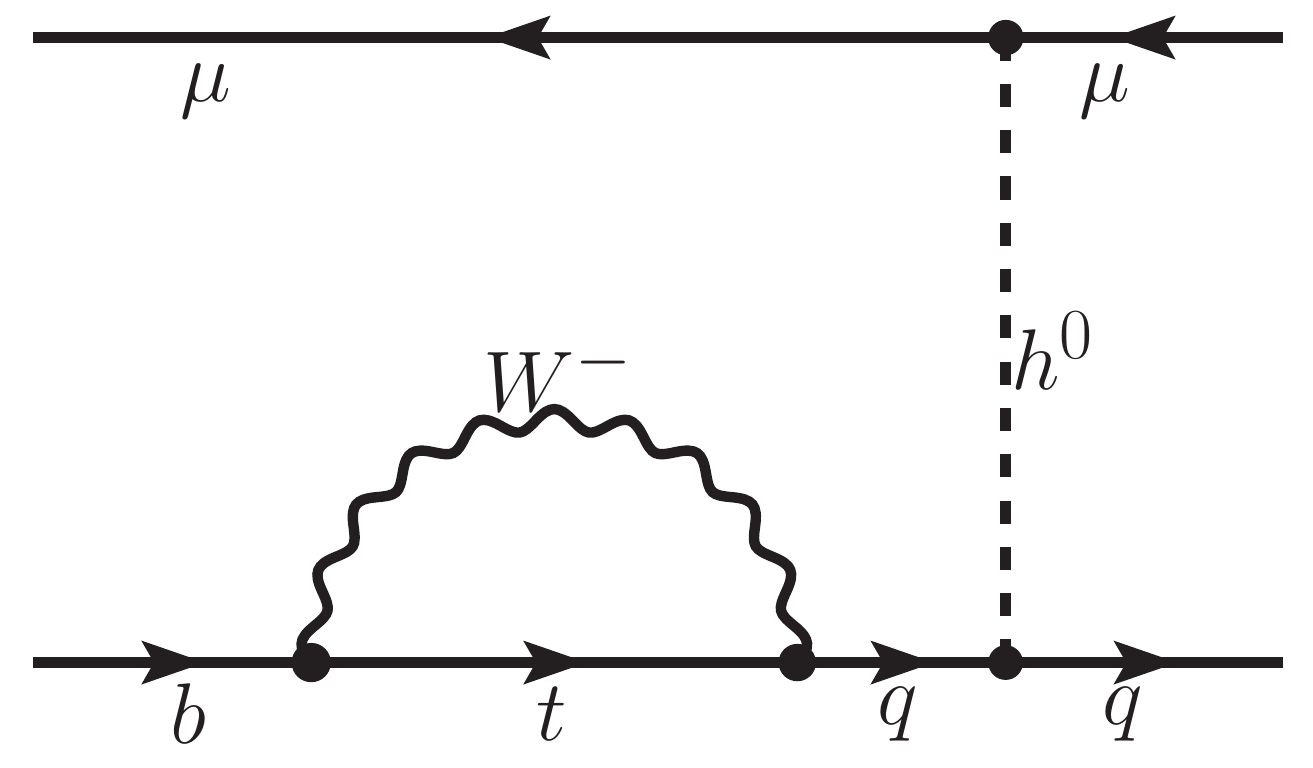}\quad \quad
\includegraphics[width=0.24\textwidth]{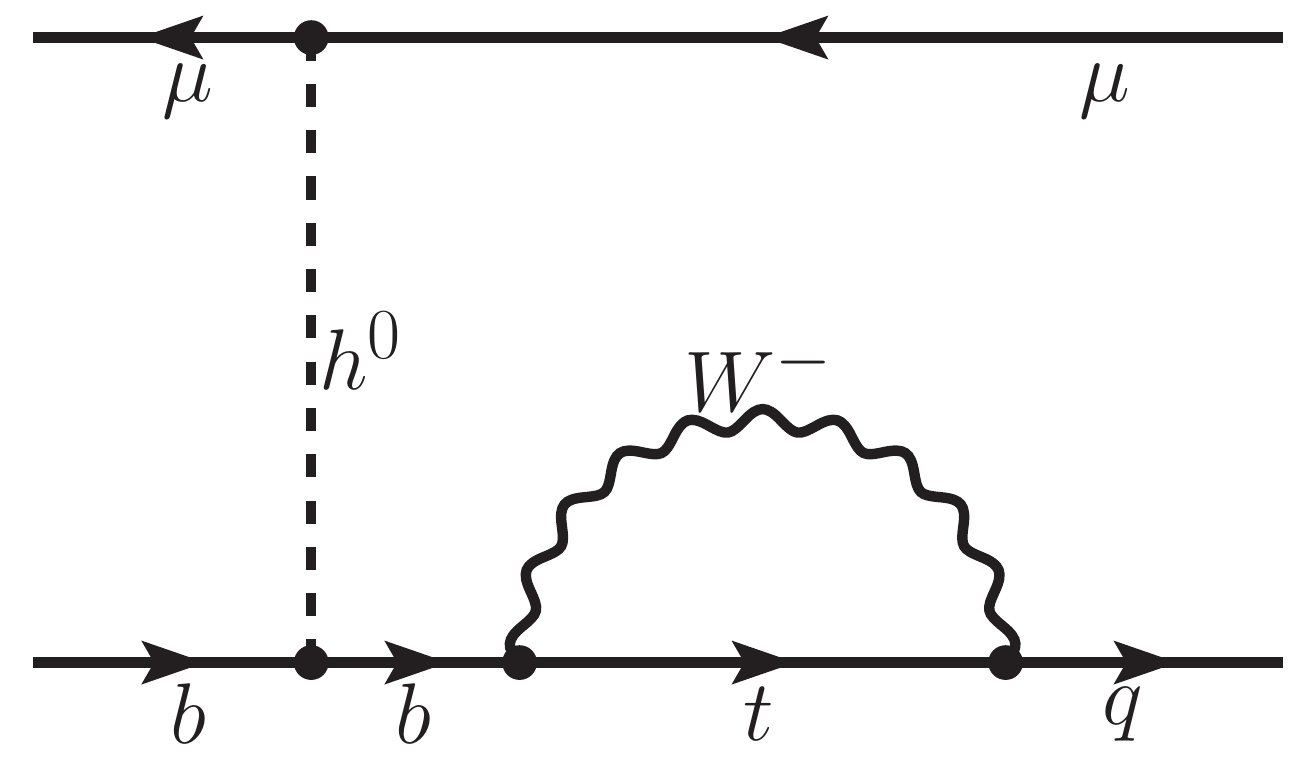}
\caption{\small SM Feynman diagrams contributing to the rare $B_q\to \mu^+ \mu^-$ decays in the unitary gauge.}
\label{btomumusm}
\end{figure}

Within the SM, the dominant contributions to the rare $B_{s,d}\rightarrow \ell^+ \ell^-$ decays originate from the $W$-box and $Z$-penguin diagrams shown in Fig.~\ref{btomumusm}, which generate the Wilson coefficient
\begin{equation}
 C^{\rm{SM}}_{10} \;=\; - \eta_Y^{\rm EW}\,\eta_Y^{\rm QCD}\,Y_0(x_t)\,,
 \label{eq:C10SM}
\end{equation}
where
\begin{equation}\label{eq:Y0}
Y_0(x_t) \, =\, \frac{x_t}{8}\left[ \frac{x_t-4}{x_t-1} + \frac{3x_t}{(x_t-1)^2}\ln x_t \right]
\end{equation}
is the one-loop function calculated firstly in Ref.~\cite{Inami:1980fz}. The factor $\eta_Y^{\rm EW}$ accounts for both the NLO EW matching corrections~\cite{Bobeth:2013tba} and the logarithmically enhanced QED corrections originating from the renormalization group evolution~\cite{Bobeth:2013uxa,Hermann:2013kca}, while $\eta_Y^{\rm QCD}$ stands for the NLO~\cite{Buchalla:1998ba,Misiak:1999yg} and next-to-next-to-leading order (NNLO)~\cite{Hermann:2013kca} QCD corrections.

As detailed in Ref.~\cite{Li:2014fea}, the $W$-box and $Z$-penguin diagrams also give rise to the pseudoscalar coefficient $C_P$. On the other hand, the scalar coefficient $C_S$ is generated exclusively from the $W$-box and Higgs-penguin diagrams shown in Fig.~\ref{btomumusm}. The complete SM expressions for $C_P$ and $C_S$ in the unitary gauge are given, respectively, by~\cite{Li:2014fea}
\begin{align} \label{eq:cpSMandcssm}
C^{\rm SM}_{P} & = C^{\rm box,\, \rm SM}_{P, \rm unitary} +C^{\rm Z\, penguin,\, \rm SM}_{P, \rm\, unitary}
\nonumber \\[0.2cm]
&\;=\; \frac{1}{24}\,\left[\frac{x_t(36 x_t^3-203 x_t^2+352 x_t-209)}{6 (x_t-1)^3} + \frac{17 x_t^4-34 x_t^3+ 4 x_t^2+23 x_t-6}{(x_t-1)^4} \,\ln x_t \right] \nonumber \\[0.2cm]
&-\frac{s_W^2}{36}\,\left[\frac{x_t(18 x_t^3-139 x_t^2+274 x_t-129)}{2 (x_t-1)^3} + \frac{24 x_t^4-33 x_t^3-45 x_t^2+50 x_t-8}{(x_t-1)^4}\, \ln x_t\right], \\[0.3cm]
C^{\rm SM}_{S} & =
C^{\rm box,\, \rm SM}_{S, \,\rm unitary}+C^{\rm h\, penguin,\, \rm SM}_{S, \,\rm unitary}
\nonumber\\[0.2cm]
&\; =\; -\frac{3x_t}{8 x_{h_{\rm SM}}} - \frac{x_t(x_t+1)}{48(x_t-1)^2} - \frac{(x_t-2)(3x_t^2-3x_t+1)}{24(x_t-1)^3}\, \ln x_t \,,
\end{align}
where $x_{h_{SM}}=m_{h^0}^2/m_W^2$, and $m_{h^0}$ is the mass of the SM Higgs boson.

\begin{figure}[t]
\centering
\includegraphics[width=0.23\textwidth]{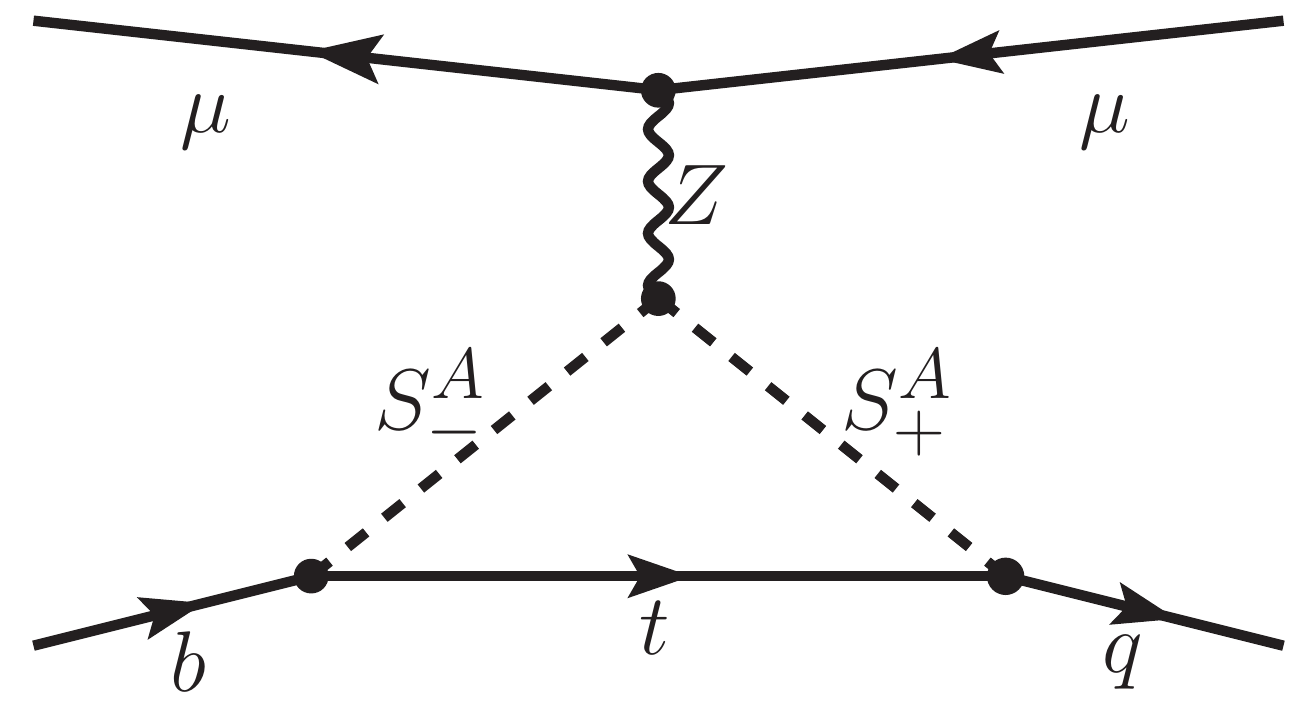}~
\includegraphics[width=0.23\textwidth]{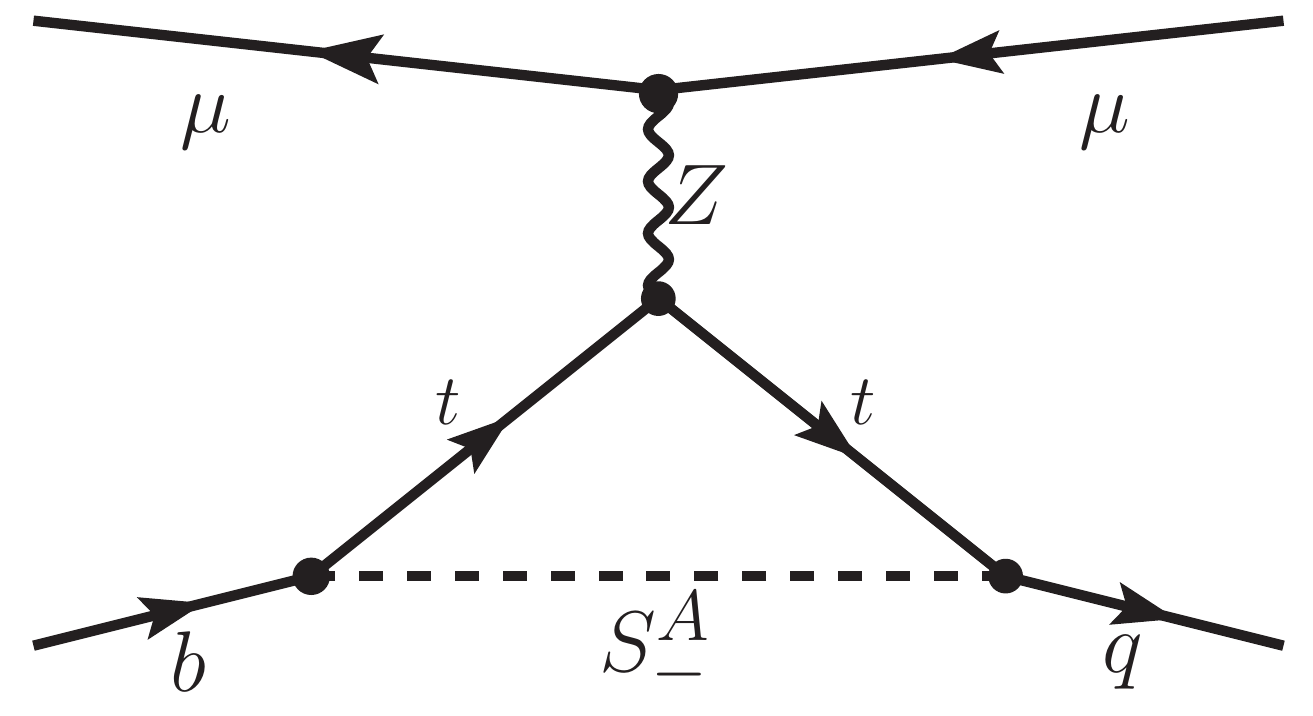}~
\includegraphics[width=0.23\textwidth]{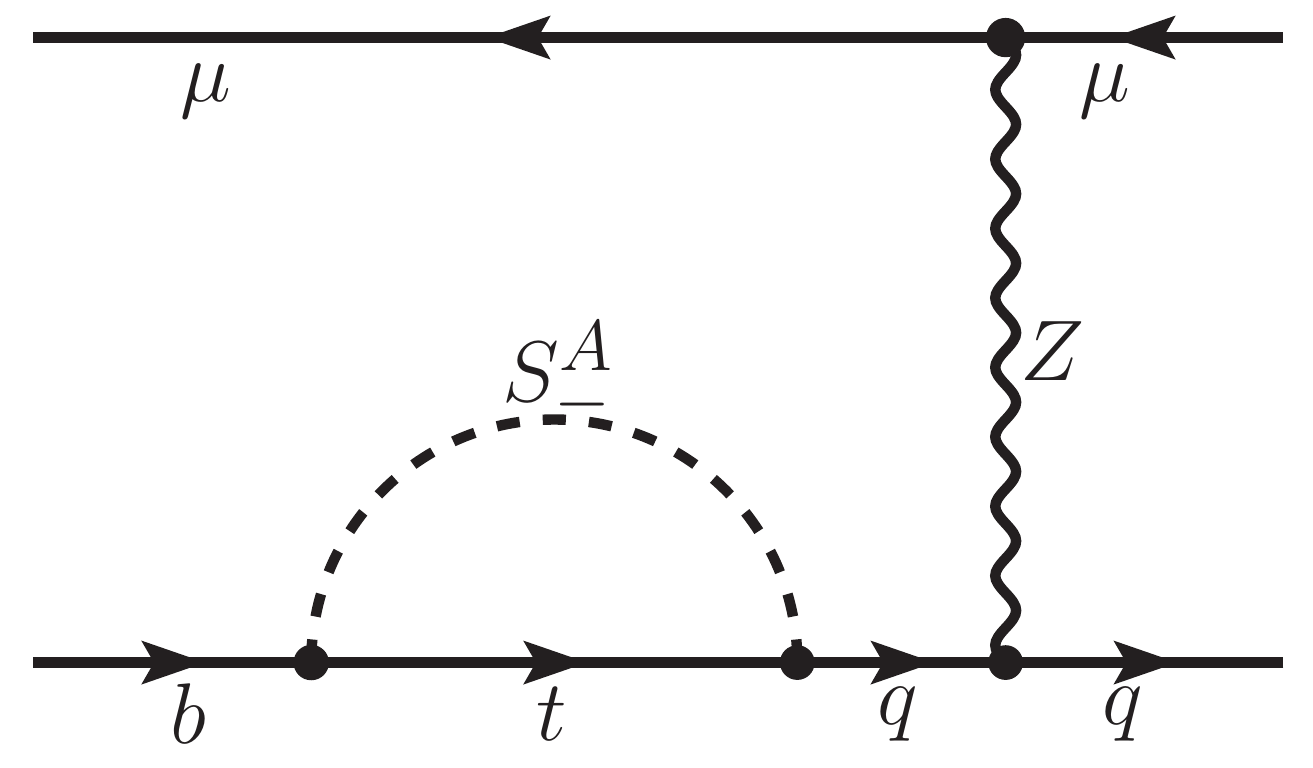}~
\includegraphics[width=0.23\textwidth]{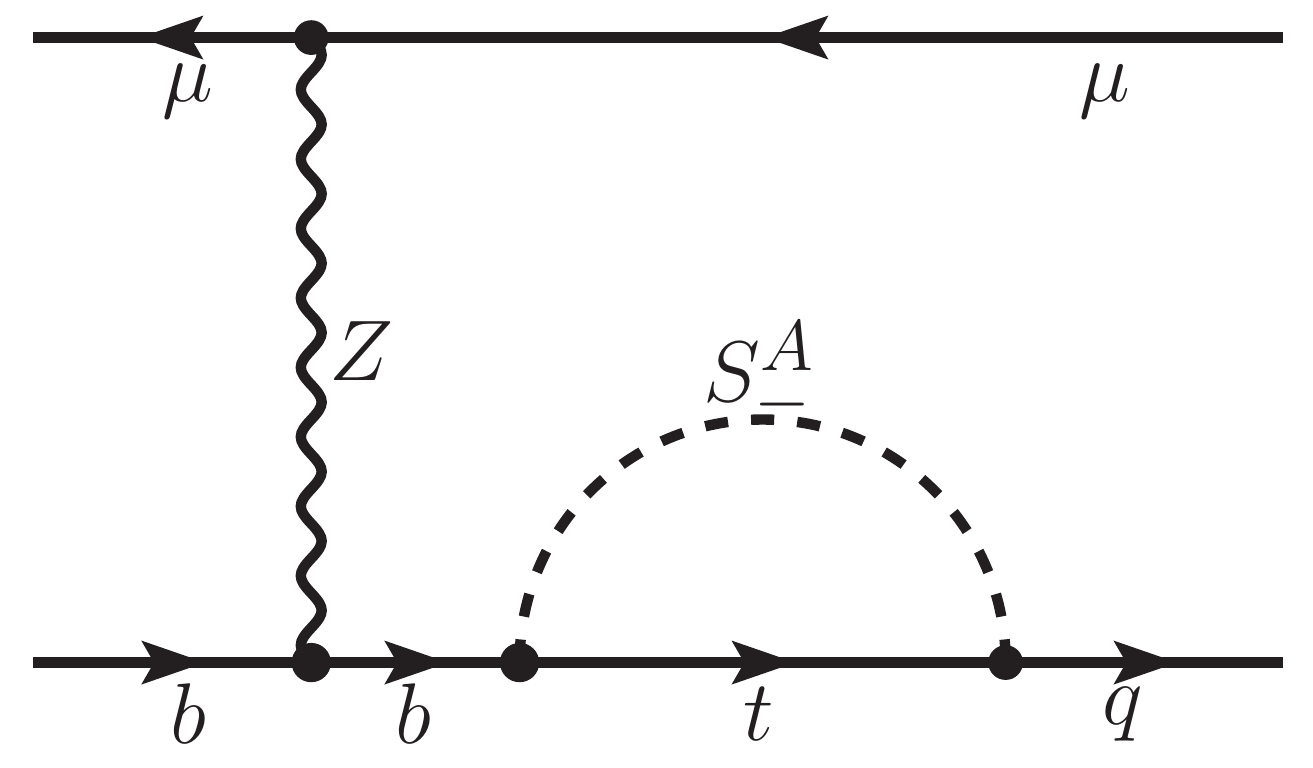}\\
\vskip 0.6cm
\includegraphics[width=0.23\textwidth]{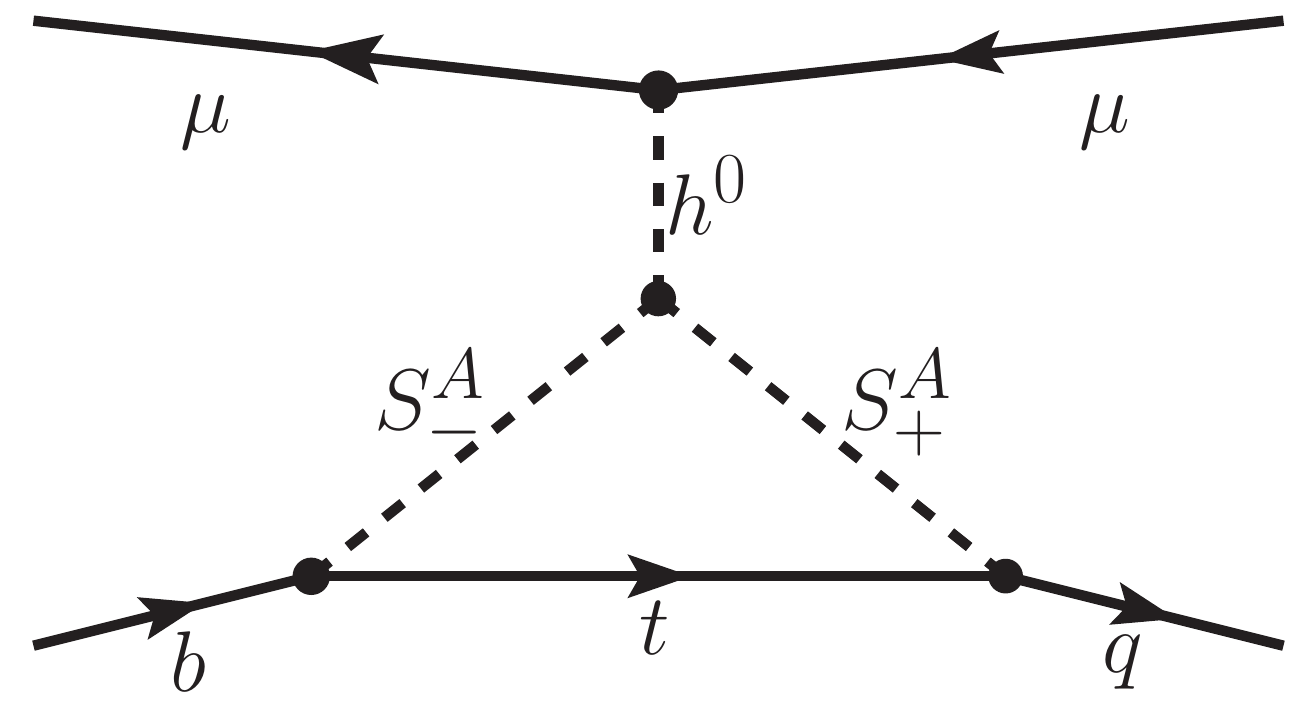}~
\includegraphics[width=0.23\textwidth]{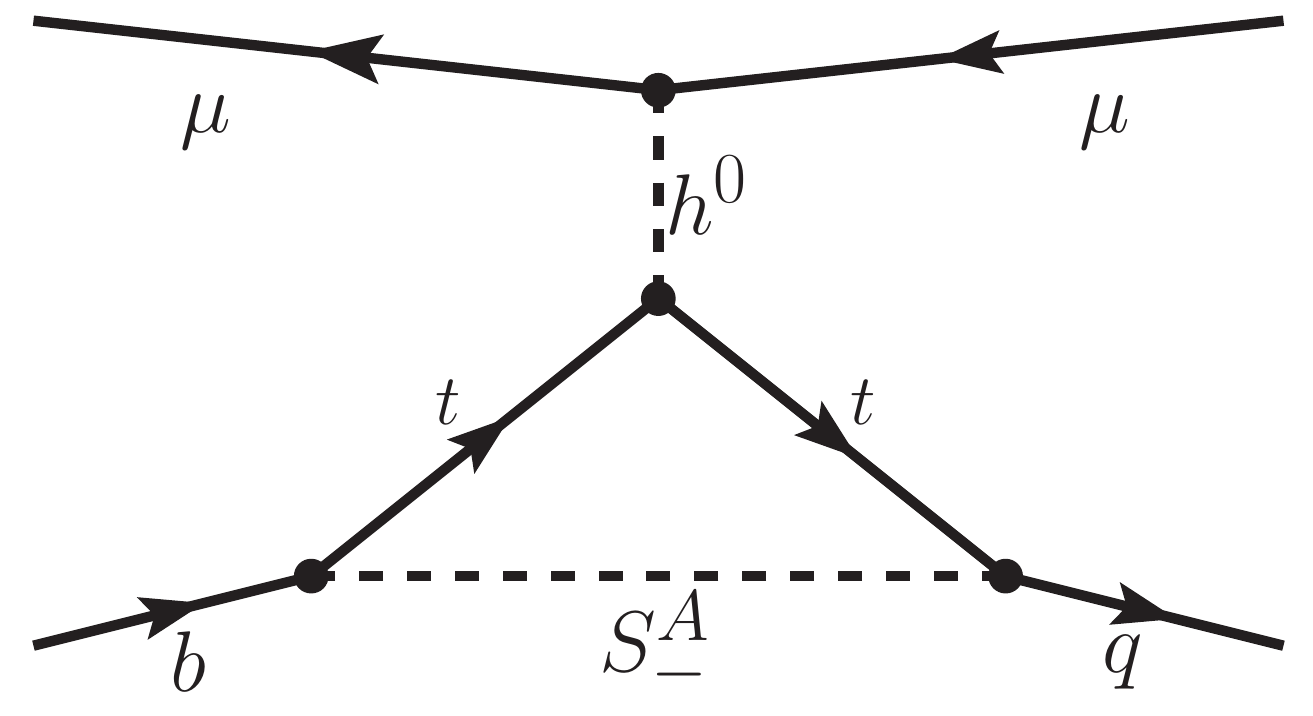}~
\includegraphics[width=0.23\textwidth]{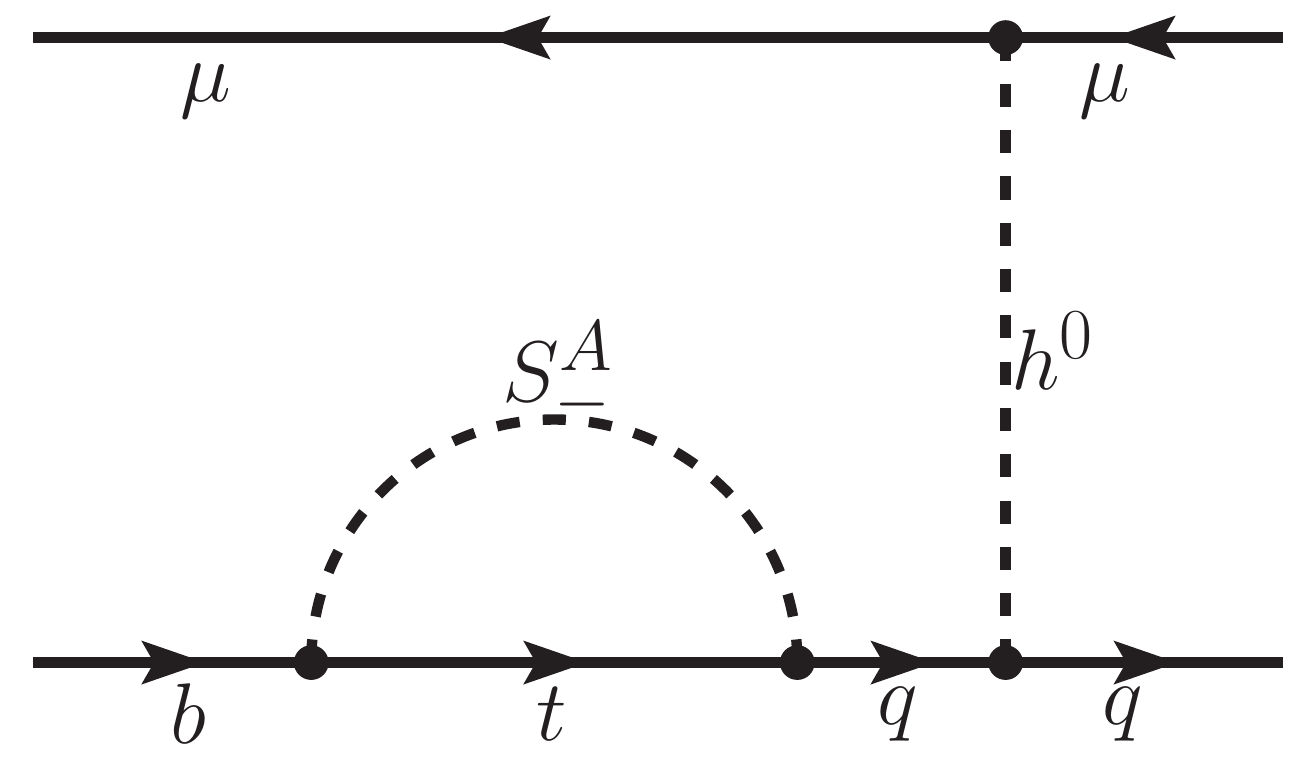}~
\includegraphics[width=0.23\textwidth]{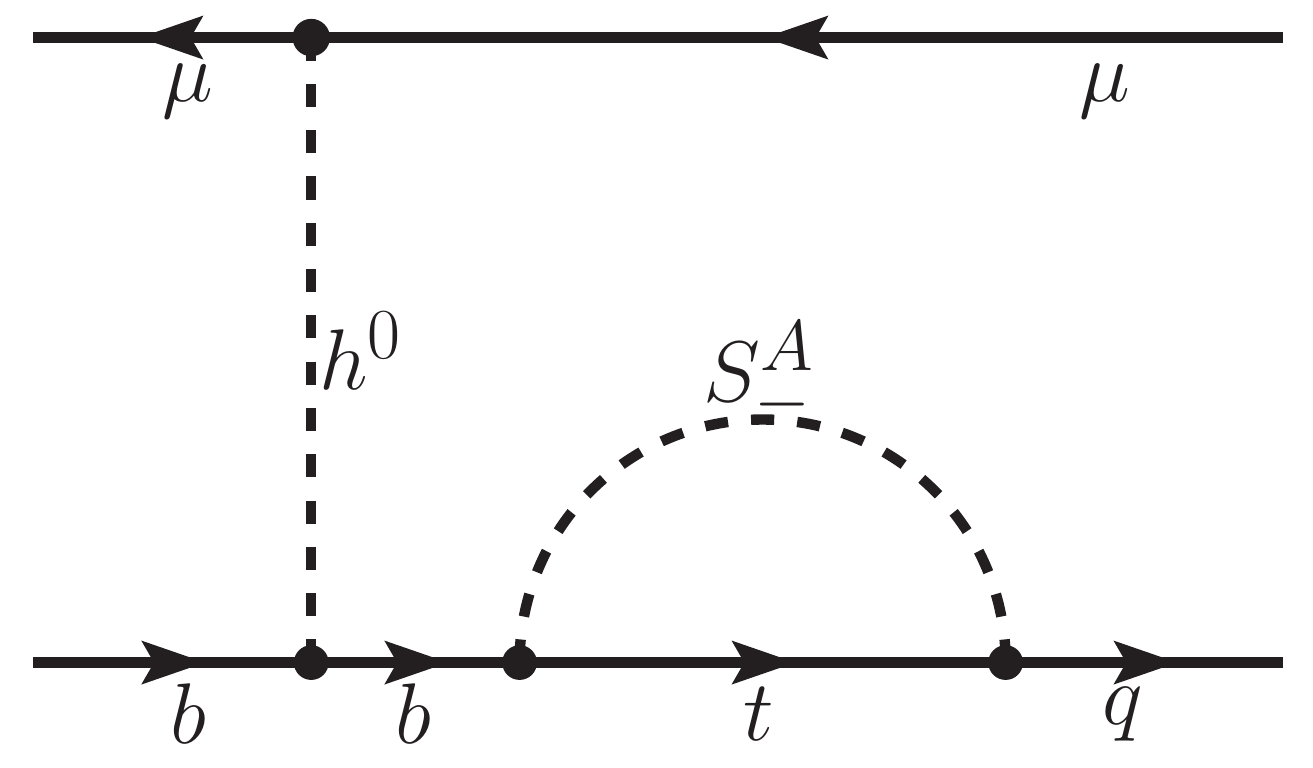}
\caption{\small Penguin diagrams contributing to the rare $B_q\to \mu^+ \mu^-$ decay within the MW model.}
\label{btomumuNP}
\end{figure}

In the MW model, the colour-octet scalars do not couple to the leptons and there is, therefore, neither additional box diagrams with charged colour-octet scalar exchanges nor penguin diagrams with neutral colour-octet scalar exchanges. The only new contributions to the rare $B_{s,d}\rightarrow \ell^+ \ell^-$ decays come from the penguin diagrams with the $Z$ boson and the SM Higgs boson exchanges, as shown in Fig.~\ref{btomumuNP}. The new $Z$-penguin diagrams involve the charged colour-octet scalar exchanges, and provide additional contributions to the Wilson coefficients $C_{10}$ and $C_P$:
\begin{align} \label{eq:ctenwise}
C_{10,{\rm unitary}}^{\rm Z \, penguin, \, NP} & = {\left| {{\eta _u}} \right|^2}\frac{{x_t^2}}{6}\left[ \frac{1}{{{x_{{S}}} - {x_t}}} + \frac{{{x_{{S}}}}}{{{{\left( {{x_{{S}}} - {x_t}} \right)}^2}}}\left( \ln {x_t} - \ln {x_S} \right) \right],\\[0.3cm]
C_{P,{\rm unitary}}^{\rm Z \, penguin, \, NP} &= \frac{{{x_t}}}{{3{{\left( {{x_S} - {x_t}} \right)}^2}}}\left\{ {{\eta _d}\eta _u^*\left[ { - \frac{{{x_t} + {x_S}}}{2} + \frac{{{x_t}{x_S}}}{{\left( {{x_S} - {x_t}} \right)}}\left( {\ln {x_S} - \ln {x_t}} \right)} \right]} \right.\nonumber\\[0.2cm]
& \left. { + \frac{{{\left| {{\eta _u}} \right|}^2}}{{6\left( {{x_S} - {x_t}} \right)}}\left[ {\frac{{x_S^2 - 8{x_S}{x_t} - 17x_t^2}}{6} + \frac{{x_t^2\left( {3{x_S} + {x_t}} \right)}}{{\left( {{x_S} - {x_t}} \right)}}\left( {\ln {x_S} - \ln {x_t}} \right)} \right]} \right\}\nonumber\\[0.2cm]
& + \frac{{2s_W^2{x_t}}}{{9{{\left( {{x_S} - {x_t}} \right)}^2}}}\left\{ {{\eta _d}\eta _u^*\left[ {\frac{{5{x_t} - 3{x_S}}}{2} + \frac{{{x_S}\left( {2{x_S} - 3{x_t}} \right)}}{{\left( {{x_S} - {x_t}} \right)}}\left( {\ln {x_S} - \ln {x_t}} \right)} \right]} \right.\nonumber\\[0.2cm]
& \hspace{-1.0cm} + \frac{{{{\left| {{\eta _u}} \right|}^2}}}{{6\left( {{x_S} - {x_t}} \right)}}\left[ { \left. {\frac{{17x_S^2 - 64{x_S}{x_t} + 71x_t^2}}{6}} \right]} \right. \left. - {\frac{{4x_S^3 - 12x_S^2{x_t} + 9{x_S}x_t^2 + 3x_t^3}}{{\left( {{x_S} - {x_t}} \right)}}\left( {\ln {x_S} - \ln {x_t}} \right)} \right\} \nonumber \\[0.2cm]
& + {\left| {{\eta _u}} \right|^2}\left( {1 - s_W^2} \right)\frac{{x_t^2}}{{3{{\left( {{x_S} - {x_t}} \right)}^2}}}\biggl[ {{x_S}\left( {\ln {x_S} - \ln {x_t}} \right) + {x_t} - {x_S}} \biggr],
\end{align}
where $x_S={m_{S^A_+}^2}/{m_W^2}$ and $m_{S^A_+}$ is the mass of the charged colour-octet scalars. While the new SM Higgs-penguin diagrams involve also the charged colour-octet scalar exchanges, they contribute only to the scalar Wilson coefficient $C_S$:
\begin{align}\label{eq:cswise}
C_{S, \, {\rm unitary}}^{\rm h \, penguin, \, NP} &=\frac{{{v}^{2}}}{m_{{{h}^{0}}}^{2}}\frac{{{x}_{t}}{{\lambda }_{1}}}{6\left( {x_S}-{{x}_{t}} \right)}\left\{ {{\eta }_{d}}\eta _{u}^{*}\left[ \frac{{{x}_{t}}}{\left( {x_S}-{{x}_{t}} \right)}\left( \ln {x_S}-\ln {{x}_{t}} \right)-1 \right] \right. \nonumber\\[0.2cm]
 & \left. +{{\left| {{\eta }_{u}} \right|}^{2}}\left[ \frac{x_{t}^{2}}{2{{\left( {x_S}-{{x}_{t}} \right)}^{2}}}\left( \ln {x_S}-\ln {{x}_{t}} \right)+\frac{{x_S}-3{{x}_{t}}}{4\left( {x_S}-{{x}_{t}} \right)} \right] \right\} \nonumber\\[0.2cm]
 & +\frac{2{{x}_{t}}}{3{{x}_{{{h}_{SM}}}}}\left\{ {{\eta }_{d}}\eta _{u}^{*}\frac{{{x}_{t}}}{\left( {x_S}-{{x}_{t}} \right)}\left[ 1-\frac{{x_S}}{\left( {x_S}-{{x}_{t}} \right)}\left( \ln {x_S}-\ln {{x}_{t}} \right) \right] \right. \nonumber\\[0.2cm]
 & \left. +{{\left| {{\eta }_{u}} \right|}^{2}}\frac{{{x}_{t}}}{2{{\left( {x_S}-{{x}_{t}} \right)}^{2}}}\left[ \frac{{x_S}+{{x}_{t}}}{2}-\frac{{x_S}{{x}_{t}}}{\left( {x_S}-{{x}_{t}} \right)}\left( \ln {x_S}-\ln {{x}_{t}} \right) \right] \right\}.
\end{align}

With Eqs.~\eqref{eq:C10SM}--\eqref{eq:cswise}, the final Wilson coefficients in the MW model can be written as
\begin{align}\label{eq:wilsontotal}
C_{10} &= C^{\rm{SM}}_{10}+C_{10, \, \rm unitary}^{\rm Z \, penguin, \, NP},\\[0.2cm]
C_S &= C^{\rm SM}_{S}+C_{S, \, \rm unitary}^{\rm h \, penguin, \, NP},\\[0.2cm]
C_P &= C^{\rm SM}_{P}+C_{P, \, \rm unitary}^{\rm Z \, penguin, \, NP}.
\end{align}
With the operators defined by Eq.~\eqref{eq:operators}, none of the above three Wilson coefficients is affected by the QCD renormalization group evolution. Together with the hadronic matrix elements,
\begin{align}
\langle 0 |\bar{q}\, \gamma_\mu \gamma_5\, b | \bar{B}_q(p) \rangle  \; = \; i f_{B_q} p_\mu\,, \qquad
\langle 0 |\bar{q}\, \gamma_5\, b | \bar{B}_q(p) \rangle  \; = \; -i f_{B_q} \frac{m_{B_q}^2}{m_b+m_q}\,,
\end{align}
the branching ratio of $B_{s,d}\rightarrow\ell^{+}\ell^{-}$ decays can be expressed as
\begin{align} \label{eq:BR}
\mathcal{B}( B_{q}\to {{\ell}^{+}}{{\ell}^{-}}) &= \frac{{{\tau }_{{{B}_{q}}}}G_{F}^{4}m_{W}^{4}}{8{{\pi }^{5}}}\,{{\left| {{V}_{tb}}V_{tq}^{*} \right|}^{2}}\left[\, {{\left| {{P}_{\rm SM}} \right|}^{2}}+{{\left| {{S}_{\rm SM}}\, \right|}^{2}} \right] f_{{{B}_{q}}}^{2}\,{{m}_{{{B}_{q}}}}\,m_{\ell}^{2}\, \sqrt{1-\frac{4m_{\ell}^{2}} {m_{{{B}_{q}}}^{2}}}\,\left[\, {{\left| P \right|}^{2}}+{{\left| S \right|}^{2}}\, \right], \nonumber\\[0.2cm]
&=\mathcal{B}( B_{q}\to {{\ell}^{+}}{{\ell}^{-}})_{\rm SM}\left[\, {{\left| P \right|}^{2}}+{{\left| S \right|}^{2}}\, \right],
\end{align}
with
\begin{align}
P_{\rm SM}=C_{10}^{\rm SM}+\frac{m_{B_q}^2}{2 m_W^2} \frac{m_b}{m_b+ m_q}C_P^{\rm SM}, \qquad
S_{\rm SM}=\sqrt{1-\frac{4 m_{\ell}^2}{m_{B_q}^2}}\,\frac{m_{B_q}^2}{2 m_W^2}\,\frac{m_b}{m_b+ m_q}\,C_S^{\rm SM}, \label{eq:psmssm}
\end{align}
and
\begin{align}
P & = \frac{1}{\sqrt{|P_{\rm SM}|^2+|S_{\rm SM}|^2}}\left[C_{10}+\frac{m_{B_q}^2}{2 m_W^2} \frac{m_b}{m_b+ m_q}C_P\right]\,\equiv\, |P|\; e^{i\varphi_P}\,,  \label{eq:pwise} \\[0.2cm]
S & = \frac{1}{\sqrt{|P_{\rm SM}|^2+|S_{\rm SM}|^2}}\,\sqrt{1-\frac{4m_{\ell}^2}{m_{B_q}^2}}\,\frac{m_{B_q}^2}{2 m_W^2}\,\frac{m_b}{m_b+ m_q}C_S \,\equiv\, |S|\; e^{i\varphi_S}\,.  \label{eq:swise}
\end{align}
While both $C_P^{\rm SM}$ and $C_S^{\rm SM}$ are negligibly small, and hence $P_{\rm SM}\doteq C_{10}^{\rm SM}$, $S_{\rm SM}\doteq0$, we keep their contributions in the numerical analysis both within the SM and in the MW model. In a more generic case, both $P$ and $S$ can carry nontrivial CP-violating phases $\varphi_P$ and $\varphi_S$. However, even in models with comparable Wilson coefficients, the contributions from ${\cal O}_{S}$ and ${\cal O}_{P}$ are suppressed by a factor of $m_{B_q}^2/m_W^2$ with respect to that from ${\cal O}_{10}$. Therefore, unless there were large enhancements for $C_S$ and $C_P$, the coefficient $C_{10}$ still provides the dominant contribution to the branching ratio.

To compare with the experimental data, the effect of $B_q-\bar B_q$ oscillations should be taken into account, and the resulting averaged time-integrated branching ratio is given by~\cite{Buras:2013uqa,DeBruyn:2012wk}
\begin{equation} \label{eq:BR_bar}
 \overline{\mathcal{B}}(B_q\to \ell^+\ell^-) \, =\, \biggl[ \frac{1+A_{\Delta\Gamma}\,y_q}{1-y_q^2}\biggr]\,\mathcal{B}(B_q\to \ell^+\ell^-)\,,
\end{equation}
where $y_q$ is related to the $B_q$-meson decay width difference $\Delta\Gamma_q$,
\begin{equation}
 y_q \,\equiv\, \frac{\Gamma^{q}_L-\Gamma^{q}_H}{\Gamma^{q}_L+\Gamma^{q}_H}\, =\, \frac{\Delta\Gamma_q}{2\Gamma_q}\,,
\end{equation}
with $\Gamma^{q}_{H(L)}$ and $\Gamma_q=\tau_{B_q}^{-1}$ denoting the heavier~(lighter)-eigenstate and the average decay widths, respectively. The time-dependent observable $A_{\Delta\Gamma}$ can be, in the absence of new CP-violating contribution to the $B_q-\bar B_q$ mixing, expressed as~\cite{DeBruyn:2012wk}
\begin{align}\label{eq:adeltgamma}
A_{\Delta\Gamma}=\frac{|P|^2\cos 2\varphi_P-|S|^2\cos 2\varphi_S}{|P|^2+|S|^2}.
\end{align}
In order to take into account the sizable $\Delta \Gamma_s$ effect, it is convenient to introduce the ratio~\cite{DeBruyn:2012wk}
\begin{align}\label{eq:ratioofexpandsm}
R \equiv \frac{\overline{{\mathcal B}}(B_s\rightarrow \ell^+\ell^-)}{{\mathcal B}(B_s\rightarrow \ell^+\ell^-)_{\text{SM}}}=\frac{1+y_s \cos 2\varphi_P}{1-y_s^2}|P|^2+\frac{1-y_s \cos 2\varphi_S}{1-y_s^2}|S|^2.
\end{align}
It should be noted that both $A_{\Delta\Gamma}$ and $R$ can be directly extracted from the experimental observables~\cite{DeBruyn:2012wk}. Because of the negligible width difference in the $B_d$ system, however, the approximation $\overline{\mathcal B}(B_d\to \ell^+\ell^-)\approx{\mathcal B}(B_d\to \ell^+\ell^-)$ will be assumed throughout this paper.

\subsection{\texorpdfstring{$\mathbf{\boldsymbol{Z\to b \bar{b}}}$}{Lg} decay}
\label{sec:Z2bbbar}

The $Z$-pole observable $R_b^0$, defined as the ratio of the partial decay widths of the $Z$ boson decaying into bottom quarks and into all quarks,
\begin{align}\label{eq:rbdefinition}
R_b^0 \equiv \frac{\Gamma(Z\rightarrow b \bar{b})}{\Gamma(Z\rightarrow \text{hadrons})}=\frac{\Gamma(Z\rightarrow b \bar{b})}{\sum\limits_{q\not{=} t}\Gamma(Z\rightarrow q\bar{q})},
\end{align}
has been measured to high accuracy by the experiments at CERN $e^+e^-$ collider (LEP) and SLAC Linear Collider (SLC)~\cite{ALEPH:2005ab}, and plays an important role in precision tests of the SM, as well as in indirect searches for NP beyond it~\cite{Baak:2014ora}.

The branching ratio $R_b^0$ is also quite sensitive to the NP models with an extended scalar sector due to the exchange of the additional charged and neutral scalars~\cite{Degrassi:2010ne,Haber:1999zh}. Specific to the MW model, the additional contribution to $R_b^0$ is dominated by the $Z$-penguin diagrams involving the charged colour-octet scalars, where the corresponding two-loop QCD corrections have been calculated for the first time by Degrassi and Slavich~\cite{Degrassi:2010ne}. Following the same notations as in Ref.~\cite{Degrassi:2010ne}, we have
\begin{align}\label{eq:rbdefinitionthrory}
\frac{1}{R_b^0} \equiv 1+ \frac{\sum\limits_{q\not=b}\left[\left(\bar{g}_L^q\right)^2+\left(\bar{g}_R^q\right)^2\right]K_q} {\left[\left(\bar{g}_L^b\right)^2+\left(\bar{g}_R^b\right)^2\right]K_b},
\end{align}
where $\bar{g}^{q}_{(L,R)}$ are the left-handed and right-handed $Zq\bar q$ couplings, and the factor $K_q$ contains the QCD, QED and quark-mass corrections. The explicit expressions for these quantities can all be found in Ref.~\cite{Degrassi:2010ne}.

\subsection{\texorpdfstring{$\mathbf{\boldsymbol{B\to X_s \gamma}}$}{Lg}, \texorpdfstring{$\mathbf{\boldsymbol{B\to K^*\gamma}}$}{Lg} and \texorpdfstring{$\mathbf{\boldsymbol{B\to \rho\gamma}}$}{Lg} decays}
\label{sec:b2sdgamma}

Both the inclusive $B\to X_{s,d} \gamma$ and the exclusive $B\to V \gamma$ ($V$ is a light vector meson) decays are induced by the quark-level $b\rightarrow s(d)\gamma$ transitions, which contribute in a significant manner to current bounds on masses and interactions of possible charged scalars in NP models~\cite{Hurth:2010tk,Hermann:2012fc}. This is because the photonic-penguin diagrams with the charged-scalar exchanges contribute to these processes at the same level as those mediated by the SM $W^{\pm}$ bosons. Besides the branching ratio $\mathcal{B}(B\to X_s \gamma)$, for which a good agreement has been observed between the experimental measurements~\cite{Agashe:2014kda,Amhis:2014hma} and the theoretical predictions~\cite{Misiak:2015xwa}, we shall consider the two isospin asymmetries~\cite{Ball:2006eu,Beneke:2001at}
\begin{align}
\Delta(K^*\gamma)\; &=\;\frac{\bar\Gamma(B^0\to K^{*0}\gamma)-\bar\Gamma(B^+\to K^{*+}\gamma)}{\bar\Gamma(B^0\to
 K^{*0}\gamma)+\bar\Gamma(B^+\to K^{*+}\gamma)}\,,\label{eq:isospin1}\\[0.2cm]
\Delta(\rho\gamma)\; &=\;\frac{\bar\Gamma(B^+\to\rho^+\gamma)}{2\bar\Gamma(B^0\to\rho^0\gamma)}-1\,,
\label{eq:isospin2}
\end{align}
which exhibit a different dependence on the model parameters compared to $\mathcal{B}(B\to X_s \gamma)$ and could, therefore, provide complementary information on the model considered~\cite{Li:2013vlx,Jung:2012vu,Mahmoudi:2009zx,B2Vg-NP}.

In the MW model, the additional contribution to $b\rightarrow s(d) \gamma$ transitions comes only from the photonic-penguin diagrams exchanged by the charged colour-octet scalars. With the approximation $m_{s,d}\to 0$, there appears no additional low-energy operator besides the SM ones, and the NP effect is driven by new additive contributions to the Wilson coefficients at the matching scale $\mu_W$, which have been calculated up to NLO by Degrassi and Slavich~\cite{Degrassi:2010ne}. Following the same notations as in Ref.~\cite{Degrassi:2010ne}, we write the matching Wilson coefficients as
\begin{align}
C^{}_i(\mu_W) =  C^{(0)}_i(\mu_W) + \delta C_{i}^{(0)}(\mu_W)
 + \frac{\alpha_s(\mu_W)}{4 \pi} \left[ C^{(1)}_i(\mu_W) +
\delta C_{i}^{(1)}(\mu_W) \right]\,,
\label{wcbtosgamma}
\end{align}
where $C^{(k)}_i(\mu_W)$ denote the SM contributions ($k=0,1$) and could be found in Refs.~\cite{WC1,WC2}, while $\delta C_{i}^{(k)}(\mu_W)$ denote the charged colour-octet scalar contributions. The explicit expressions for $\delta C_{i}^{(k)}(\mu_W)$ both at the LO and at the NLO could be found in Ref.~\cite{Degrassi:2010ne}. The evolution of the Wilson coefficients from the matching scale $\mu_W$ down to the low-energy scale $\mu_b$, which is the same as in the SM, can be found in Refs.~\cite{WC1,WC2}. The charged colour-octet scalar effects on the observables ${\mathcal B} (B\rightarrow X_s \gamma)$, $\Delta(\rho\gamma)$ and $\Delta(K^*\gamma)$ have been studied in Ref.~\cite{Li:2013vlx}, which will be followed in this paper.

\section{Numerical results and discussions}
\label{sec:result}

\begin{table}[t]
\begin{center}
\caption{\label{bigtable} \small Central values and uncertainties for the input parameters used in our analysis. The assumptions entering the stage II estimates are described in the text, and the entries ``id" refer to the value in the same row as in the ``2014" column. }
\vspace{0.2cm}
\doublerulesep 0.8pt \tabcolsep 0.18in
\begin{tabular}{l|cc|cc}
\hline\hline
            &  2014                        &   & Stage~II     &  \\
\hline
$|V_{us}|$ $(K_{\ell 3})$    &  $0.2258\pm 0.0008\pm 0.0012$  & \cite{Charles:2013aka,Charles:2004jd} & $0.22494\pm 0.0006$   & \cite{Charles:2013aka} \\
$|V_{cb}|\times 10^3$  &  $41.15 \pm 0.33 \pm 0.59$  & \cite{Charles:2013aka,Charles:2004jd} &
  $42.3\pm 0.3$& \cite{Aushev:2010bq}\\
$|V_{ub}|\times 10^3$  &  $3.75 \pm 0.14 \pm 0.26$ & \cite{Charles:2013aka,Charles:2004jd} &
  $3.56\pm 0.08$& \cite{Aushev:2010bq} \\
$\gamma$  &  $(68.0^{+8.0}_{-8.5})^\circ$   &  \cite{Charles:2013aka,Charles:2004jd} &  $(67.1\pm 1)^\circ$  &  \cite{Bediaga:2012py,Aushev:2010bq}\\
\hline
$f_{B_s}$ [GeV]   &  $0.228\pm 0.008$   & \cite{Carrasco:2013zta,Aoki:2013ldr}  & $0.232\pm 0.001$  & \cite{Charles:2013aka,Blum:2013mhx}\\
$f_{B_d}$ [GeV]   &  $0.189\pm 0.008$   & \cite{Carrasco:2013zta,Aoki:2013ldr} & $0.193\pm 0.001$  & \cite{Charles:2013aka,Blum:2013mhx}\\
$f_{B_s}/f_{B_d}$   &  $1.206\pm 0.024$    & \cite{Carrasco:2013zta,Aoki:2013ldr}  & $1.205\pm 0.005$ & \cite{Charles:2013aka,Blum:2013mhx}\\
\hline
$\bar{m}_c$  &  $1.275\pm 0.025$  & \cite{Agashe:2014kda} & $1.286\pm 0.010$ & \cite{Charles:2013aka}\\
$\bar{m}_b$  &  $4.15\pm 0.03$  & \cite{Agashe:2014kda} & id & \cite{Agashe:2014kda}\\
$m_t^{pole}$ &  $173.21\pm 0.51\pm0.71$  & \cite{Agashe:2014kda}   & id &\cite{Agashe:2014kda}\\
$m_{h^0}$    & $125.7\pm0.4$  & \cite{Agashe:2014kda}   & id &\cite{Agashe:2014kda}\\
\hline
$\sin^2\theta_W$          & $0.23126\pm0.00005$   &  \cite{Agashe:2014kda} & id & \cite{Agashe:2014kda}\\
$\alpha_s(m_Z)$  &  $0.1185\pm 0.0006$    &  \cite{Agashe:2014kda} & id & \cite{Agashe:2014kda}\\
\hline\hline
\end{tabular}
\end{center}
\end{table}

With the theoretical framework at hand and the input parameters collected in Table~\ref{bigtable}, we now present and discuss our numerical results in this section. The stage II projection in the table refers to an epoch with $50~{\rm fb}^{-1}$ LHCb and $50~{\rm ab}^{-1}$ Belle II data, which corresponds probably to the middle of 2020s at the earliest~\cite{Charles:2013aka}. Estimates of future experimental and theoretical uncertainties are taken mainly from Refs.~\cite{Bediaga:2012py,Aushev:2010bq,Charles:2013aka,Blum:2013mhx}.

\subsection{SM predictions and experimental data}
\label{sec:smanddata}

With the input parameters collected in Table~\ref{bigtable}, our predictions for the observables studied in this paper are listed in Table~\ref{observable}, where the current best measurements~\cite{Agashe:2014kda,Amhis:2014hma,CMS:2014xfa} together with the precision expected in stage II~\cite{Bediaga:2012py,Aushev:2010bq,Charles:2013aka} are also given. The theoretical errors are obtained by varying each input parameter within the corresponding range and adding the individual error in quadrature. We assume that the central values of the future measurements in stage II remain the same as the current ones except for $\Delta(\rho\gamma)$. For the latter, it is observed that the current data disagrees with the SM expectation at more than $2\sigma$ level, and we assume therefore the central value of the future measurement to coincide with the SM expectation for $\Delta(\rho\gamma)$.

\begin{table}
\begin{center}
\caption{\label{observable} \small SM predictions and experimental measurements for the observables discussed in this paper. The entries ``id" refer to the value in the same row as in the ``2014" column. }
\vspace{0.2cm}
\doublerulesep 0.8pt \tabcolsep 0.05in
\begin{tabular}{l|cccc|cccc}
\hline\hline
Observable            &  \multicolumn{4}{c|}   {Exp. data}              & \multicolumn{4}{c}   {SM prediction}      \\
\cline{2-9}                    & 2014  & &  Stage~II &    & 2014 & & Stage~II &  \\
\hline
$\Delta m_{B_s^0}$ $[{\rm ps}^{-1}]$ & $17.761\pm 0.022$ &  \cite{Agashe:2014kda} &    id    &   &   $17.568_{-1.503}^{+1.541}$        &                &    $17.589^{+0.431}_{-0.429}$      &                 \\
$\Delta m_{B_d^0}$ $[{\rm ps}^{-1}]$ & $0.510\pm 0.003$ &  \cite{Agashe:2014kda} &    id    &   &   $0.533_{-0.085}^{+0.086}$        &                &    $0.515^{+0.016}_{-0.016}$      &                 \\
\hline
$\overline{ {\mathcal B}}({{{B}}_{s}}\to \mu^+\mu^-)$ $[10^{-9}]$ & $2.8_{-0.6}^{+0.7}$ & \cite{CMS:2014xfa} &    $2.8\pm 0.3$ & \cite{Bediaga:2012py} &   $3.46^{+0.33}_{-0.30}$        &                &    $3.79_{-0.22}^{+0.16}$      &                 \\
$\overline{ {\mathcal B}}({{{B}}_{d}}\to \mu^+\mu^-)$ $[10^{-10}]$ & $3.9_{-1.4}^{+1.6}$ & \cite{CMS:2014xfa} &    $3.9\pm 1.4$ & \cite{Bediaga:2012py}&   $0.99^{+0.16}_{-0.16}$        &                &    $1.07_{-0.06}^{+0.05}$      &                 \\
\hline
${\mathcal B}(B\to X_s \gamma)$ $[10^{-4}]$ & $3.43\pm 0.22$ & \cite{Amhis:2014hma} &  $3.43\pm 0.21$   & \cite{Aushev:2010bq} &  $3.14_{-0.27}^{+0.26}$        &                &    $3.11_{-0.26}^{+0.25}$      &                 \\
$ \Delta(K^*\gamma)$ $[10^{-2}]$ & $5.2\pm 2.6$ & \cite{Amhis:2014hma} &  $5.2\pm 0.3$   & \cite{Aushev:2010bq} &  $4.15^{+2.57}_{-2.57}$        &                &    $4.20_{-2.60}^{+2.61}$      &                 \\
$ \Delta(\rho\gamma)$ $[10^{-2}]$ & $-46\pm 17$ & \cite{Amhis:2014hma} &  $-7.4\pm 1.7$   & \cite{Aushev:2010bq} &  $-8.29_{-7.42}^{+6.63}$        &                &    $-7.38_{-5.71}^{+5.47}$      &                 \\
\hline
$R_b^0$ $[10^{-3}]$  & $216.29\pm0.66$ & \cite{Agashe:2014kda} &    id &  &  $215.86_{-0.01}^{+0.01}$     &                &    id      &                 \\
\hline\hline
\end{tabular}
\end{center}
\end{table}

Taking into account the theoretical and experimental uncertainties, one can see that the SM predictions presented here agree with the corresponding experimental data within $2\sigma$ error bars, and hence strong constraints on the model parameters are expected from these observables.

\subsection{Procedure in the numerical analysis}
\label{sec:procedure}

The flavour observables discussed in this paper involve only four parameters of the MW model: $\eta_u, \eta_d, m_{S^A_{+}}$ and $\lambda_1$. As discussed in Sec.~\ref{sec:MWmodel}, the parameter $\lambda_1$ is real due to the Hermiticity of the scalar potential~\cite{Manohar:2006ga}. For the Yukawa coupling parameters $\eta_u$ and $\eta_d$, we shall consider the following two cases:
\begin{description}
\item [ (i) ] both of $\eta_u$ and $\eta_d$ are real, which is referred as ``real couplings" throughout this paper. Unlike in the case of the usual 2HDMs~\cite{Branco:2011iw}, these two parameters are assumed to be independent from each other.

\item [ (ii) ] both $\eta_u$ and $\eta_d$ are complex, which is referred as ``complex couplings" in the following. In this case, the independent Yukawa coupling parameters can be chosen as the magnitudes $|\eta_u|$ and $|\eta_d|$, and the relative phase $\theta$ defined through $\eta_u^* \eta_d = |\eta_u^* \eta_d| e^{i\theta}$.
\end{description}
Constraints on the model parameters have been analyzed both theoretically by requiring the perturbative unitarity and the vacuum stability~\cite{He:2013tla}, as well as phenomenologically by satisfying the current flavour, Higgs and EW precision data~\cite{Manohar:2006ga,Li:2013vlx,Degrassi:2010ne,Heo:2008sr,Gresham:2007ri,
Carpenter:2011yj}. It is important to mention that the current experimental searches based on the dijet events by the ATLAS~\cite{Aad:2011fq} and CMS~\cite{CMS:2012yf} collaborations exclude already the existence of colour-octet scalars with mass below $\sim2$~TeV, while the four-jet searches at the LHC do not observe any signal at a much lower mass region~\cite{Aad:2011yh,Chatrchyan:2013izb}. However, if the colour-octet scalar decays into more than two light quarks, top-quark and jet, or $t\bar t$, the existing bounds on the colour-octet scalar mass would be different and even masses of order $400$~GeV could not be excluded, as noticed in \cite{Manohar:2006ga,Gresham:2007ri}. In addition, all these mass bounds depend on some assumptions about the underlying NP models. In this paper, therefore, we shall specify the model parameters within the following ranges:
\begin{equation}
\lambda_1\in [-40,40], \;\; m_{S^A_+}\in [300,1000]\,\text{GeV}, \;\; \eta_u\in [-3,3], \;\; \eta_d\in [-100,100]
\label{scenario i value}
\end{equation}
in the case of real couplings, and
\begin{equation}
\lambda_1\in [-40,40], \;\; m_{S^A_+}\in [300,1000]\,\text{GeV}, \;\; |\eta_u|\in [0,3], \;\; |\eta_d|\in [0,100], \;\; \theta\in[-180^{\circ},180^{\circ}]
\label{scenario ii value}
\end{equation}
in the case of complex couplings. Then, we make an extensive random scan for these model parameters over the above constrained ranges to generate the initial samples.

For the numerical analysis, we adopt the following procedure: for each sample generated from the parameter ranges specified by Eqs.~\ref{scenario i value} and \ref{scenario ii value}, we give the theoretical prediction for an observable together with the corresponding theoretical uncertainty. If the obtained theoretical range for the observable has overlap with the $2\sigma$ range of the experimental data, the sample is regarded to be allowed. Following this procedure, we get the final survived ranges for the model parameters. To incorporate the theoretical uncertainty, we use the statistical treatment based on frequentist statistics and Rfit scheme~\cite{Hocker:2001xe}, which has been implemented in the CKMfitter package~\cite{Charles:2004jd}. Here the main observation is that, while the experimental data yields approximatively a Gaussian distribution, the theoretical calculation for an observable does not. The latter depends on a set of input parameters like form factors, decay constants and Gegenbauer moments etc., for which no probability distribution is known. The Rfit scheme assumes no particular distribution for the theory parameters, only that they are constrained to certain allowed ranges with an equal weighting, irrespective of how close they are from the edges of the allowed range. In addition, for simplicity, the relative theoretical uncertainty is assumed to be constant at each point in the parameter space. This is a reasonable assumption, since the main theoretical uncertainties are due to the hadronic input parameters, common to both the SM and the NP contributions. Therefore, the theoretical range for an observable at each point is obtained by varying each input parameter within its respective allowed range and then adding the individual uncertainty in quadrature. This procedure has already been adopted in Refs.~\cite{Li:2013vlx,Jung:2012vu,Cheng:2014ova}.

In order to investigate the capability of the MW model in explaining the flavour physics data listed in Table~\ref{observable} and to check the validity of the samples satisfying all the experimental constraints, we construct the $\chi^2$ function following the method proposed in Ref.~\cite{Mahmoudi:2009zx,Hurth:2013ssa}
\begin{align}
\chi^2\displaystyle =
\sum_{i\in ({ obs.})} \frac{(O_i^{\rm exp} - O_i^{\rm MW})^2}{(\sigma_i^{\rm exp})^2 + (\sigma_i^{\rm MW})^2}\;,
\label{eq:chi2abs}
\end{align}
where $O_i^{\rm exp}$ and $O_i^{\rm MW}$ denote the central values of the experimental measurements and the theoretical predictions for an observable $i$, with $\sigma_i^{\rm exp}$ and $\sigma_i^{\rm MW}$ being the corresponding experimental and theoretical errors, respectively. Using the method, one can obtain the $\chi^2$ value for each point in the parameter space, and then the minimal $\chi^2$ value.

\subsection{\texorpdfstring{$\mathbf{\boldsymbol{B_{s,d}-\bar{B}_{s,d}}}$}{Lg} mixings within the MW model}
\label{sec:Bmixing_result}

As mentioned already in Sec.~\ref{sec:Bmixing}, the $B_{s,d}-\bar{B}_{s,d}$ mixings within the MW model involve only the four Wilson coefficients $\tilde{C}_{VLL}^{SM}$, $ \tilde{C}_{VLL}^{NP}$, $ \tilde{C}_{SRR1}^{NP}$ and $ \tilde{C}_{SRR2}^{NP}$. Taking $m_{S^A_+}=500~{\rm GeV}$ as a benchmark, we obtain numerically:
\begin{align}
\frac{\tilde{C}_{VLL}^{NP}\left( {{\mu }_{b}} \right)}{\tilde{C}_{VLL}^{SM}\left( {{\mu }_{b}} \right)} & =  0.05{{\left| {{\eta }_{u}} \right|}^{2}}+0.02{{\left| {{\eta }_{u}} \right|}^{4}}+{{\left( \frac{{{m}_{b}}}{{{m}_{W}}} \right)}^{2}}\cdot {{10}^{-3}}\Bigl[ 1.02\eta _{u}^{*}{{\eta }_{d}}+2.35{{\left| {{\eta }_{u}} \right|}^{2}}+0.22{{\left| {{\eta }_{u}} \right|}^{4}} \Bigr],
\end{align}
\begin{align}
\frac{\tilde{C}_{SRR,1}^{NP}\left( {{\mu }_{b}} \right)}{\tilde{C}_{VLL}^{SM}\left( {{\mu }_{b}} \right)} & = {{\left( \frac{{{m}_{b}}}{{{m}_{W}}} \right)}^{2}}\cdot {{10}^{-2}} \Bigl[ 1.05\eta _{u}^{*}{{\eta }_{d}}-0.52{{\left| {{\eta }_{u}} \right|}^{2}}+0.14{{\left( \eta _{u}^{*}{{\eta }_{d}} \right)}^{2}}-0.14\eta _{u}^{*}{{\eta }_{d}}{{\left| {{\eta }_{u}} \right|}^{2}}+0.03{{\left| {{\eta }_{u}} \right|}^{4}} \Bigr],
\end{align}
\begin{align}
\frac{\tilde{C}_{SRR,2}^{NP}\left( {{\mu }_{b}} \right)}{\tilde{C}_{VLL}^{SM}\left( {{\mu }_{b}} \right)} & = {{\left( \frac{{{m}_{b}}}{{{m}_{W}}} \right)}^{2}}\cdot {{10}^{-4}}\Bigl[3.22{{\left| {{\eta }_{u}} \right|}^{2}}-6.55\eta_{u}^{*}{{\eta }_{d}} -1.85{{\left( \eta _{u}^{*}{{\eta }_{d}} \right)}^{2}}+1.85\eta _{u}^{*}{{\eta }_{d}}{{\left| {{\eta }_{u}} \right|}^{2}}-0.36{{\left| {{\eta }_{u}} \right|}^{4}} \Bigr],
\end{align}
for the short-distance Wilson coefficients, and
\begin{align}
\langle B_{s} |\mathcal{H}_{\rm eff}^{|\Delta B|=2}| \bar{B}_{s}\rangle &= \biggl\{ 5.77-0.22i+\left( 0.29-0.01i \right){{\left| {{\eta }_{u}} \right|}^{2}}+{{10}^{-2}} \cdot \left( 8.77-0.34i \right){{\left| {{\eta }_{u}} \right|}^{4}} \nonumber\\[0.2cm]
& +{{\left( \frac{{{m}_{b}}}{{{m}_{W}}} \right)}^{2}}\cdot {{10}^{-2}}\Bigl[ -\left( 4.27-0.17i \right)\eta _{u}^{*}{{\eta }_{d}}+\left( 3.75-0.15i \right){{\left| {{\eta }_{u}} \right|}^{2}} \nonumber\\[0.2cm]
& \hspace{1.0cm} -\left( 0.51-0.02i \right){{\left( \eta _{u}^{*}{{\eta }_{d}} \right)}^{2}}+\left( 0.51-0.02i \right)\eta _{u}^{*}{{\eta }_{d}}{{\left| {{\eta }_{u}} \right|}^{2}} \nonumber\\[0.2cm]
& \hspace{1.0cm} +0.03{{\left| {{\eta }_{u}} \right|}^{4}} \Bigr] \biggr\} \times {{10}^{-12}}~{\rm GeV}\,,
\end{align}
for the off-diagonal matrix element. From these results, we make the following observations:
\begin{itemize}
\item The theoretical expressions for the short-distance Wilson coefficients are independent of the parameter $\lambda_1$, and hence the mixing observable $\Delta m_{B^0_s}$ puts no constraints on the trilinear scalar coupling $g_{S^A_+S^B_-h^0}$.

\item The Wilson coefficients $\tilde{C}_{SRR,1}^{NP}$ and $\tilde{C}_{SRR,2}^{NP}$ are suppressed by a factor of $m_b^2/m_W^2$ with respect to $\tilde{C}_{VLL}^{NP}$. So the dominant NP contribution to the mass difference $\Delta m_{B^0_s}$ comes from the operator $Q_1^{VLL}$ and is always constructive to the SM one.

\item Except for the case with much smaller $\eta_u$, the terms containing $\eta_u^*\eta_d$ in the off-diagonal matrix element are suppressed by a factor of $m_b^2/m_W^2$. So the constraint on the parameter $\eta_d$ from the mixing observable $\Delta m_{B_{s}^0}$ alone is quite weak.

\item With the $\eta_u^*\eta_d$ terms neglected, the off-diagonal matrix element and hence the mass difference $\Delta m_{B_{s}^0}$ depend on the parameter $\eta_u$ only via $|\eta_u|$. So the case with complex couplings can be inferred trivially from that with real couplings.
\end{itemize}

\begin{figure}[t]
\centering
\includegraphics[width=0.9\textwidth]{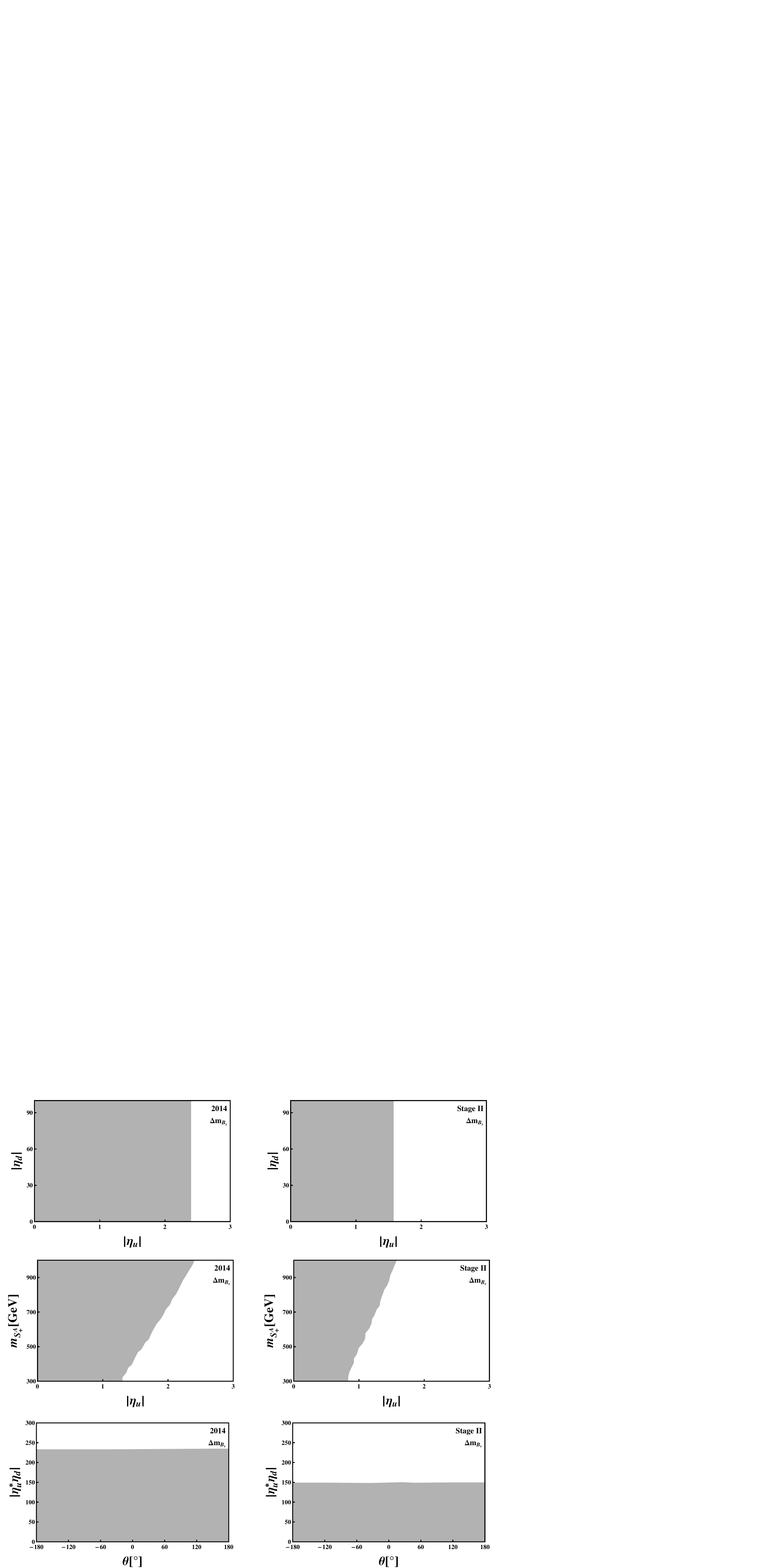}
\caption{\small Bounds on the model parameters from $\Delta m_{B_s^0}$ in the case of complex couplings. }
\label{constraint from bsbsmixng complex}
\end{figure}

The bounds on the model parameters derived from $\Delta m_{B^0_s}$ are shown in Fig.~\ref{constraint from bsbsmixng complex} in the case of complex couplings. As expected, the regions with large $|\eta_u|$ are already excluded due to the good agreement between the SM prediction and the experimental measurement for $\Delta m_{B^0_s}$. The current bounds on the model parameters can be significantly improved by the stage II projection. There are, however, almost no constraints from $\Delta m_{B^0_s}$ on the coupling $|\eta_d|$ and the phase $\theta$. Except for being a little bit weaker, the bounds on the model parameters derived from $\Delta m_{B_{d}^0}$ are similar to that from $\Delta m_{B_{s}^0}$, and will not be shown here. 

\subsection{\texorpdfstring{$\mathbf{\boldsymbol{B_{s,d} \rightarrow \mu^+\mu^-}}$}{Lg} within the MW model}
\label{sec:bsd2mumu_result}

The $B_s\to \mu^+\mu^-$ decay has been observed with a statistical significance exceeding $6\sigma$ and, furthermore, a $3\sigma$ evidence for the $B_d\to \mu^+\mu^-$ decay is obtained by the CMS and LHCb experiments~\cite{CMS:2014xfa}. Both measurements are statistically compatible with the SM predictions and place, therefore, stringent constraints on physics beyond the SM~\cite{Li:2014fea,Buras:2013uqa}. We now investigate the constraints on the model parameters from the branching ratios of these two processes.

As mentioned already in Sec.~\ref{sec:bsd2mumu}, the rare leptonic B-meson decays are dominated by the Wilson coefficient $C_{10}$ both within the SM and in the MW model. To improve the accuracy of the SM prediction, we use the fitting formula for $C_{10}^{\rm SM}$ given by Eq.~(4) of Ref.~\cite{Bobeth:2013uxa}, which has been transformed to our convention for the effective weak Hamiltonian~\cite{Li:2014fea}
\begin{equation} \label{eq:C10SMfit}
C^{\rm SM}_{10} = -0.9604 \left[\frac{M_t}{173.1~\mathrm{GeV}}\right]^{1.52} \left[\frac{\alpha_s(M_Z)}{0.1184}\right]^{-0.09} + 0.0224 \left[\frac{M_t}{173.1~\mathrm{GeV}}\right]^{0.89}\, \left[\frac{\alpha_s(M_Z)}{0.1184}\right]^{-0.09}.
\end{equation}
As the NP contribution to $C_{10}$ in the MW model involves only the parameter $\eta_u$ and the mass $m_{S^A_+}$, stringent bounds on them are expected from these processes. For a benchmark value $m_{S^A_+}=500~{\rm GeV}$, the coefficients $P$ and $S$ read numerically
\begin{align}
P &= -1.003-0.138|\eta_u|^2 + \frac{m_{B_s}^2}{m_W^2}\Bigl(0.586+0.003 \eta_d \eta_u^* +0.068 |\eta_u|^2\,\Bigr)\label{eq:numericalbstomumup},\\[0.2cm]
S &= \frac{m_{B_s}^2}{m_W^2}\biggl[-0.306 - \lambda_1 \Bigl(0.020 \eta_d \eta_u^* - 0.006 |\eta_u|^2\,\Bigr)-0.075 \eta_d \eta_u^* +0.008 |\eta_u|^2\,\biggr]\label{eq:numericalbstomumus}\,,
\end{align}
from which, we make the following two observations:
\begin{itemize}
\item In Eq.~\eqref{eq:numericalbstomumup}, the term proportional to $|\eta_u|^2$ corresponds to the NP contribution to $C_{10}$, and is not suppressed by the factor $m_{B_s}^2/m_W^2$. It contributes always constructively to $C_{10}^{\rm SM}$, irrespectively of whether $\eta_u$ is real or complex.

\item The terms involving $\eta_d$ and $\lambda_1$ are all suppressed by a factor of $m_{B_s}^2/m_W^2$ and their effects are, therefore, negligibly small unless when $\eta_d$ or $\lambda_1$ is abnormally large.
\end{itemize}

\begin{figure}[t]
\centering
\includegraphics[width=0.9\textwidth]{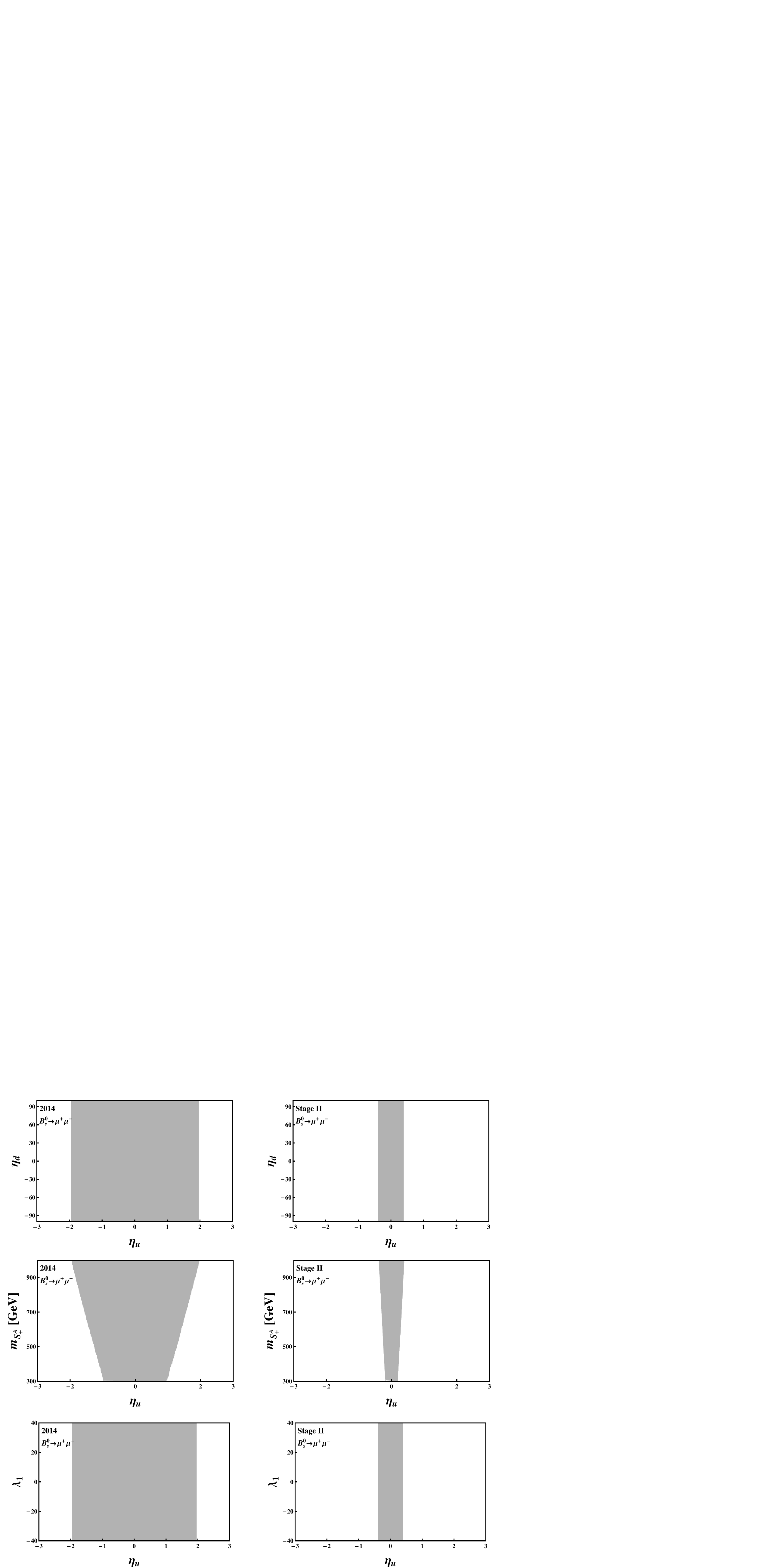}
\caption{\small Constraints on the model parameters from $\overline{ {\mathcal B}}({{{B}}_{s}}\to \mu^+\mu^-)$ in the case of real couplings.}
\label{constraint from bstomumu real}
\end{figure}

\begin{figure}[ht]
\centering
\includegraphics[width=0.9\textwidth]{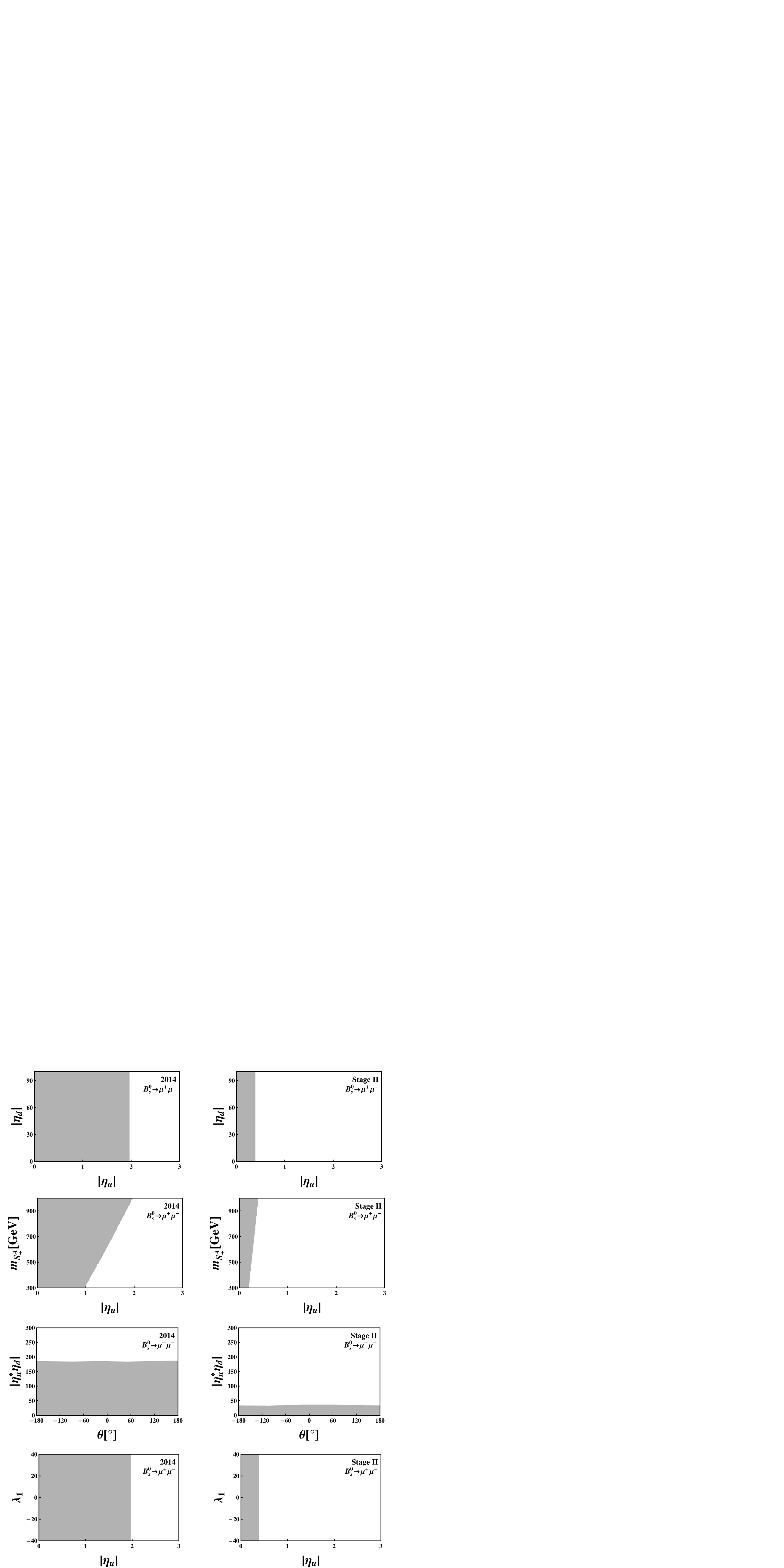}
\caption{\small Constraints on the model parameters from $\overline{ {\mathcal B}}({{{B}}_{s}}\to \mu^+\mu^-)$ in the case of complex couplings. }
\label{constraint from bstomumu complex}
\end{figure}

Under the constraints from the branching ratio $\overline{\mathcal B} (B_s\to \mu^+\mu^-)$, the allowed parameter spaces are shown in Figs.~\ref{constraint from bstomumu real} and \ref{constraint from bstomumu complex}, corresponding to the case of real and complex couplings, respectively. As expected, the observable can provide strong bounds on $|\eta_u|$ and $m_{S^A_+}$, but not on $\eta_d$ and $\lambda_1$, irrespective of whether they are real or complex. In the stage II epoch, as both the experimental and the theoretical uncertainties will become much smaller and, furthermore, due to the constructive interference between the SM and the NP contributions, the parameter spaces allowed by the observable $\overline{\mathcal B} (B_s\to \mu^+\mu^-)$ will be reduced significantly.

\begin{figure}[t]
\centering
\includegraphics[width=0.9\textwidth]{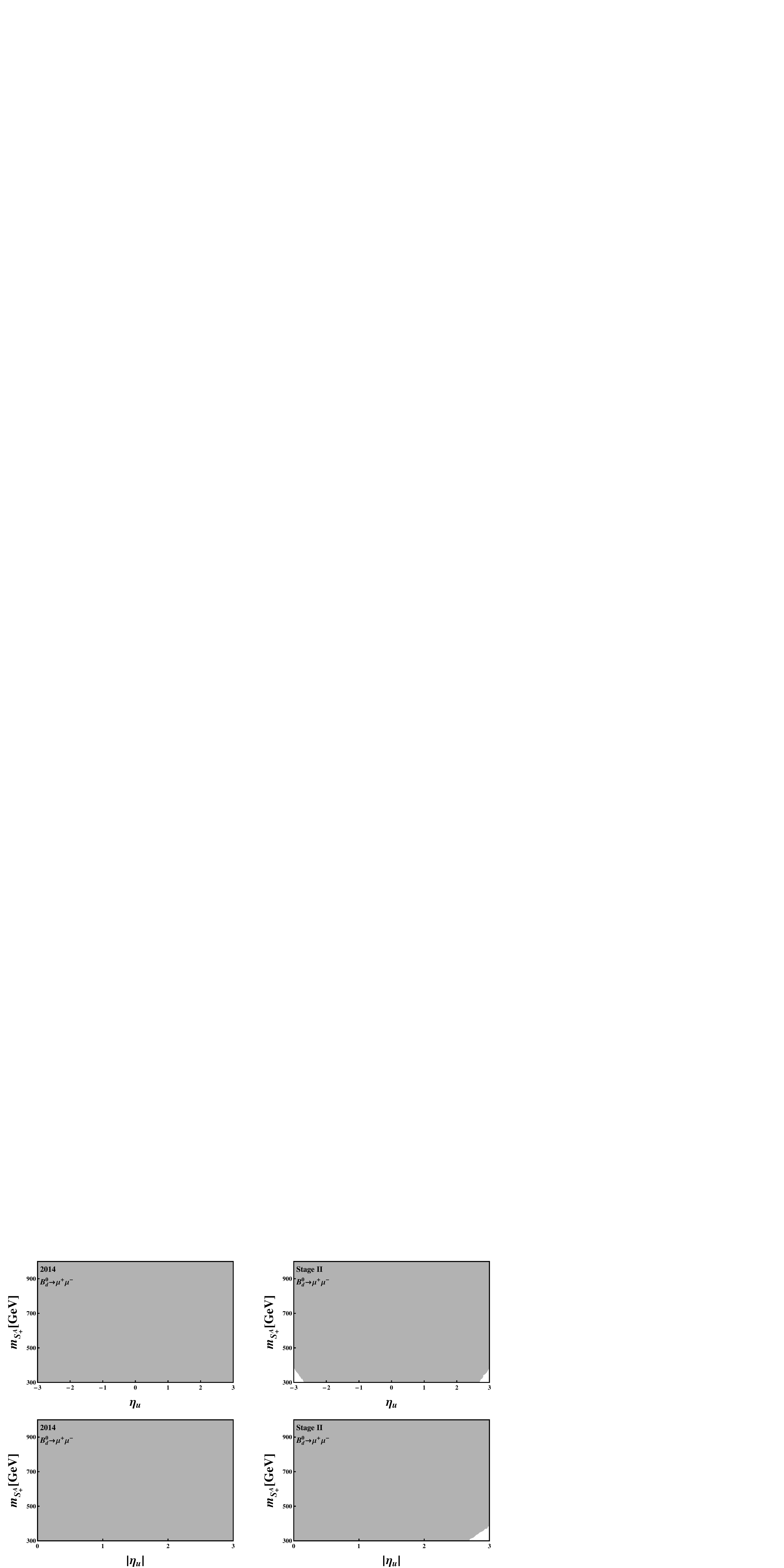}
\caption{\small Constraints on the model parameters from $\overline{ {\mathcal B}}({{{B}}_{d}}\to \mu^+\mu^-)$ in the case of real (the upper panel) and complex ( the lower panel) couplings.}
\label{constraint from bdtomumu real and complex}
\end{figure}

As shown in Fig.~\ref{constraint from bdtomumu real and complex}, the constraints from $\overline{\mathcal B} (B_d\to \mu^+\mu^-)$ are much weaker compared to that from $\overline{\mathcal B} (B_s\to \mu^+\mu^-)$, which is understandable due to the following two facts:
\begin{itemize}
\item Besides the much larger error bars, the experimental data on $\overline{\mathcal B}(B_d\to \mu^+\mu^-)$ is almost 2$\sigma$ above the corresponding SM prediction. An opposite behaviour is, however, observed for $\overline{\mathcal B} (B_s\to \mu^+\mu^-)$.

\item For both of these two processes, due to the constructive interference  between the SM and NP contributions, the NP effect can only enhance the theoretical prediction.
\end{itemize}

\subsection{\texorpdfstring{$\mathbf{\boldsymbol{R_b^0}}$}{Lg} within the MW model}
\label{sec:Z2bbar_result}

\begin{figure}[t]
\centering
\includegraphics[width=0.95\textwidth]{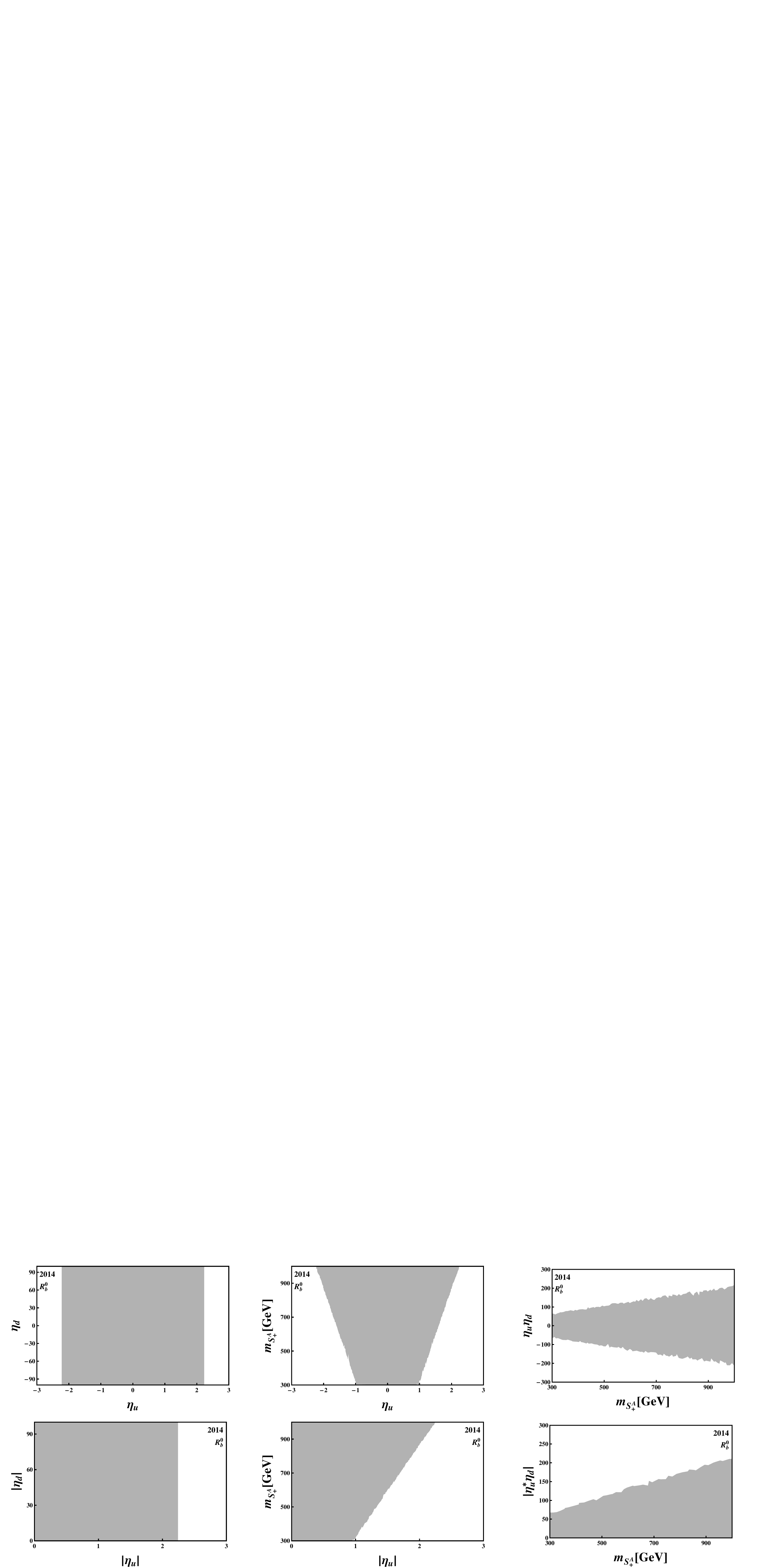}
\caption{\small Constraints on the model parameters from $R_b^0$ in the case of real (the upper panel) and complex (the lower panel) couplings. }
\label{constraint from ztobbbar complex}
\end{figure}

For the $Z\rightarrow b\bar{b}$ decay, the NP effect resides only in the left-handed and right-handed $Zb\bar{b}$ couplings ${\bar{g}}_{L,R}^b$ and, in the assumption of negligible oblique corrections due to the second scalar doublet, only the parameters $\eta_u$, $\eta_d$ and $m_{S^A_+}$ are involved~\cite{Degrassi:2010ne}. Taking $m_{S^A_+}=500~{\rm GeV}$ as a benchmark, we get numerically
\begin{align}
\hspace{-0.35cm}({\bar{g}}_{L}^b)^2 = 0.18-5.5\times 10^{-4}|\eta_u|^2+2.34\times 10^{-7}|\eta_u|^4-5.5\times 10^{-10}|\eta_d|^2-2.9\times 10^{-10}|\eta_u\eta_d|^2, \label{eq:glb}
\end{align}
\begin{align}
\hspace{-0.35cm}({\bar{g}}_{R}^b)^2 = 0.006-1.9\times 10^{-7}|\eta_u|^2-3.1\times 10^{-8}|\eta_u|^4-2.6\times 10^{-8}|\eta_d|^2+1.9\times 10^{-11}|\eta_u\eta_d|^2. \label{eq:grb}
\end{align}
Together with Eq.~\eqref{eq:rbdefinitionthrory}, the following features are observed:
\begin{itemize}
\item The ratio $R_b^0$ is sensitive to the parameter $\eta_u$, but not to $\eta_d$. This is due to the fact that the terms controlled by $\eta_d$ are suppressed by a factor of $m_b^2/m_t^2$ with respect to the ones controlled by $\eta_u$.

\item The NP effect is dominated by the $|\eta_u|^2$ terms and contributes destructively to the SM one. So the contribution from the MW model will always decrease the value of $R_b^0$.

\item As both of the couplings ${\bar{g}}_{L}^b$ and ${\bar{g}}_{R}^b$ depend only on the magnitudes of $\eta_u$ and $\eta_d$, no information on the phase $\theta$ can be inferred from the ratio $R_b^0$.
\end{itemize}

In Fig.~\ref{constraint from ztobbbar complex}, we present the constraints on the model parameters from $R_b^0$. As expected, there are almost no constraints on $\eta_d$, irrespective of whether it is real or complex. In both cases, only small values for $|\eta_u|$ are allowed by the current data. It is also noted that the constraints in stage II remain almost the same as that in 2014 and are, therefore, not shown here.

\subsection{\texorpdfstring{$\mathbf{\boldsymbol{\mathcal B(B \to X_s \gamma)}}$}{Lg}, \texorpdfstring{$\mathbf{\boldsymbol{\Delta(K^*\gamma)}}$}{Lg} and \texorpdfstring{$\mathbf{\boldsymbol{\Delta(\rho\gamma)}}$}{Lg} within the MW model}
\label{sec:b2sdgamma_result}

\begin{figure}[t]
\centering
\includegraphics[width=0.85\textwidth]{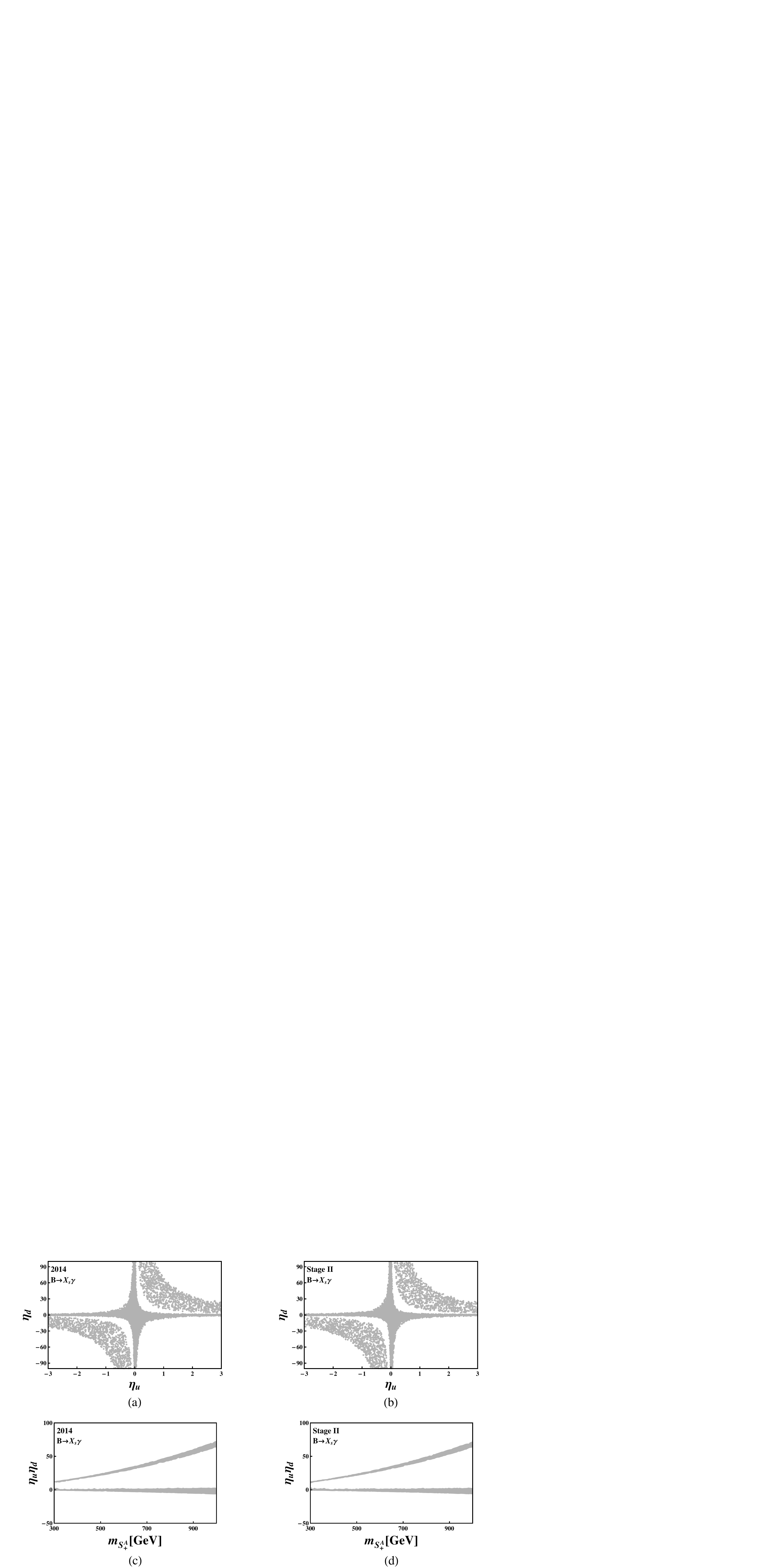}
\caption{\small Constraints on the model parameters from $B\rightarrow X_s \gamma$ in the case of real couplings.}
\label{constraint from btosgamma real}
\end{figure}

\begin{figure}[t]
\centering
\includegraphics[width=0.85\textwidth]{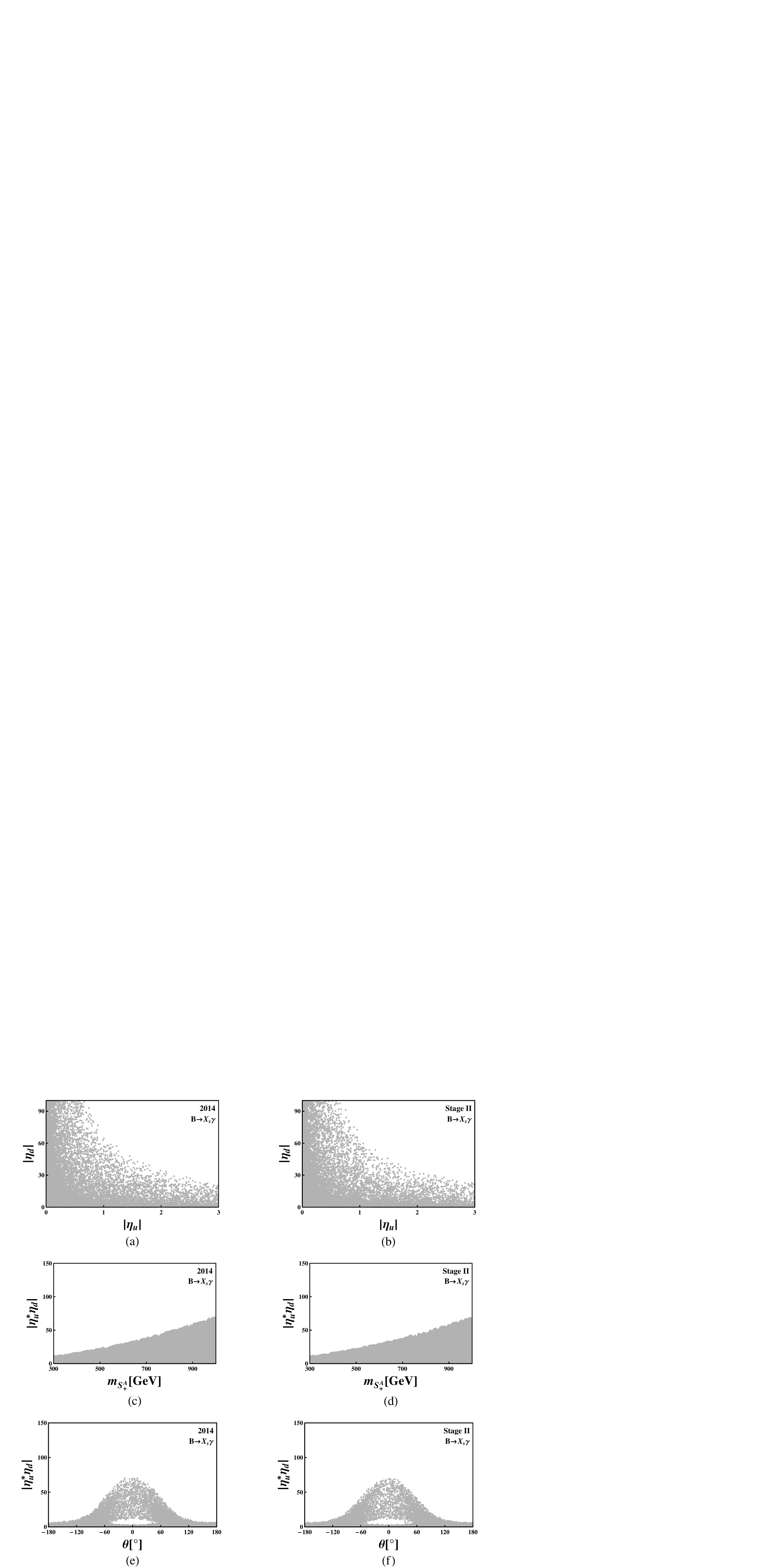}
\caption{\small Constraints on the model parameters from $B\rightarrow X_s \gamma$ in the case of complex couplings. }
\label{constraint from btosgamma complex}
\end{figure}

The dominant contributions to the radiative $b\rightarrow s(d) \gamma$ transitions in the MW model are related to the effective Wilson coefficients of dipole operators $C_{7}^{\rm eff}(\mu_b)$ and $C_{8}^{\rm eff}(\mu_b)$. In the leading logarithmic approximation and taking $m_{S^A_+}= 500 $ GeV, we get numerically
\begin{align}
C_7^{\rm eff}(\mu_b)/C_{7, \rm SM}^{\rm eff} (\mu_b) &= 1 -0.11 \eta_u^* \eta_d +0.012 |\eta_u|^2, \label{eq:numerical C7} \\[0.2cm]
C_8^{\rm eff}(\mu_b)/C_{8, \rm SM}^{\rm eff}(\mu_b) &=1 +0.13 \eta_u^* \eta_d  -0.022 |\eta_u|^2. \label{eq:numerical C8}
\end{align}
It is observed that the dominant NP contribution comes from the term proportional to the combination $\eta_u^* \eta_d$ when $\eta_u$ and $\eta_d$ are comparable. In the LO approximation, the branching ratio ${\mathcal B}(B\rightarrow X_s \gamma)$ is known to be proportional to $|C_7^{\rm eff}(\mu_b)|^2$. This observable can, therefore, provide stringent constraints on the combination $\eta_u^* \eta_d$. This is shown in Figs.~\ref{constraint from btosgamma real} and \ref{constraint from btosgamma complex}, corresponding to the case of real and complex couplings, respectively. Based on these plots, we make the following observations:
\begin{itemize}
\item As shown in Fig.~\ref{constraint from btosgamma real}, in the case of real couplings, there are two allowed regions under the constraint from ${\mathcal B}(B \to X_s \gamma)$. The region close to the axes corresponds to the case when the NP contribution is small and constructive to the SM one. The other region, in which large and same-sign values for $\eta_u$ and $\eta_d$ are allowed simultaneously, corresponds to the case when the NP contribution is destructive to the SM one and makes the coefficient $C_7^{\rm eff}(\mu_b)$ sign-flipped. It is also noted that, the regions with simultaneously large values for $\eta_u$ and $\eta_d$ but with opposite signs are already excluded.

\item As shown in Fig.~\ref{constraint from btosgamma complex}(a-b), the current data on ${\mathcal B}(B \to X_s \gamma)$ also gives strong constraints on the parameters $|\eta_u|$ and $|\eta_d|$ in the case of complex couplings. Furthermore, the interference between the SM and NP contributions depends on the phase $\theta$, as shown in Fig.~\ref{constraint from btosgamma complex}(e-f). When $\theta \approx \pm 180^\circ$, the interference is constructive and only a small region with smaller $|\eta_u^* \eta_d|$ remains. While for $\theta \approx \pm 0^\circ$, the interference becomes destructive and there exist two allowed regions, corresponding to the case with relatively small NP influence~(the lower region) and the case when the NP contribution is about twice the size of the SM one~(the upper region), respectively.

\item As shown in Figs.~\ref{constraint from btosgamma real}(c-d) and \ref{constraint from btosgamma complex}(c-d), the allowed values for the combination $\eta_u \eta_d$ (in the case of real couplings) and $|\eta_u^* \eta_d|$ (in the case of complex couplings) increase with the mass of the charged colour-octet scalars, as they should be.
\end{itemize}

\begin{figure}[t]
\centering
\includegraphics[width=0.99\textwidth]{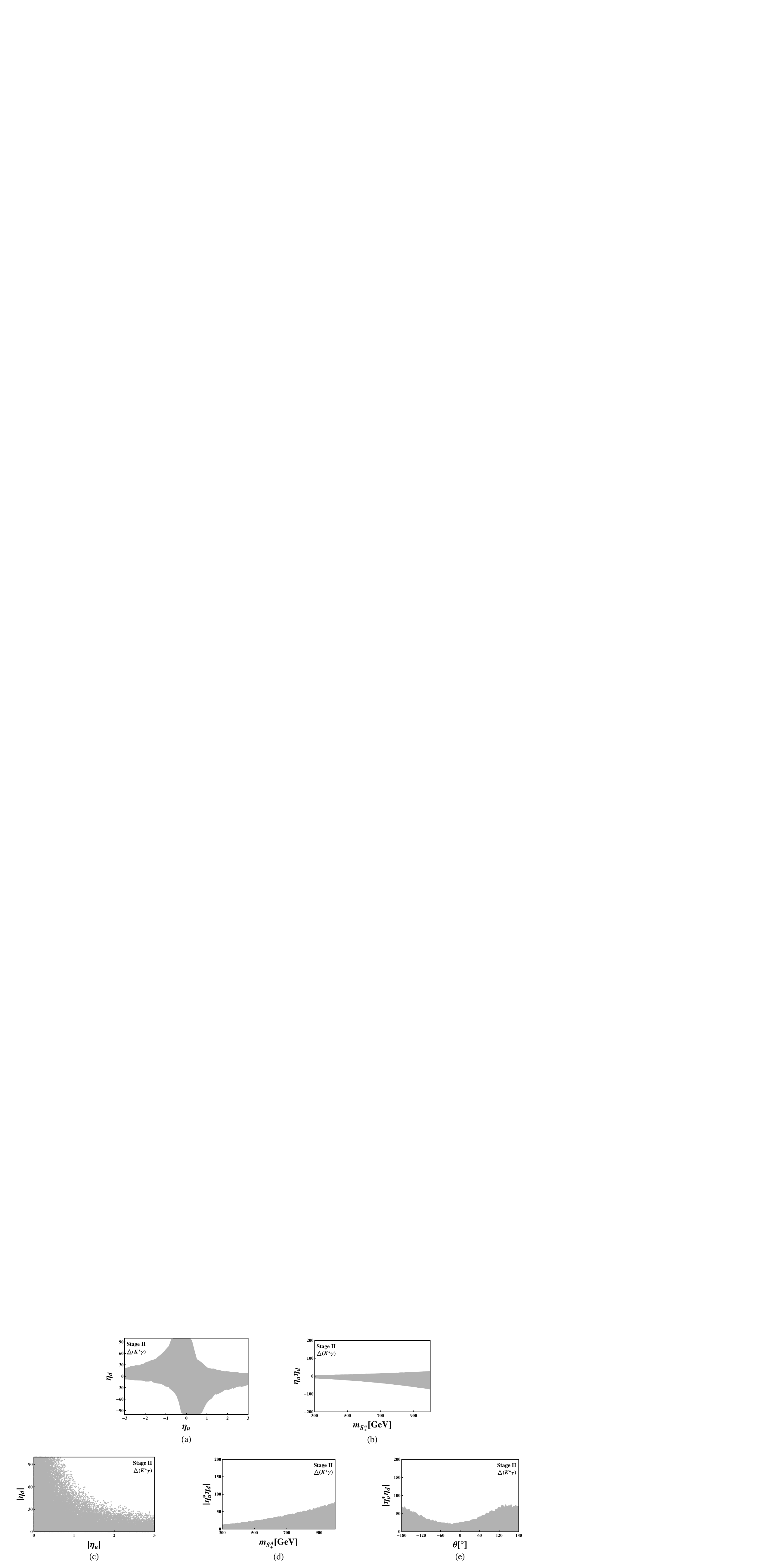}
\caption{\small Constraints on the model parameters from $\Delta(K^*\gamma)$: (a--b) for real couplings, while (c--e) for complex couplings. }
\label{constraint from btokstargamma real and complex}
\end{figure}

\begin{figure}[t]
\centering
\includegraphics[width=0.99\textwidth]{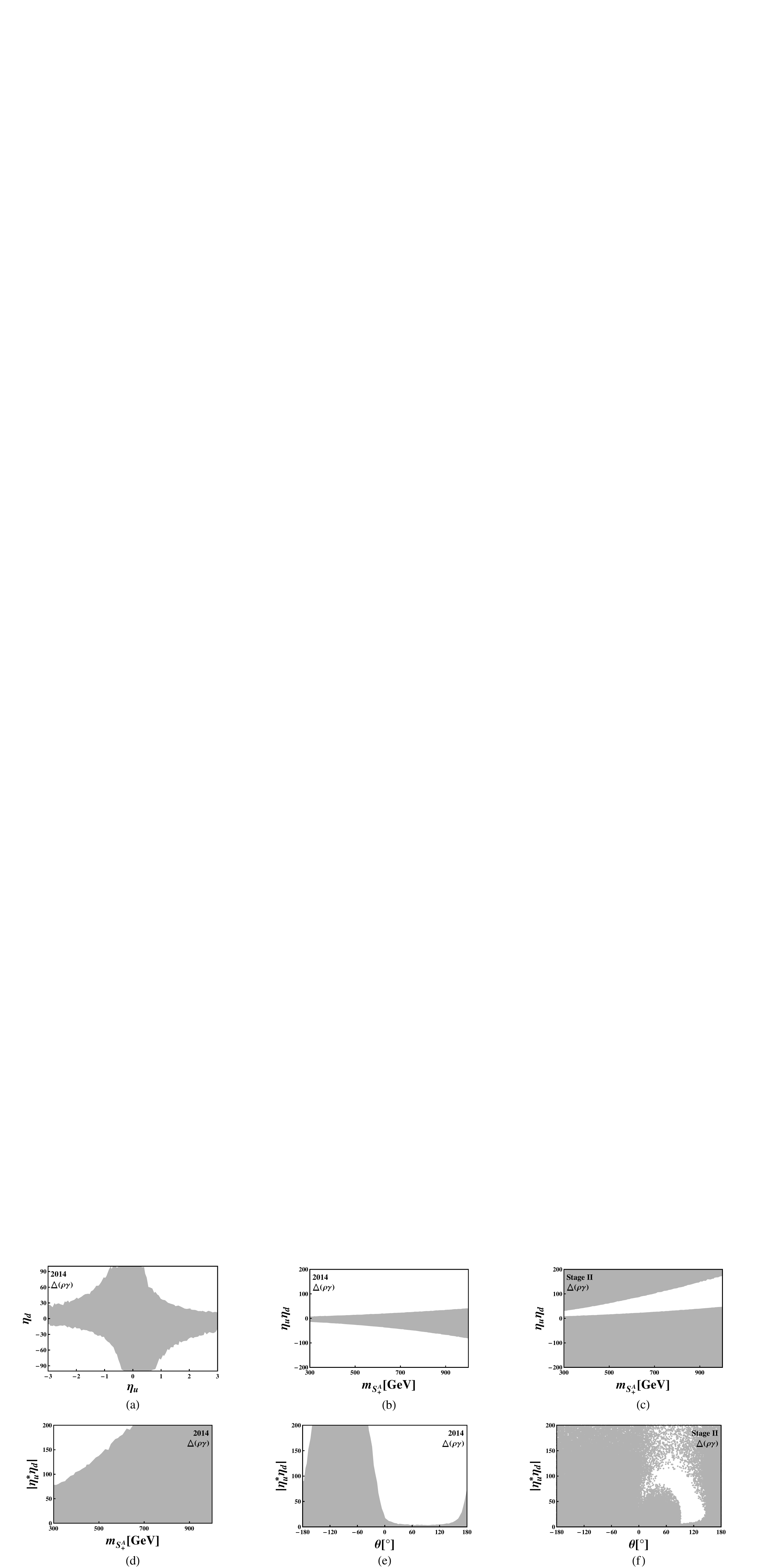}
\caption{\small Constraints on the model parameters from $\Delta(\rho\gamma)$: (a--c) for real couplings, while (d--f) for complex couplings. }
\label{constraint from btorhogamma real and complex}
\end{figure}

For the exclusive $B\rightarrow V \gamma$ decays, the branching ratios could not provide further constraints on the model parameters compared to that of the inclusive $B \to X_s \gamma$ decay. However, the two isospin asymmetries defined by Eqs.~\eqref{eq:isospin1} and \eqref{eq:isospin2}, which show a different dependence on the model parameters from that of the branching ratios, may provide complementary constraints~\cite{Li:2013vlx}. Under the constraints from $\Delta(K^*\gamma)$ and $\Delta(\rho\gamma)$, we show in Figs.~\ref{constraint from btokstargamma real and complex} and \ref{constraint from btorhogamma real and complex} the allowed regions for the model parameters. From these plots, we make the following observations:
\begin{itemize}
\item As for $\Delta(K^*\gamma)$, the 2014 data has almost no constraints on the model parameters and are, therefore, not shown here. In the stage II projection, however, because of the reduced uncertainties in both experimental measurements and theoretical predictions, the constraints from this observable become significant, which results in an upper bound on the combination $|\eta_u^*\eta_d|$, as shown in Figs.~\ref{constraint from btokstargamma real and complex}(b) and \ref{constraint from btokstargamma real and complex}(d). Strong correlations are also observed in the $\eta_u-\eta_d$, $m_{S^A_+}-\eta_u\eta_d$ (in the case of real couplings), as well as in the $|\eta_u| -|\eta_d|$, $m_{S^A_+}-|\eta_u^* \eta_d|$ and $\theta-|\eta_u^* \eta_d|$ (in the case of complex couplings) planes, as shown in Figs.~\ref{constraint from btokstargamma real and complex}(a), \ref{constraint from btokstargamma real and complex}(b), \ref{constraint from btokstargamma real and complex}(c), \ref{constraint from btokstargamma real and complex}(d) and \ref{constraint from btokstargamma real and complex}(e), respectively.

\item As for $\Delta(\rho\gamma)$, the 2014 data provides already an upper bound on $|\eta_u\eta_d|$ in the case of real couplings, as shown in Fig.~\ref{constraint from btorhogamma real and complex}(b). Moreover, under the constraint of the 2014 data, strong correlations exist in the $\eta_u-\eta_d$, $m_{S^A_+}-\eta_u\eta_d$ (in the case of real couplings), as well as in the $m_{S^A_+}-|\eta_u^* \eta_d|$ and $\theta-|\eta_u^*\eta_d|$ (in the case of complex couplings) planes, which are plotted in Figs.~\ref{constraint from btorhogamma real and complex}(a), \ref{constraint from btorhogamma real and complex}(b), \ref{constraint from btorhogamma real and complex}(d) and \ref{constraint from btorhogamma real and complex}(e), respectively. In the stage II epoch, however, only weak bounds on the model parameters can be obtained, as shown in Figs.~\ref{constraint from btorhogamma real and complex}(c) and \ref{constraint from btorhogamma real and complex}(f). This is because the allowed range for NP contribution is much larger than that in 2014, when the central value of the future measurement is chosen to coincide with the SM prediction and the interference between the SM and NP contributions is constructive.

\item As can be seen from Figs.~\ref{constraint from btosgamma complex}(e-f), Fig.~\ref{constraint from btokstargamma real and complex}(e) and Fig.~\ref{constraint from btorhogamma real and complex}(e-f), the branching ratio ${\mathcal B}(B\to X_s\gamma)$ and the two isospin asymmetries $\Delta(K^*\gamma)$ and $\Delta(\rho\gamma)$ exhibit a different dependence on the phase $\theta$. So a combined constraint from these observables should be more stringent, which will be discussed later~\cite{Li:2013vlx}.
\end{itemize}

\subsection{Combined analysis within the MW model}
\label{sec:combined constraints}

\begin{figure}[t]
\centering
\includegraphics[width=0.99\textwidth]{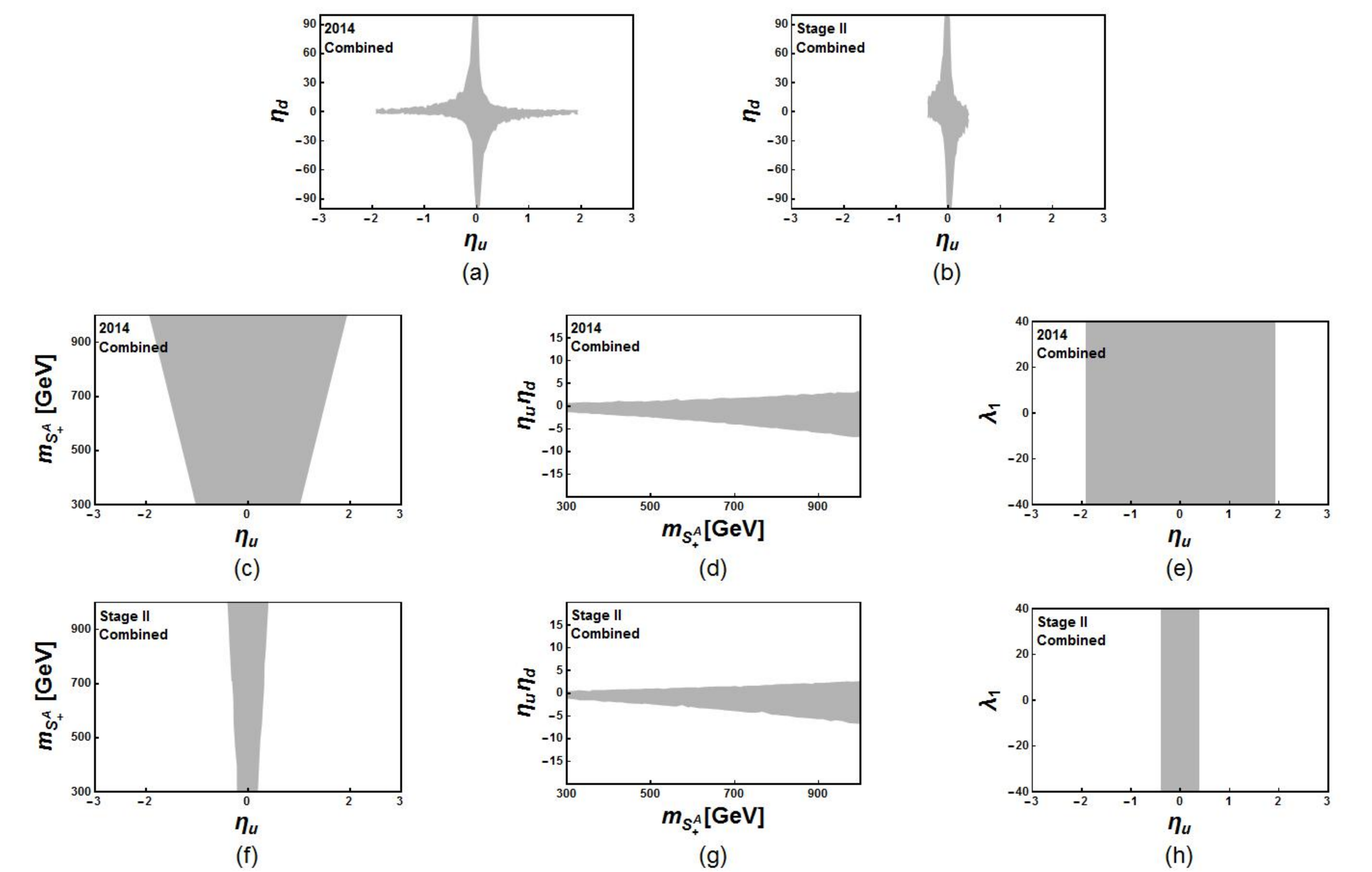}
\caption{\small Combined constraints on the model parameters in the case of real couplings.}
\label{combined constraint  real}
\end{figure}

\begin{figure}[t]
\centering
\includegraphics[width=0.99\textwidth]{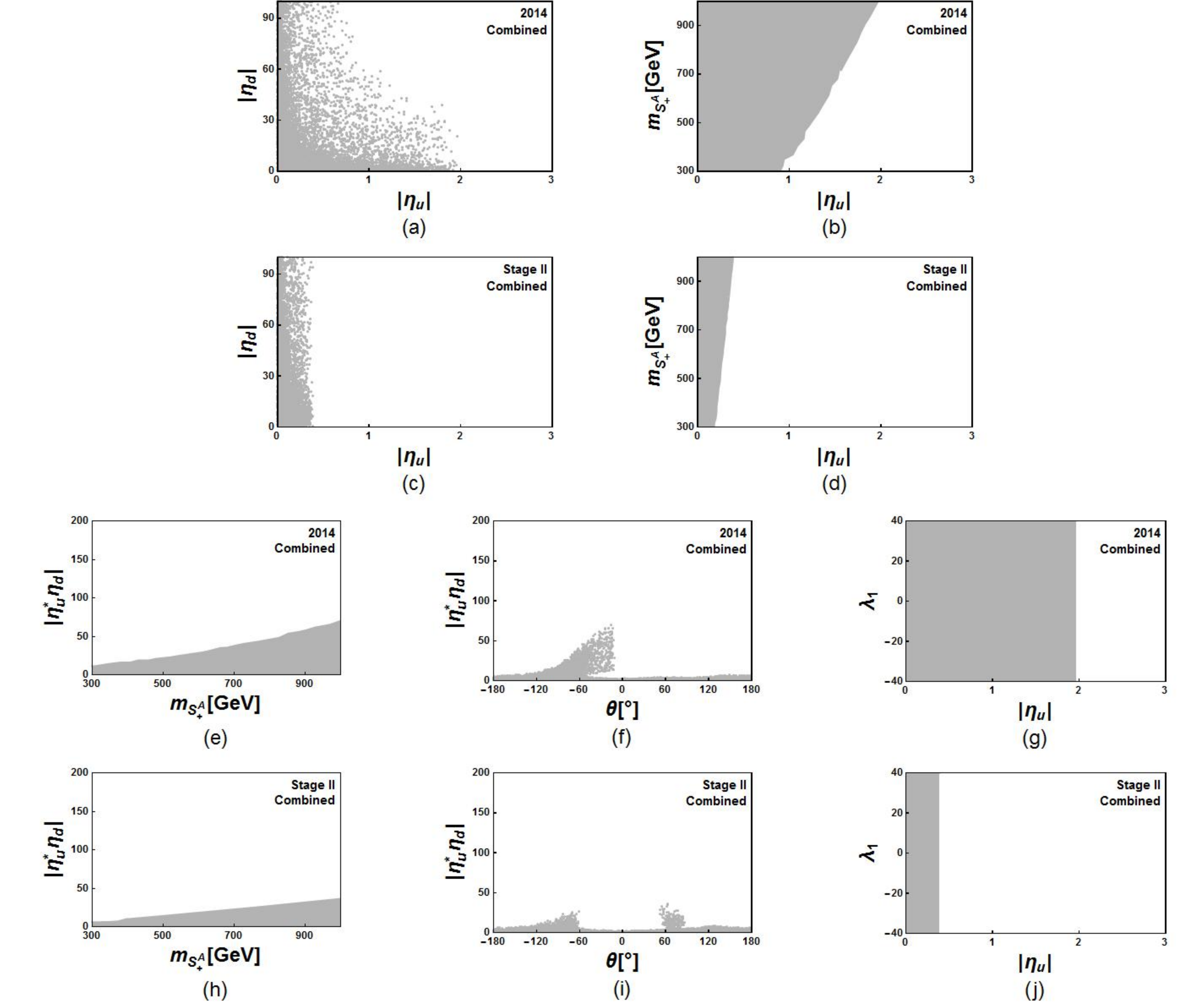}
\caption{\small Combined constraints on the model parameters in the case of complex couplings.}
\label{combined constraint  complex}
\end{figure}

Combining all the constraints from the observables discussed in the previous subsections, the final survived parameter spaces of the MW model are shown in Fig.~\ref{combined constraint  real} and ~\ref{combined constraint  complex}, corresponding to the case of real and complex couplings, respectively. From these plots, we make the following observations:
\begin{description}
\item[I).] {\bf In the case of real couplings}

\begin{itemize}
 \item The parameter $\eta_u$ is strongly bounded by $\Delta m_{{B}_{s,d}^0}$, $\overline{ {\mathcal B}}({{{B}}_{s,d}}\to \mu^+\mu^-)$ and $R^0_b$, with large values for it being already excluded. The most stringent constraint on $\eta_u$ comes from the branching ratio $\overline{ {\mathcal B}}({{{B}}_{s}}\to \mu^+\mu^-)$.

 \item All the observables considered put no bounds on the individual parameter $\eta_d$. However, the combination $\eta_u\eta_d$ is significantly bounded by $\mathcal B(B \to X_s \gamma)$, $\Delta(K^*\gamma)$ and $\Delta(\rho\gamma)$, with only small values for it being allowed, as shown in Figs.~\ref{combined constraint  real}(d) and \ref{combined constraint  real}(g). Strong constraints on the combination $\eta_u\eta_d$ can also be seen in the correlation between $\eta_u$ and $\eta_d$, as shown in Figs.~\ref{combined constraint  real}(a) and \ref{combined constraint  real}(b).

 \item There is no limit on the individual parameter $\lambda_1$, which can be seen from Figs.~\ref{combined constraint  real}(e) and \ref{combined constraint  real}(h). This is mainly because the parameter $\lambda_1$ is relevant only to $\overline{ {\mathcal B}}({{{B}}_{s,d}}\to \mu^+\mu^-)$ and the terms containing $\lambda_1$ are, however, suppressed by a factor of $m_{B_{s,d}}^2/{m_W^2}$ as discussed in Sec.~\ref{sec:bsd2mumu_result}.

 \item Bounds on the mass of the charged colour-octet scalars are always accompanied by the Yukawa coupling parameters $\eta_u$, $\eta_d$ or their combination, which results in strong correlations between $\eta_u$ and $m_{S^A_+}$ or between $\eta_u\eta_d$ and $m_{S^A_+}$, as shown in Figs.~\ref{combined constraint  real}(c-d) and
     \ref{combined constraint  real}(f-g). It is also noted that the observables $\mathcal B(B \to X_s \gamma)$, $\Delta(K^*\gamma)$ and $\Delta(\rho\gamma)$ provide strong correlations only in the $\eta_u \eta_d - m_{S^A_+}$ plane, while the observables $\Delta m_{{B}_{s,d}^0}$, $\overline{ {\mathcal B}}({{{B}}_{s,d}}\to \mu^+\mu^-)$ and $R^0_b$ have significant effects only on the $\eta_u -m_{S^A_+}$ plane.
\end{itemize}
\end{description}

\begin{description}
\item[II).] {\bf In the case of complex couplings}

\begin{itemize}
 \item Similar to the case of real couplings, the magnitude $|\eta_u|$ is strongly bounded by $\Delta m_{{B}_{s,d}^0}$, $\overline{ {\mathcal B}}({{{B}}_{s,d}}\to \mu^+\mu^-)$ and $R^0_b$, with large values for it being already excluded. The most stringent constraint on $|\eta_u|$ comes also from $\overline{ {\mathcal B}}({{{B}}_{s}}\to \mu^+\mu^-)$.

 \item No bounds on the parameter $|\eta_d|$ are provided from the observables discussed in this paper. However, the combination $|\eta_u^*\eta_d|$ is significantly bounded by $\mathcal B(B \to X_s \gamma)$, $\Delta(K^*\gamma)$ and $\Delta(\rho\gamma)$, as shown in Figs.~\ref{combined constraint  complex}(e-f) and \ref{combined constraint  complex}(h-i). The strong constraints on $|\eta_u^*\eta_d|$ from these observables can also be seen in the correlations between $|\eta_u|$ and $|\eta_d|$ shown in Figs.~\ref{combined constraint  complex}(a) and \ref{combined constraint  complex}(c).

 \item As can be seen from Figs.~\ref{combined constraint  complex}(g) and \ref{combined constraint  complex}(j), there is no limit on the parameter $\lambda_1$, which is due to the same reason as in the case of real couplings. It implies also that there is no correlation between $|\eta_u|$ and $\lambda_1$ under the combined constraints.

 \item There exist strong correlations between $|\eta_u|$ and $m_{S^A_+}$ or between $|\eta_u^*\eta_d|$ and $m_{S^A_+}$, which can be seen from Figs.~\ref{combined constraint  complex}(b), \ref{combined constraint  complex}(d-e) and \ref{combined constraint  complex}(h). Here complementary constraints on the model parameters from the observables $\mathcal B(B \to X_s \gamma)$, $\Delta(K^*\gamma)$ and $\Delta(\rho\gamma)$ are crucial.

  \item As the phase $\theta$ is always associated with the combination $|\eta_u^*\eta_d|$, strong correlations between them are observed in the $\theta-|\eta_u^*\eta_d|$ plane, which can be seen from Figs.~\ref{combined constraint  complex}(f) and \ref{combined constraint  complex}(i).
\end{itemize}
\end{description}

We keep the initial samples generated within the ranges specified by Eqs.~\eqref{scenario i value} and \eqref{scenario ii value}, and calculate the corresponding $\chi^2$ associated with each of the samples. It is found that, for the 2014 data, $\chi^2/d.o.f.$ is $10$ within the SM, while the minimal value for it equals $9.5$ (for real couplings) and $6.8$ (for complex couplings) in the MW model. Here $d.o.f.$ denotes the number of free parameters in a considered model and, specific to the MW model, it is $4$ in the case of real and $5$ in the case of complex couplings, respectively. It is noted that $\chi^2_{min}$ for the case of complex couplings is much smaller than that for the case of real couplings. This is mainly due to the different values of $d.o.f.$ in the two cases. It is, therefore, concluded that the MW model is well suited to explain the current data on the observables discussed in the paper, with the resulting $\chi^2$ being significantly smaller than the SM counterpart.

\begin{figure}[t]
\centering
\includegraphics[width=0.45\textwidth]{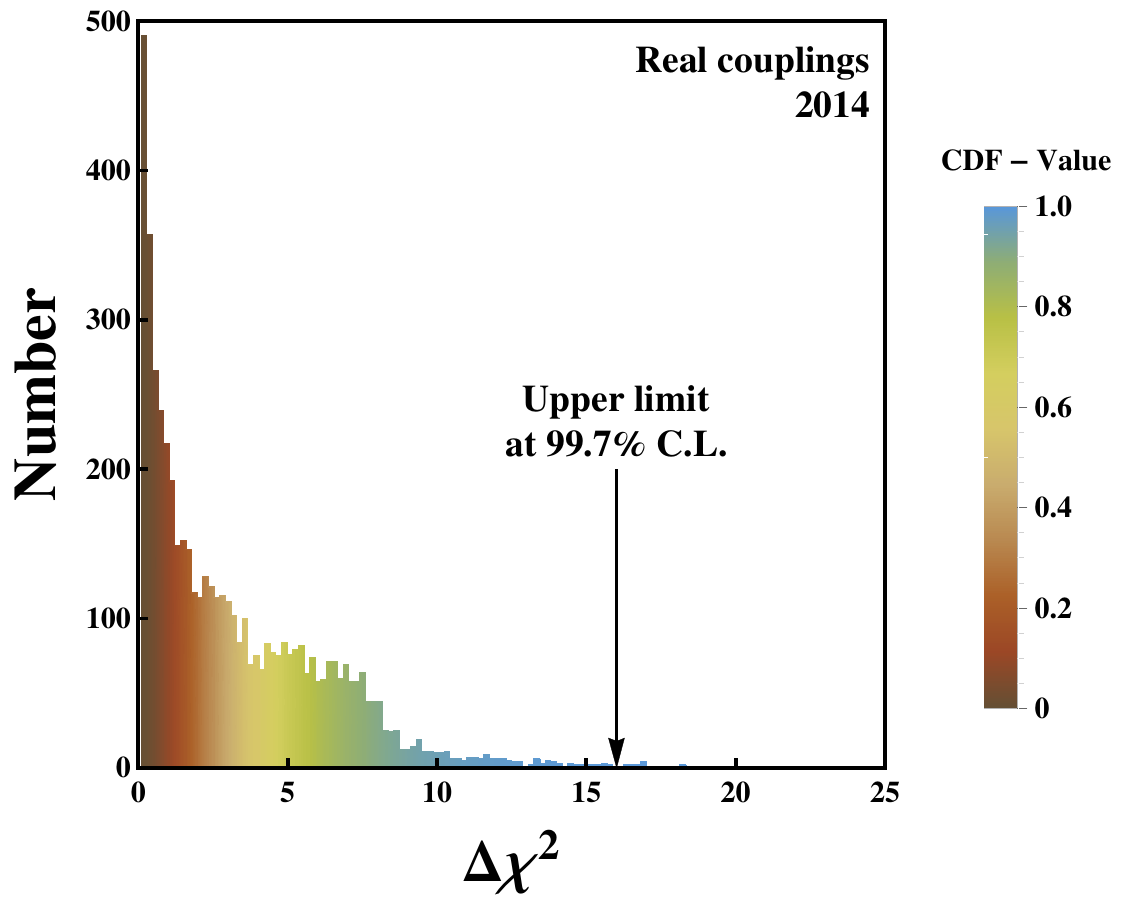}\qquad \quad
\includegraphics[width=0.45\textwidth,origin=c]{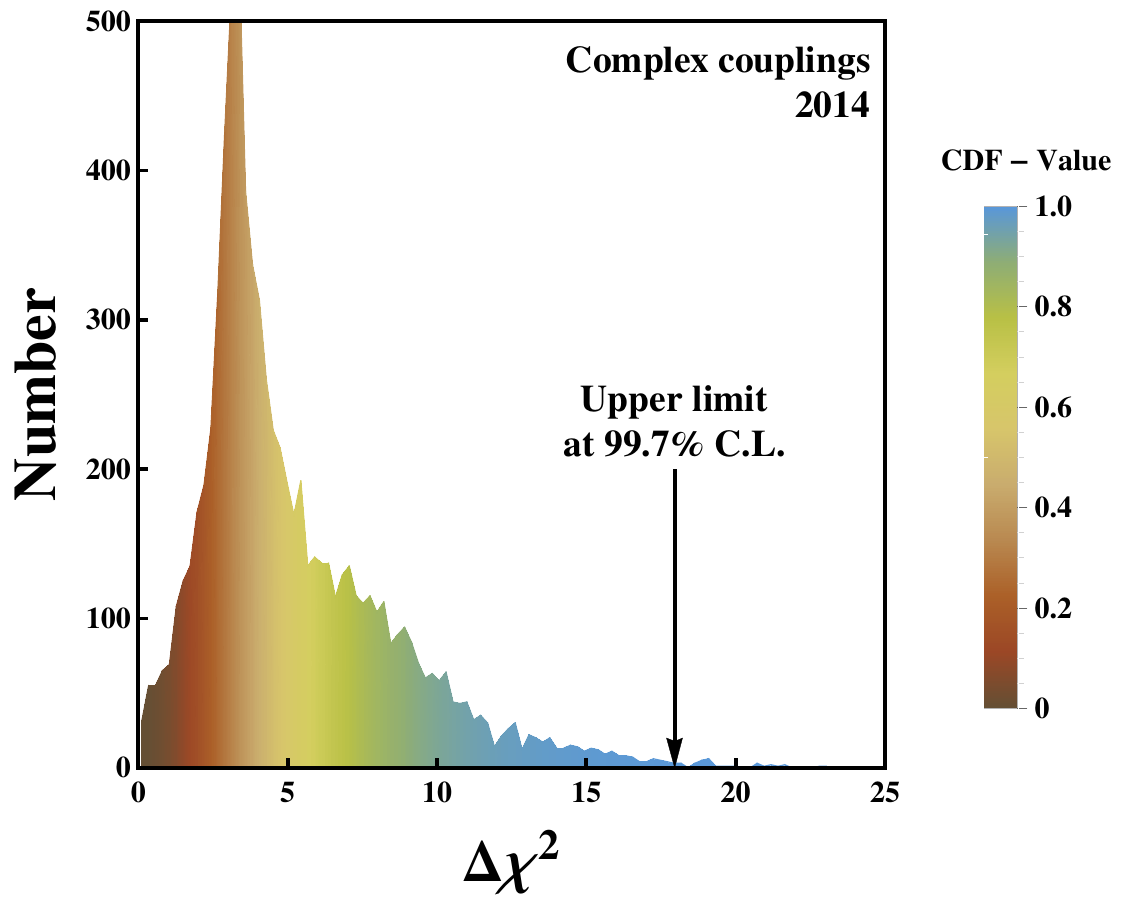}
\caption{\small The $\Delta \chi^2$ distribution of the data samples allowed by the experimental data on the observables discussed in the paper. The different colours correspond to different values for the cumulative distribution function.}
\label{chisqnumber}
\end{figure}

To explore the validity of the final data samples obtained by imposing all the constraints in the way mentioned in Sec.~\ref{sec:procedure}, we calculate the $\Delta \chi^2$ associated with each point in the samples and show in Fig.~\ref{chisqnumber} the $\Delta \chi^2$ distribution of the points. It is concluded that almost all the points lie in the $99.7\%$ confidence intervals, indicating that the procedure adopted in this paper is reliable and conservative.

\subsection{Correlations between different observables in the MW model}
\label{sec:correlation}

\begin{figure}[t]
\centering
\includegraphics[width=0.45\textwidth]{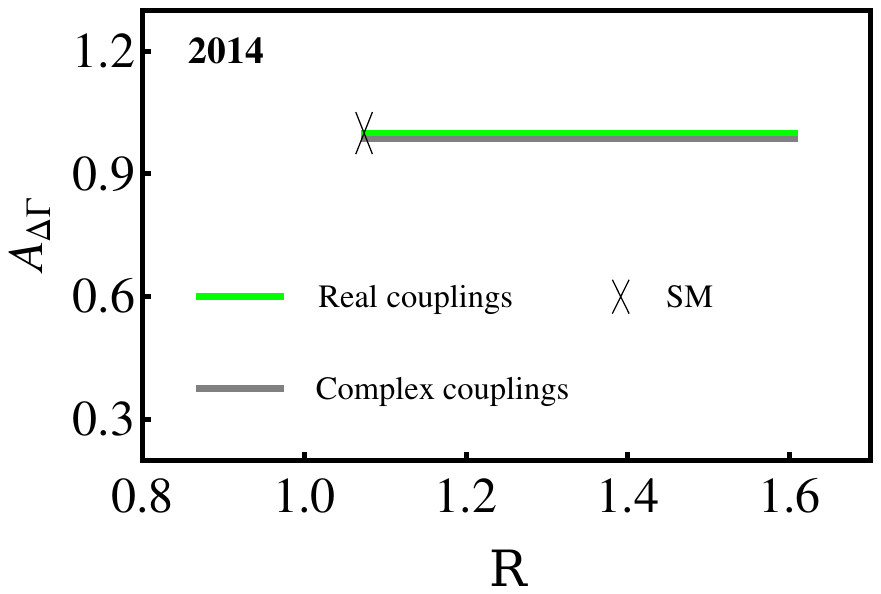}\qquad \quad
\includegraphics[width=0.45\textwidth]{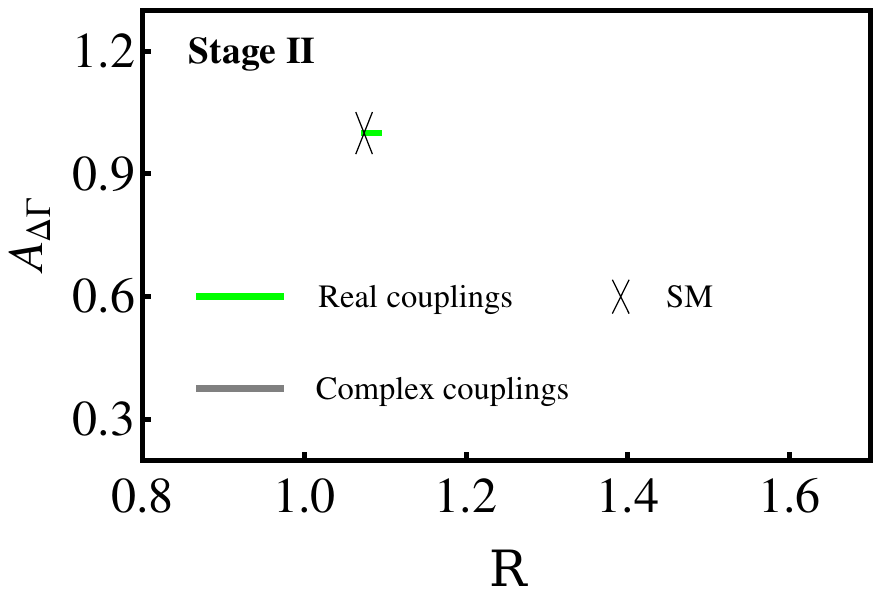}
\caption{\small Correlation plots in the $R-A_{\Delta\Gamma}$ plane under the combined constraints. The allowed regions are shown by green~(dark grey) and grey~(light grey) points, which are obtained in the case of real and complex couplings, respectively. }
  \label{fig:corr:r and adeltgamma}
\end{figure}

Up to now, only $\overline{ {\mathcal B}}({{{B}}_{s,d}}\to \mu^+\mu^-)$ have been measured among the observables for ${{{B}}_{s,d}}\to \mu^+\mu^-$ decays. In order to gain further insights into the model parameters from the future measurements, we present in Fig.~\ref{fig:corr:r and adeltgamma} the correlations between the observables $A_{\Delta\Gamma}$ and $R$ defined respectively by Eqs.~\eqref{eq:adeltgamma} and \eqref{eq:ratioofexpandsm}, in the survived parameter spaces under the constraints discussed in previous subsections.

\begin{figure}[t]
\centering
\includegraphics[width=0.99\textwidth]{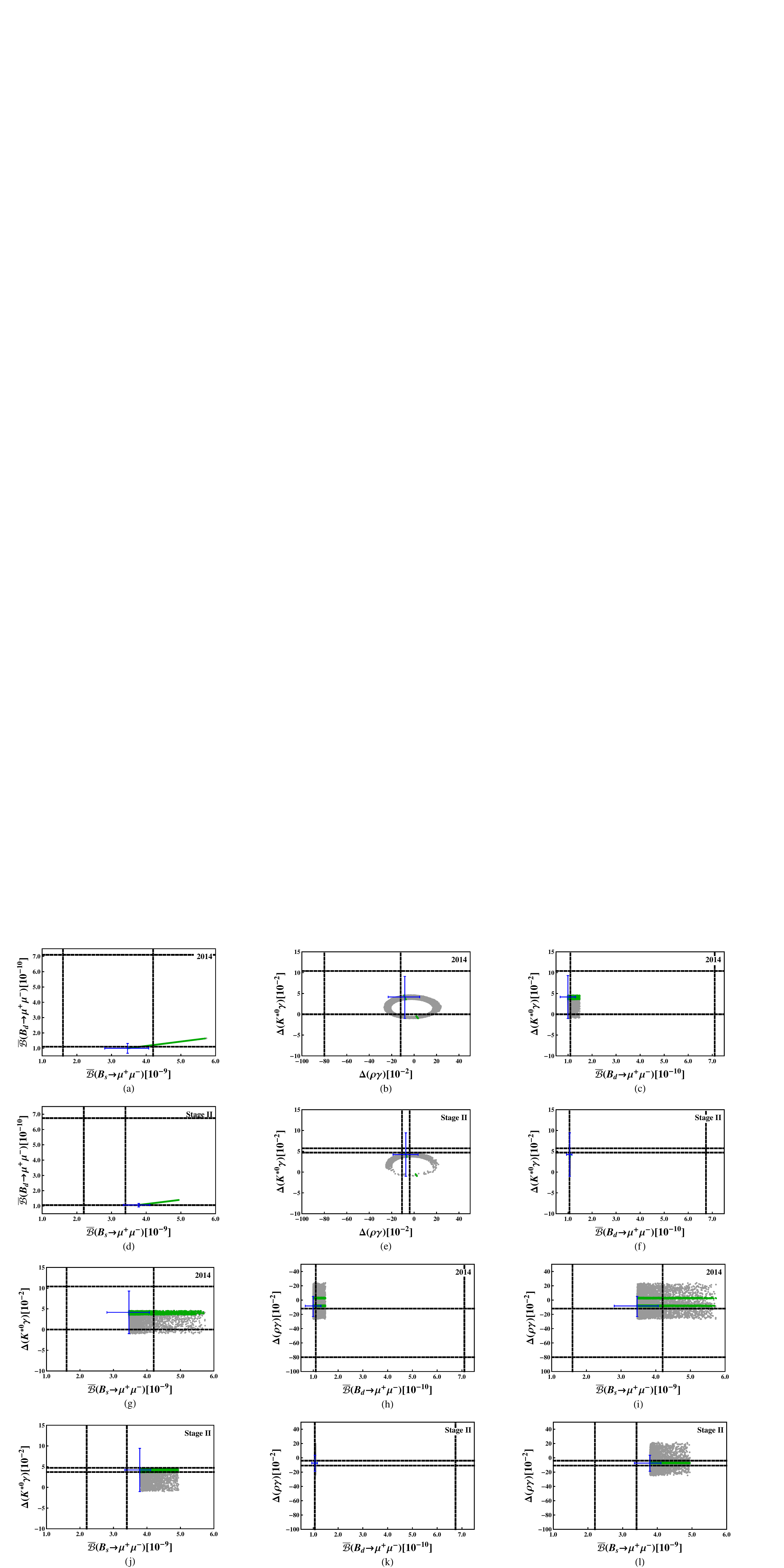}
\caption{\small Correlation plots among $\overline{ {\mathcal B}}({{{B}}_{s,d}}\to \mu^+\mu^-)$, $\Delta(K^*\gamma)$ and $\Delta(\rho\gamma)$. The dashed lines denote the experimental data with $2\sigma$ error bars, while the SM predictions with the corresponding $2\sigma$ range are shown by the blue~(dark) cross. This cross is also applied to each point to account for the theoretical uncertainty at that point. The other captions are the same as in Fig.~\ref{fig:corr:r and adeltgamma}. }
\label{fig:corr:all observables}
\end{figure}

As some of the observables discussed in this paper still suffer large uncertainties, we also investigate the correlations among them under the constraints from the other observables, which is shown in Fig.~\ref{fig:corr:all observables}. For simplicity, we do not consider the theoretical uncertainty at each point in the parameter space. Given the assumption that the relative theoretical uncertainties of these observables are constant at each point in the parameters space, we show the SM predictions with the corresponding theoretical range in these plots, which can be applied to each point to account for the theoretical uncertainty at that point.

From the correlation plots shown in Figs.~\ref{fig:corr:r and adeltgamma} and \ref{fig:corr:all observables}, we make the following observations:
\begin{itemize}
\item In the MW model, large deviations from the SM prediction for the ratio $R$ are still allowed. The observable $A_{\Delta\Gamma}$ remains, however, almost independent of the NP effect, because the NP contribution to $A_{\Delta\Gamma}$ is suppressed by a factor of $m_{B_q}^2/m_W^2$. In the stage II projection, the allowed range for $R$ becomes much smaller due to the stringent constraint from $\overline{ {\mathcal B}}({{{B}}_{s}}\to \mu^+\mu^-)$, which has already been discussed in Sec.~\ref{sec:bsd2mumu_result}. It is also noted that the NP effect always increases the value of $R$ with respect to the SM prediction.

\item Under the constraints from $\Delta m_{{B}_{s,d}^0}$, $\mathcal B(B \to X_s \gamma)$, $R^0_b$, $\Delta(K^*\gamma)$ and $\Delta(\rho\gamma)$, the correlation between $\overline{ {\mathcal B}}({{{B}}_{s}}\to \mu^+\mu^-)$ and $\overline{ {\mathcal B}}({{{B}}_{d}}\to \mu^+\mu^-)$ is very strong, as shown in Figs.~\ref{fig:corr:all observables}(a) and \ref{fig:corr:all observables}(d). The behaviour of the correlation is independent of whether the parameters are real or complex, because both of these two observables are dominated by the terms proportional to $|\eta_u|^2$, as discussed already in Sec.~\ref{sec:bsd2mumu_result}.

\item Under the constraints from $\Delta m_{{B}_{s,d}^0}$, $\mathcal B(B \to X_s \gamma)$, $R^0_b$ and $\overline{ {\mathcal B}}({{{B}}_{s,d}}\to \mu^+\mu^-)$, the observable $\Delta(K^*\gamma)$ is strongly correlated with $\Delta(\rho\gamma)$, as shown in Figs.~\ref{fig:corr:all observables}(b) and \ref{fig:corr:all observables}(e). This is mainly because both of them are sensitive to the value of $\eta_u\eta_d$ (for real couplings), and $|\eta_u^*\eta_d|$ and $\theta$ (for complex couplings). The correlation between $\Delta(K^*\gamma)$ and $\Delta(\rho\gamma)$ in the case of real couplings is, however, different from that for complex couplings.

\item As discussed already in Sec.~\ref{sec:bsd2mumu_result}, the observable $\overline{ {\mathcal B}}({{{B}}_{s}}\to \mu^+\mu^-)$ in the stage II projection can provide stringent bounds on $|\eta_u|$, restricting the theoretical prediction for $\overline{ {\mathcal B}}({{{B}}_{d}}\to \mu^+\mu^-)$ to a very narrow range under the constraint from $\overline{ {\mathcal B}}({{{B}}_{s}}\to \mu^+\mu^-)$, which can be clearly seen from  Figs.~\ref{fig:corr:all observables}(c), \ref{fig:corr:all observables}(f), \ref{fig:corr:all observables}(h) and \ref{fig:corr:all observables}(k).

\item As the observables $\Delta(K^*\gamma)$, $\Delta(\rho\gamma)$ and $\overline{ {\mathcal B}}({{{B}}_{s,d}}\to \mu^+\mu^-)$ exhibit a quite different dependence on the model parameters, the parameter space allowed by $\Delta(K^*\gamma)$ or $\Delta(\rho\gamma)$ is complementary to that allowed by $\overline{ {\mathcal B}}({{{B}}_{s,d}}\to \mu^+\mu^-)$. This fact results in mild correlations between either of $\Delta(K^*\gamma)$ and $\Delta(\rho\gamma)$ and either of $\overline{ {\mathcal B}}({{{B}}_{s}}\to \mu^+\mu^-)$ and $\overline{ {\mathcal B}}({{{B}}_{d}}\to \mu^+\mu^-)$, as shown in Figs.~\ref{fig:corr:all observables}(c), \ref{fig:corr:all observables}(f) and \ref{fig:corr:all observables}(g-l).
\end{itemize}

\subsection{Comparison with collider constraints within the MW model}
\label{subsec:collier constraint}

The phenomenological aspects of the MW model at hadron colliders, including the single and pair productions of these colour-octet scalars, their decays, as well as their implications in Higgs data have been studies extensively in the literature~\cite{Manohar:2006ga,Gresham:2007ri,Carpenter:2011yj}. It should be noted that the LHC signatures of the colour-octet scalars depend strongly on the model parameters, such as the masses of the colour-octet scalars, as well as the values of $\eta_u$, $\eta_d$ and $\lambda_i$. As noticed already in Refs.~\cite{Manohar:2006ga,Gresham:2007ri}, if the colour-octet scalar decays into more than two light quarks, top-quark and jet, or $t\bar{t}$, the existing bounds on the colour-octet scalar mass from ATLAS and CMS would be different and even masses of order $400$~GeV could not be excluded.

The low-energy processes considered by us, on the other hand, involve only the parameters $\eta_u$, $\eta_d$, $m_{S^A_+}$ and $\lambda_1$. It is found that these low-energy flavour observables could put stronger bounds on the parameter $\eta_u$ and the combination $\eta_u\eta_d$, especially with the observables achievable with $50~{\rm fb}^{-1}$ of LHCb and $50~{\rm ab}^{-1}$ of Belle-II data. Furthermore, under these indirect constraints, strong correlations are observed between $\eta_u$ and $m_{S^A_+}$ and between $\eta_u\eta_d$ and $m_{S^A_+}$. All these indirect constraints can provide very important and complementary information on the model parameters, which is useful for the direct searches of the colour-octet scalars at hadron colliders.

\section{Conclusion}
\label{sec:conclusion}

In this paper, we have performed a complete one-loop computation of the short-distance Wilson coefficients for $B_{s,d}-\bar{B}_{s,d}$ mixings and $B_{s,d}\to\ell^+\ell^-$ decays within the MW model. It is found that, in order to get a gauge-independent result, the external momenta of the heavy quarks inside the mesons should be taken into account, and the heavy-quark masses should be kept up to the second order.

Based on these calculations, combined constraints on the model parameters have been derived from the current flavour data, including the $B_{s,d}-{\bar{B}}_{s,d}$ mixings, $B_{s,d}\to\mu^+\mu^-$, $B\to X_s\gamma$, $B\to K^{\ast}\gamma$, $B\to\rho\gamma$, and $Z\to b \bar{b}$ decays. The future sensitivity to the model has also been explored in the observables achievable with $50~{\rm fb}^{-1}$ of LHCb and $50~{\rm ab}^{-1}$ of Belle-II data. Our main conclusions are summarized as follows:
\begin{itemize}
\item The flavour observables exhibit a different dependence on the model parameters and could, therefore, provide complementary bounds on the parameter space. The observables $\Delta m_{{B}_{s,d}^0}$, $\overline{ {\mathcal B}}({{{B}}_{s,d}}\to \mu^+\mu^-)$ and $R^0_b$ are shown to be only sensitive to the parameter $\eta_u$. Furthermore, there exist strong correlations between $\eta_u$ and $m_{S^A_+}$ or between $|\eta_u|$ and $m_{S^A_+}$. On the other hand, the observables $\mathcal B(B \to X_s \gamma)$, $\Delta(K^*\gamma)$ and $\Delta(\rho\gamma)$ are found to be only sensitive to the combination $\eta_u^*\eta_d$. Strong correlations between $\eta_u\eta_d$ and $m_{S^A_+}$ or between $|\eta_u^*\eta_d|$ and either of $m_{S^A_+}$ and $\theta$ are also observed.

\item The correlation between the observables $A_{\Delta\Gamma}$ and $R$ in the $B_s \to \mu^+ \mu^-$ decay, which is characterized by the sizable decay width difference of $B_s$ system, is investigated within the survived parameter space from all the constraints. While large deviations from the SM prediction are still allowed for the ratio $R$, the observable $A_{\Delta\Gamma}$ remains almost independent of the NP contribution. It is also noted that the NP effect always increases the ratio $R$ with respect to the SM prediction.

\item We have calculated the $\chi^2$ value associated with each of the generated samples and concluded that the MW model in the case of complex couplings is more suitable to explain the current flavour data discussed in this paper than in the case of real couplings. It is also concluded that the MW model in both cases is capable of explaining the current data, with the resulting $\chi^2_{min}$ being significantly smaller than that of the SM.

\item Since some of the observables still possess large uncertainties, we have also investigated correlations among the observables $\Delta(K^*\gamma)$, $\Delta(\rho\gamma)$ and $\overline{ {\mathcal B}}({{{B}}_{s,d}}\to \mu^+\mu^-)$, within the allowed parameter space. Some of the correlations are strong and could provide further insights into the MW model, once more precise experimental measurements and theoretical predictions are available in the future.
\end{itemize}

In summary, the results obtained in this paper are helpful to deepen our understanding of the MW model, and could be complementary to the analysis performed at the high-energy frontier of the LHC. With the promising progress expected from the LHCb and Belle-II, and the improved theoretical predictions, more detailed information on the model can be inferred from these low-energy processes.

\section*{Acknowledgements}

The work was supported by the National Natural Science Foundation of China (NSFC) under contract Nos. 11005032, 11225523, 11221504 and 11435003. X.~Li is also supported by the Scientific Research Foundation for the Returned Overseas Chinese Scholars, State Education Ministry, by the Open Project Program of State Key Laboratory of Theoretical Physics, Institute of Theoretical Physics, Chinese Academy of Sciences, China~(No.Y4KF081CJ1), and by the self-determined research funds of CCNU from the colleges' basic research and operation of MOE~(CCNU15A02037). X.~Cheng and X.~Zhang are supported by the CCNU-QLPL Innovation Fund (QLPL201308 and QLPL2015P01).

\begin{appendix}

\section{Relevant coefficients for \texorpdfstring{$B_{s,d}-{\bar{B}}_{s,d}$}{Lg} mixings}
\label{appendix:1}

Here we present the explicit expressions for $C_{VLL}^{NP,ct}$, $C_{VLL}^{NP,tt}$, $C_{VLL}^{NP,cc}$, $C_{SRR,1}^{NP,ct}$, $C_{SRR,1}^{NP,tt}$, $C_{SRR,1}^{NP,cc}$, $C_{SRR,2}^{NP,ct}$, $C_{SRR,2}^{NP,tt}$ and $C_{SRR,2}^{NP,cc}$ appearing in Eq.~\eqref{bsbsmixingcvllsm:1}:
\begin{align}
C_{VLL}^{NP,k} \,=\, &{\eta _d}\eta _u^*{\left| {{\eta _u}} \right|}^2 C_{VLL,{\eta _d}\eta _u^*{{\left| {{\eta _u}} \right|}^2}}^{NP,k}
+  {{\left( {{\eta _d}\eta _u^*} \right)}^2} C_{VLL,{{\left( {{\eta _d}\eta _u^*} \right)}^2}}^{NP,k}
+ {\eta _d}\eta _u^* C_{VLL,{\eta _d}\eta _u^*}^{NP,k}\nonumber\\[0.1cm]
&+ {{\left| {{\eta _u}} \right|}^4} C_{VLL,{{\left| {{\eta _u}} \right|}^4}}^{NP,k}
+ {{\left| {{\eta _u}} \right|}^2} C_{VLL,{{\left| {{\eta _u}} \right|}^2}}^{NP,k} \,,\\[0.2cm]
C_{SRR,1}^{NP,k} \,=\, &{\eta _d}\eta _u^*{\left| {{\eta _u}} \right|}^2 C_{SRR1,{\eta _d}\eta _u^*{{\left| {{\eta _u}} \right|}^2}}^{NP,k}
+  {{\left( {{\eta _d}\eta _u^*} \right)}^2} C_{SRR1,{{\left( {{\eta _d}\eta _u^*} \right)}^2}}^{NP,k}+ {\eta _d}\eta _u^* C_{SRR1,{\eta _d}\eta _u^*}^{NP,k}\nonumber\\[0.1cm]
&+ {{\left| {{\eta _u}} \right|}^4} C_{SRR1,{{\left| {{\eta _u}} \right|}^4}}^{NP,k}
+ {{\left| {{\eta _u}} \right|}^2} C_{SRR1,{{\left| {{\eta _u}} \right|}^2}}^{NP,k} \,,\\[0.2cm]
C_{SRR,2}^{NP,k} \,=\, &{\eta _d}\eta _u^*{\left| {{\eta _u}} \right|}^2 C_{SRR2,{\eta _d}\eta _u^*{{\left| {{\eta _u}} \right|}^2}}^{NP,k}
+  {{\left( {{\eta _d}\eta _u^*} \right)}^2} C_{SRR2,{{\left( {{\eta _d}\eta _u^*} \right)}^2}}^{NP,k}+ {\eta _d}\eta _u^* C_{SRR2,{\eta _d}\eta _u^*}^{NP,k}\nonumber\\[0.1cm]
&+ {{\left| {{\eta _u}} \right|}^4} C_{SRR2,{{\left| {{\eta _u}} \right|}^4}}^{NP,k}
+ {{\left| {{\eta _u}} \right|}^2} C_{SRR2,{{\left| {{\eta _u}} \right|}^2}}^{NP,k} \,,
\end{align}
with $k=ct,tt,cc$, and
\begin{align}
C_{VLL,{\eta _d}\eta _u^*{{\left| {{\eta _u}} \right|}^2}}^{NP,ct} = 0, \hspace{2.0cm}
C_{VLL,{{\left( {{\eta _d}\eta _u^*} \right)}^2}}^{NP,ct} = 0,
\end{align}
\begin{align}
C_{VLL,{\eta _d}\eta _u^*}^{NP,ct} &= \frac{{2{x_b}{x_c}{x_t}\left( {{x_S}{x_t} - 2x_S^2 + {x_t}} \right)}}{{3{{\left( {{x_S} - 1} \right)}^2}\left( {{x_S} - 2{x_c}} \right){{\left( {{x_S} - {x_t}} \right)}^2}}}\ln {x_S} + \frac{{2{x_b}{x_c}{x_t}}}{{3\left( {{x_c}\left( {{x_S}\left( {2{x_t} + 1} \right) + 2{x_t}} \right) - {x_S}{x_t}} \right)}}\ln {x_c}\nonumber\\[0.2cm]
& \hspace{-0.5cm} + \frac{{2{x_b}{x_c}{x_t}\left( {{x_S} + \left( {{x_t} - 2} \right){x_t}} \right)}}{{3{{\left( {{x_t} - 1} \right)}^2}\left( {{x_t} - {x_c}} \right){{\left( {{x_S} - {x_t}} \right)}^2}}}\ln {x_t} - \frac{{2{x_b}{x_c}{x_t}\left( {{x_S}{x_t} - x_S^2 + {x_t} - 1} \right)}}{{3\left( {{x_S} - 1} \right)\left( {{x_t} - 1} \right)\left( {{x_c}\left( {{x_S} + 1} \right) - {x_S}} \right)\left( {{x_S} - {x_t}} \right)}},
\end{align}
\begin{align}
C_{VLL,{{\left| {{\eta _u}} \right|}^4}}^{NP,ct} & =  - \frac{{11{x_c}{x_S}{x_t}\left( {{x_b}{x_S}\left( {{x_S} - 3{x_t}} \right) - 3{x_t}{{\left( {{x_S} - {x_t}} \right)}^2}} \right)}}{{54\left( {4{x_c} - {x_S}} \right){{\left( {{x_S} - {x_t}} \right)}^4}}}\ln {x_S}\nonumber\\[0.2cm]
& + \frac{{11{x_c}x_t^2\left( {{x_b}{x_S}\left( {{x_S} - 3{x_t}} \right) - 3{x_t}{{\left( {{x_S} - {x_t}} \right)}^2}} \right)}}{{54\left( {3{x_c} - {x_t}} \right){{\left( {{x_S} - {x_t}} \right)}^4}}}\ln {x_t}\nonumber\\[0.2cm]
& + \frac{{11{x_c}x_t^2\left( {{x_b}\left( { - 22{x_S}{x_t} + 5x_S^2 + 5x_t^2} \right) - 18{x_S}{{\left( {{x_S} - {x_t}} \right)}^2}} \right)}}{{324{{\left( {{x_S} - {x_t}} \right)}^3}\left( {{x_c}\left( {2{x_S} + 3{x_t}} \right) - {x_S}{x_t}} \right)}},
\end{align}
\begin{align}
C_{VLL,{{\left| {{\eta _u}} \right|}^2}}^{NP,ct} & = {f_1} \left( {{x_c},{x_S},{x_b},{x_t}} \right) \Bigl[ {{f_2}\left( {{x_c},{x_S},{x_b},{x_t}} \right) + {f_3}\left( {{x_c},{x_S},{x_b},{x_t}} \right) + {f_4}\left( {{x_c},{x_S},{x_b},{x_t}} \right)} \Bigr] \ln {x_S}\nonumber\\[0.2cm]
& \hspace{-0.5cm} + {f_5} \left( {{x_c},{x_S},{x_b},{x_t}} \right) \Bigl[ {{f_6}\left( {{x_c},{x_S},{x_b},{x_t}} \right)+ {f_7}\left( {{x_c},{x_S},{x_b},{x_t}} \right) + {f_8}\left( {{x_c},{x_S},{x_b},{x_t}} \right)} \Bigr] \ln {x_t}\nonumber\\[0.2cm]
& \hspace{-0.5cm} - {f_9}\left( {{x_c},{x_S},{x_b},{x_t}} \right) \Bigl[ {{f_{10}}\left( {{x_c},{x_S},{x_b},{x_t}} \right) + {f_{11}}\left( {{x_c},{x_S},{x_b},{x_t}} \right) + {f_{12}}\left( {{x_c},{x_S},{x_b},{x_t}} \right)} \Bigr],
\end{align}
\begin{align}
C_{VLL,{\eta _d}\eta _u^*{{\left| {{\eta _u}} \right|}^2}}^{NP,tt} = 0,\hspace{2.0cm}
C_{VLL,{{\left( {{\eta _d}\eta _u^*} \right)}^2}}^{NP,tt} = 0,
\end{align}
\begin{align}
C_{VLL,{\eta _d}\eta _u^*}^{NP,tt} & = \frac{{4{x_b}x_t^2\left( {{x_t} - x_S^2} \right)}}{{3{{\left( {{x_S} - 1} \right)}^2}{{\left( {{x_S} - {x_t}} \right)}^3}}} \ln {x_S} + \frac{{4{x_b}x_t^2\left( {2{x_S}{x_t} - x_S^2 + x_t^3 - 3x_t^2 + {x_t}} \right)}}{{3{{\left( {{x_t} - 1} \right)}^3}{{\left( {{x_S} - {x_t}} \right)}^3}}}\ln {x_t}\nonumber\\[0.2cm]
&  + \frac{{2{x_b}{x_t}\left( {x_S^2\left( {{x_t} + 1} \right) - {x_S}{{\left( {{x_t} + 1} \right)}^2} + {x_t}\left( {2x_t^2 - 3{x_t} + 3} \right)} \right)}}{{3\left( {{x_S} - 1} \right){{\left( {{x_t} - 1} \right)}^2}{{\left( {{x_S} - {x_t}} \right)}^2}}},
\end{align}
\begin{align}
C_{VLL,{{\left| {{\eta _u}} \right|}^4}}^{NP,tt} &= \frac{{11x_t^2\left( {{x_b}\left( { - 3x_S^2{x_t} - 3{x_S}x_t^2 + x_S^3 + x_t^3} \right) - 6{x_S}{x_t}{{\left( {{x_S} - {x_t}} \right)}^2}} \right)}}{{54{{\left( {{x_S} - {x_t}} \right)}^5}}}\ln {x_S}\nonumber\\[0.2cm]
& - \frac{{11x_t^2\left( {{x_b}\left( { - 3x_S^2{x_t} - 3{x_S}x_t^2 + x_S^3 + x_t^3} \right) - 6{x_S}{x_t}{{\left( {{x_S} - {x_t}} \right)}^2}} \right)}}{{54{{\left( {{x_S} - {x_t}} \right)}^5}}}\ln {x_t}\nonumber\\[0.2cm]
& - \frac{{11x_t^2\left( {{x_b}\left( { - 22{x_S}{x_t} + 5x_S^2 + 5x_t^2} \right) - 9{{\left( {{x_S} - {x_t}} \right)}^2}\left( {{x_S} + {x_t}} \right)} \right)}}{{162{{\left( {{x_S} - {x_t}} \right)}^4}}},
\end{align}
\begin{align}
C_{VLL,{{\left| {{\eta _u}} \right|}^2}}^{NP,tt} &= {f_{13}}\left( {{x_c},{x_S},{x_b},{x_t}} \right) \Bigl[ {{f_{14}}\left( {{x_c},{x_S},{x_b},{x_t}} \right) + {f_{15}}\left( {{x_c},{x_S},{x_b},{x_t}} \right) + {f_{16}}\left( {{x_c},{x_S},{x_b},{x_t}} \right)} \Bigr] \ln {x_S}\nonumber\\[0.2cm]
& - {f_{17}}\left( {{x_c},{x_S},{x_b},{x_t}} \right) \Bigl[ {{f_{22}}\left( {{x_c},{x_S},{x_b},{x_t}} \right) + {f_{23}}\left( {{x_c},{x_S},{x_b},{x_t}} \right)} \Bigr] \ln {x_t}\nonumber\\
& + {f_{24}}\left( {{x_c},{x_S},{x_b},{x_t}} \right) \Bigl[ {{f_{30}}\left( {{x_c},{x_S},{x_b},{x_t}} \right) + {f_{31}}\left( {{x_c},{x_S},{x_b},{x_t}} \right)} \Bigr],
\end{align}
\begin{align}
C_{VLL,{\eta _d}\eta _u^*{{\left| {{\eta _u}} \right|}^2}}^{NP,cc} = 0,\hspace{1.0cm}
C_{VLL,{{\left( {{\eta _d}\eta _u^*} \right)}^2}}^{NP,cc} = 0,\hspace{1.0cm}
C_{VLL,{{\left| {{\eta _u}} \right|}^4}}^{NP,cc}= 0,
\end{align}
\begin{align}
C_{VLL,{\eta _d}\eta _u^*}^{NP,cc} =  - \frac{{2{x_b}{x_c}}}{{3\left( {2{x_c}\left( {{x_S} + 1} \right) - {x_S}} \right)}},\qquad
C_{VLL,{{\left| {{\eta _u}} \right|}^2}}^{NP,cc} =  - \frac{{4{x_b}{x_c}}}{{9\left( {3{x_c}\left( {{x_S} + 1} \right) - {x_S}} \right)}},
\end{align}
\begin{align}
C_{SRR1,{\eta _d}\eta _u^*{{\left| {{\eta _u}} \right|}^2}}^{NP,ct} & = \frac{{19{x_b}{x_c}{x_S}{x_t}}}{{18\left( {2{x_c} - {x_S}} \right){{\left( {{x_S} - {x_t}} \right)}^2}}}\ln {x_S} - \frac{{19{x_b}{x_c}x_t^2}}{{18\left( {{x_c} - {x_t}} \right){{\left( {{x_S} - {x_t}} \right)}^2}}}\ln {x_t} \nonumber\\[0.2cm]
& - \frac{{19{x_b}{x_c}{x_t}}}{{18\left( {{x_c} - {x_S}} \right)\left( {{x_S} - {x_t}} \right)}},
\end{align}
\begin{align}
C_{SRR1,{{\left( {{\eta _d}\eta _u^*} \right)}^2}}^{NP,ct} &= - \frac{{19{x_b}{x_c}{x_S}{x_t}}}{{18\left( {2{x_c} - {x_S}} \right){{\left( {{x_S} - {x_t}} \right)}^2}}}\ln {x_S} + \frac{{19{x_b}{x_c}x_t^2}}{{18\left( {{x_c} - {x_t}} \right){{\left( {{x_S} - {x_t}} \right)}^2}}}\ln {x_t}  \nonumber\\[0.2cm]
& + \frac{{19{x_b}{x_c}{x_t}}}{{18\left( {{x_c} - {x_S}} \right)\left( {{x_S} - {x_t}} \right)}},
\end{align}
\begin{align}
C_{SRR1,{\eta _d}\eta _u^*}^{NP,ct} & = \frac{{5{x_b}{x_c}{x_S}{x_t}\left( {{x_S}\left( {{x_t} + 2} \right) - 3{x_t}} \right)}}{{6{{\left( {{x_S} - 1} \right)}^2}\left( {{x_S} - 2{x_c}} \right){{\left( {{x_S} - {x_t}} \right)}^2}}}\ln {x_S} + \frac{{5{x_b}{x_c}x_t^2\left( {{x_S}\left( {{x_t} - 2} \right) + \left( {3 - 2{x_t}} \right){x_t}} \right)}}{{6{{\left( {{x_t} - 1} \right)}^2}\left( {{x_t} - {x_c}} \right){{\left( {{x_S} - {x_t}} \right)}^2}}}\ln {x_t}\nonumber\\[0.2cm]
& - \frac{{5{x_b}{x_c}{x_S}{x_t}\left( {{x_S} - 2{x_t} + 1} \right)}}{{6\left( {{x_S} - 1} \right)\left( {{x_t} - 1} \right)\left( {{x_c}\left( {{x_S} + 1} \right) - {x_S}} \right)\left( {{x_S} - {x_t}} \right)}},
\end{align}
\begin{align}
C_{SRR1,{{\left| {{\eta _u}} \right|}^4}}^{NP,ct} & =  - \frac{{19{x_b}{x_c}x_S^2{x_t}\left( {{x_S} - 3{x_t}} \right)}}{{108\left( {4{x_c} - {x_S}} \right){{\left( {{x_S} - {x_t}} \right)}^4}}} \ln {x_S} + \frac{{19{x_b}{x_c}{x_S}x_t^2\left( {{x_S} - 3{x_t}} \right)}}{{108\left( {3{x_c} - {x_t}} \right){{\left( {{x_S} - {x_t}} \right)}^4}}}\ln {x_t} \nonumber\\[0.2cm]
& + \frac{{19{x_b}{x_c}x_t^2\left( { - 22{x_S}{x_t} + 5x_S^2 + 5x_t^2} \right)}}{{648{{\left( {{x_S} - {x_t}} \right)}^3}\left( {{x_c}\left( {2{x_S} + 3{x_t}} \right) - {x_S}{x_t}} \right)}},
\end{align}
\begin{align}
C_{SRR1,{{\left| {{\eta _u}} \right|}^2}}^{NP,ct} &= {f_{32}}\left( {{x_c},{x_S},{x_b},{x_t}} \right) \ln {x_S} - {f_{33}}\left( {{x_c},{x_S},{x_b},{x_t}} \right) {f_{34}}\left( {{x_c},{x_S},{x_b},{x_t}} \right) \ln {x_t}\nonumber\\[0.2cm]
& + {f_{35}}\left( {{x_c},{x_S},{x_b},{x_t}} \right) \Bigl[ {{f_{36}}\left( {{x_c},{x_S},{x_b},{x_t}} \right) + {f_{37}}\left( {{x_c},{x_S},{x_b},{x_t}} \right)} \Bigr],
\end{align}
\begin{align}
C_{SRR1,{\eta _d}\eta _u^*{{\left| {{\eta _u}} \right|}^2}}^{NP,tt} =  - \frac{{19{x_b}x_t^2\left( {{x_S} + {x_t}} \right)}}{{18{{\left( {{x_S} - {x_t}} \right)}^3}}}\ln {x_S} + \frac{{19{x_b}x_t^2\left( {{x_S} + {x_t}} \right)}}{{18{{\left( {{x_S} - {x_t}} \right)}^3}}}\ln {x_t} + \frac{{19{x_b}x_t^2}}{{9{{\left( {{x_S} - {x_t}} \right)}^2}}},
\end{align}
\begin{align}
C_{SRR1,{{\left( {{\eta _d}\eta _u^*} \right)}^2}}^{NP,tt} = \frac{{19{x_b}x_t^2\left( {{x_S} + {x_t}} \right)}}{{18{{\left( {{x_S} - {x_t}} \right)}^3}}}\ln {x_S} - \frac{{19{x_b}x_t^2\left( {{x_S} + {x_t}} \right)}}{{18{{\left( {{x_S} - {x_t}} \right)}^3}}}\ln {x_t} - \frac{{19{x_b}x_t^2}}{{9{{\left( {{x_S} - {x_t}} \right)}^2}}},
\end{align}
\begin{align}
C_{SRR1,{\eta _d}\eta _u^*}^{NP,tt} &= \frac{{5{x_b}{x_S}x_t^2\left( {{x_S}\left( {{x_t} + 1} \right) - 2{x_t}} \right)}}{{3{{\left( {{x_S} - 1} \right)}^2}{{\left( {{x_S} - {x_t}} \right)}^3}}}\ln {x_S} + \frac{{5{x_b}x_t^2\left( { - 2{x_S}{x_t} + x_S^2 - \left( {{x_t} - 2} \right)x_t^3} \right)}}{{3{{\left( {{x_t} - 1} \right)}^3}{{\left( {{x_S} - {x_t}} \right)}^3}}}\ln {x_t}\nonumber\\[0.2cm]
& + \frac{{5{x_b}x_t^2\left( {x_S^2\left( {{x_t} - 3} \right) + {x_S}\left( { - 3x_t^2 + 6{x_t} + 1} \right) + \left( {{x_t} - 3} \right){x_t}} \right)}}{{6\left( {{x_S} - 1} \right){{\left( {{x_t} - 1} \right)}^2}{{\left( {{x_S} - {x_t}} \right)}^2}}},
\end{align}
\begin{align}
C_{SRR1,{{\left| {{\eta _u}} \right|}^4}}^{NP,tt} &= \frac{{19{x_b}x_t^2\left( { x_S^3 + x_t^3 - 3x_S^2{x_t} - 3{x_S}x_t^2} \right)}}{{108{{\left( {{x_S} - {x_t}} \right)}^5}}}\ln {x_S} - \frac{{19{x_b}x_t^2\left( { x_S^3 + x_t^3 - 3x_S^2{x_t} - 3{x_S}x_t^2 } \right)}}{{108{{\left( {{x_S} - {x_t}} \right)}^5}}}\ln {x_t}\nonumber\\[0.2cm]
& - \frac{{19{x_b}x_t^2\left( { - 22{x_S}{x_t} + 5x_S^2 + 5x_t^2} \right)}}{{324{{\left( {{x_S} - {x_t}} \right)}^4}}},
\end{align}
\begin{align}
C_{SRR1,{{\left| {{\eta _u}} \right|}^2}}^{NP,tt} &= {f_{38}}\left( {{x_c},{x_S},{x_b},{x_t}} \right) \ln {x_S} + {f_{39}}\left( {{x_c},{x_S},{x_b},{x_t}} \right) \ln {x_t} \nonumber\\[0.2cm]
& \hspace{-1.0cm} - {f_{40}}\left( {{x_c},{x_S},{x_b},{x_t}} \right) \Bigl[ {{f_{41}}\left( {{x_c},{x_S},{x_b},{x_t}} \right) + {f_{42}}\left( {{x_c},{x_S},{x_b},{x_t}} \right) + {f_{43}}\left( {{x_c},{x_S},{x_b},{x_t}} \right)} \Bigr],
\end{align}
\begin{align}
& C_{SRR1,{\eta _d}\eta _u^*{{\left| {{\eta _u}} \right|}^2}}^{NP,cc} = 0, \; C_{SRR1,{{\left( {{\eta _d}\eta _u^*} \right)}^2}}^{NP,cc} = 0, \; C_{SRR1,{\eta _d}\eta _u^*}^{NP,cc} = 0, \; C_{SRR1,{{\left| {{\eta _u}} \right|}^4}}^{NP,cc} = 0, \; C_{SRR1,{{\left| {{\eta _u}} \right|}^2}}^{NP,cc} = 0,
\end{align}
\begin{align}
& C_{SRR2,{\eta _d}\eta _u^*}^{NP,ct} =  - \frac{3}{{20}}C_{SRR1,{\eta _d}\eta _u^*}^{NP,ct}, \quad
  C_{SRR2,{{\left| {{\eta _u}} \right|}^4}}^{NP,ct} =  - \frac{{21}}{{76}}C_{SRR1,{{\left| {{\eta _u}} \right|}^4}}^{NP,ct}, \quad
  C_{SRR2,{{\left| {{\eta _u}} \right|}^2}}^{NP,ct} =  - \frac{3}{{20}}C_{SRR1,{{\left| {{\eta _u}} \right|}^2}}^{NP,ct}, \nonumber\\[0.2cm]
& C_{SRR2,{\eta _d}\eta _u^*{{\left| {{\eta _u}} \right|}^2}}^{NP,ct} =  - \frac{{21}}{{76}}C_{SRR1,{\eta _d}\eta _u^*{{\left| {{\eta _u}} \right|}^2}}^{NP,ct}, \quad
C_{SRR2,{{\left( {{\eta _d}\eta _u^*} \right)}^2}}^{NP,ct} =  - \frac{{21}}{{76}}C_{SRR1,{{\left( {{\eta _d}\eta _u^*} \right)}^2}}^{NP,ct},
\end{align}
\begin{align}
& C_{SRR2,{\eta _d}\eta _u^*}^{NP,tt} =  - \frac{3}{{20}}C_{SRR1,{\eta _d}\eta _u^*}^{NP,tt}, \quad
  C_{SRR2,{{\left| {{\eta _u}} \right|}^4}}^{NP,tt} =  - \frac{{21}}{{76}}C_{SRR1,{{\left| {{\eta _u}} \right|}^4}}^{NP,tt}, \quad
  C_{SRR2,{{\left| {{\eta _u}} \right|}^2}}^{NP,tt} =  - \frac{3}{{20}}C_{SRR1,{{\left| {{\eta _u}} \right|}^2}}^{NP,tt}, \nonumber\\[0.2cm]
& C_{SRR2,{\eta _d}\eta _u^*{{\left| {{\eta _u}} \right|}^2}}^{NP,tt} =  - \frac{{21}}{{76}}C_{SRR1,{\eta
  _d}\eta _u^*{{\left| {{\eta _u}} \right|}^2}}^{NP,tt}, \quad
  C_{SRR2,{{\left( {{\eta _d}\eta _u^*} \right)}^2}}^{NP,tt} =  - \frac{{21}}{{76}}C_{SRR1,{{\left( {{\eta _d}\eta _u^*} \right)}^2}}^{NP,tt},
\end{align}
\begin{align}
C_{SRR2,{\eta _d}\eta _u^*{{\left| {{\eta _u}} \right|}^2}}^{NP,cc} = 0,\,
C_{SRR2,{{\left( {{\eta _d}\eta _u^*} \right)}^2}}^{NP,cc} = 0,\,
C_{SRR2,{\eta _d}\eta _u^*}^{NP,cc} = 0,\,
C_{SRR2,{{\left| {{\eta _u}} \right|}^4}}^{NP,cc} = 0,\,
C_{SRR2,{{\left| {{\eta _u}} \right|}^2}}^{NP,cc} = 0.
\end{align}
Here $x_b=\frac{m_{b}^2}{m_W^2}$, $x_c=\frac{m_{c}^2}{m_W^2}$ and the functions $f_j \left( {{x_c},{x_S},{x_b},{x_t}} \right)$ are defined, respectively, as
\begin{align}
{f_1}\left( {{x_c},{x_S},{x_b},{x_t}} \right)& = \frac{{ - {x_c}{x_S}{x_t}}}{{9{{\left( {{x_S} - 1} \right)}^3}\left( {{x_S} - 3{x_c}} \right){{\left( {{x_S} - {x_t}} \right)}^3}}},\\[0.2cm]
{f_2}\left( {{x_c},{x_S},{x_b},{x_t}} \right) &=  - 6x_S^5 + 12x_S^4\left( {{x_t} + 3} \right) - 3x_S^3\left( {{x_b}\left( {{x_t} - 6} \right) + 2\left( {x_t^2 + 12{x_t} + 9} \right)} \right),\\[0.2cm]
{f_3}\left( {{x_c},{x_S},{x_b},{x_t}} \right) &= x_S^2\left( {{x_b}\left( {x_t^2 - 8{x_t} + 2} \right) + 12\left( {3x_t^2 + 9{x_t} + 2} \right)} \right),\\[0.2cm]
{f_4}\left( {{x_c},{x_S},{x_b},{x_t}} \right) &= {x_S}{x_t}\left( {{x_b}\left( {6{x_t} - 41} \right) - 6\left( {9{x_t} + 8} \right)} \right) + {x_t}\left( {{x_b}\left( {13{x_t} + 12} \right) + 24{x_t}} \right),\\[0.2cm]
{f_5}\left( {{x_c},{x_S},{x_b},{x_t}} \right) &= \frac{{{x_c}x_t^2}}{{9{{\left( {{x_t} - 1} \right)}^3}\left( {{x_t} - 3{x_c}} \right){{\left( {{x_S} - {x_t}} \right)}^3}}},\\[0.2cm]
{f_6}\left( {{x_c},{x_S},{x_b},{x_t}} \right) &= x_S^2\left( {{x_b}\left( {x_t^2 + 7{x_t} + 2} \right) - 6\left( {{x_t} - 4} \right){{\left( {{x_t} - 1} \right)}^2}} \right),\\[0.2cm]
{f_7}\left( {{x_c},{x_S},{x_b},{x_t}} \right) &= {x_S}{x_t}\left( {12\left( {{x_t} - 4} \right){{\left( {{x_t} - 1} \right)}^2} - {x_b}\left( {3x_t^2 + 12{x_t} + 5} \right)} \right),\\[0.2cm]
{f_8}\left( {{x_c},{x_S},{x_b},{x_t}} \right) &= {x_t}\left( {{x_b}\left( {21x_t^2 - 23{x_t} + 12} \right) - 6\left( {{x_t} - 4} \right){{\left( {{x_t} - 1} \right)}^2}{x_t}} \right),\\[0.2cm]
{f_9}\left( {{x_c},{x_S},{x_b},{x_t}} \right) &= \frac{{2{x_b}{x_c}{x_t}}}{{9{{\left( {{x_S} - 1} \right)}^2}{{\left( {{x_t} - 1} \right)}^2}{{\left( {{x_S} - {x_t}} \right)}^2}\left( {2{x_c}{x_S}{x_t} + 2{x_c}{x_S} + 2{x_c}{x_t} - {x_S}{x_t}} \right)}},\\[0.2cm]
{f_{10}}\left( {{x_c},{x_S},{x_b},{x_t}} \right) &= x_S^4\left( {4x_t^2 - 15{x_t} + 6} \right) + x_S^3\left( { - 9x_t^3 + 17x_t^2 + 19{x_t} - 12} \right),\\[0.2cm]
{f_{11}}\left( {{x_c},{x_S},{x_b},{x_t}} \right) &= x_S^2\left( {4x_t^4 + 24x_t^3 - 70x_t^2 + 21{x_t} + 6} \right),\\[0.2cm]
{f_{12}}\left( {{x_c},{x_S},{x_b},{x_t}} \right) &= {x_S}{x_t}\left( { - 17x_t^3 + 21x_t^2 + 16{x_t} - 15} \right) + 3{\left( {{x_t} - 1} \right)^2}x_t^2,\\[0.2cm]
{f_{13}}\left( {{x_c},{x_S},{x_b},{x_t}} \right) &= \frac{{2{x_S}x_t^2}}{{9{{\left( {{x_S} - 1} \right)}^3}{{\left( {{x_S} - {x_t}} \right)}^4}}},\\[0.2cm]
{f_{14}}\left( {{x_c},{x_S},{x_b},{x_t}} \right) &= 3x_S^5 - 6x_S^4\left( {{x_t} + 3} \right) + 3x_S^3\left( {{x_b}\left( {{x_t} - 3} \right) + x_t^2 + 12{x_t} + 9} \right),\\[0.2cm]
{f_{15}}\left( {{x_c},{x_S},{x_b},{x_t}} \right) &=  - x_S^2\left( {{x_b}\left( {x_t^2 + 10{x_t} + 1} \right) + 6\left( {3x_t^2 + 9{x_t} + 2} \right)} \right),\\[0.2cm]
{f_{16}}\left( {{x_c},{x_S},{x_b},{x_t}} \right) &= 3{x_S}{x_t}\left( {{x_b}\left( {{x_t} + 13} \right) + 9{x_t} + 8} \right) - 12{x_t}\left( {{x_b}\left( {{x_t} + 1} \right) + {x_t}} \right),\\[0.2cm]
{f_{17}}\left( {{x_c},{x_S},{x_b},{x_t}} \right) &= \frac{{2x_t^2}}{{9{{\left( {{x_t} - 1} \right)}^4}{{\left( {{x_S} - {x_t}} \right)}^4}}},\\[0.2cm]
{f_{18}}\left( {{x_c},{x_S},{x_b},{x_t}} \right) &= x_S^3\left( {{x_b}\left( {9{x_t} + 1} \right) + 3\left( {x_t^2 - 2{x_t} + 4} \right){{\left( {{x_t} - 1} \right)}^2}} \right),\\[0.2cm]
{f_{19}}\left( {{x_c},{x_S},{x_b},{x_t}} \right) &=  - 3x_S^2{x_t}\left( {{x_b}\left( {9{x_t} + 1} \right) + \left( {2x_t^2 - {x_t} + 8} \right){{\left( {{x_t} - 1} \right)}^2}} \right),\\[0.2cm]
{f_{20}}\left( {{x_c},{x_S},{x_b},{x_t}} \right)& = 3{x_S}{x_t}\left( {{x_b}\left( {x_t^4 - 7x_t^3 + 24x_t^2 - 12{x_t} + 4} \right) + {x_t}{{\left( {x_t^2 + {x_t} - 2} \right)}^2}} \right),\\[0.2cm]
{f_{21}}\left( {{x_c},{x_S},{x_b},{x_t}} \right) &=  - x_t^4\left( {{x_b}\left( {x_t^2 - 3{x_t} + 12} \right) + 9{{\left( {{x_t} - 1} \right)}^2}} \right),\\[0.2cm]
{f_{22}}\left( {{x_c},{x_S},{x_b},{x_t}} \right) &= {f_{18}}\left( {{x_c},{x_S},{x_b},{x_t}} \right) + {f_{19}}\left( {{x_c},{x_S},{x_b},{x_t}} \right),\\[0.2cm]
{f_{23}}\left( {{x_c},{x_S},{x_b},{x_t}} \right) &= {f_{20}}\left( {{x_c},{x_S},{x_b},{x_t}} \right) + {f_{21}}\left( {{x_c},{x_S},{x_b},{x_t}} \right),\\[0.2cm]
{f_{24}}\left( {{x_c},{x_S},{x_b},{x_t}} \right) &= \frac{{{x_t}}}{{27{{\left( {{x_S} - 1} \right)}^2}{{\left( {{x_t} - 1} \right)}^3}{{\left( {{x_S} - {x_t}} \right)}^3}}},\\[0.2cm]
{f_{25}}\left( {{x_c},{x_S},{x_b},{x_t}} \right) &= {x_b}x_S^4\left( {5x_t^3 + 2x_t^2 + 65{x_t} - 12} \right) - 2{x_b}x_S^3\left( {11x_t^4 - 11x_t^3 + 85x_t^2 + 47{x_t} - 12} \right),\\[0.2cm]
{f_{26}}\left( {{x_c},{x_S},{x_b},{x_t}} \right) &= {x_b}x_S^2\left( {5x_t^5 + 130x_t^4 - 246x_t^3 + 574x_t^2 - 91{x_t} - 12} \right),\\[0.2cm]
{f_{27}}\left( {{x_c},{x_S},{x_b},{x_t}} \right) &=  - 2{x_b}{x_S}{x_t}\left( {17x_t^4 - 11x_t^3 + 49x_t^2 + 95{x_t} - 30} \right),\\[0.2cm]
{f_{28}}\left( {{x_c},{x_S},{x_b},{x_t}} \right) &=  + {x_b}x_t^2\left( { - 31x_t^3 + 110x_t^2 - 43{x_t} + 24} \right),\\[0.2cm]
{f_{29}}\left( {{x_c},{x_S},{x_b},{x_t}} \right)& =  - 18{\left( {{x_S} - 1} \right)^2}{\left( {{x_S} - {x_t}} \right)^2}{\left( {{x_t} - 1} \right)^2}\left( {{x_t} - 4} \right){x_t},\\[0.2cm]
{f_{30}}\left( {{x_c},{x_S},{x_b},{x_t}} \right) &= {f_{25}}\left( {{x_c},{x_S},{x_b},{x_t}} \right) + {f_{26}}\left( {{x_c},{x_S},{x_b},{x_t}} \right) + {f_{27}}\left( {{x_c},{x_S},{x_b},{x_t}} \right),\\[0.2cm]
{f_{31}}\left( {{x_c},{x_S},{x_b},{x_t}} \right) &= {f_{28}}\left( {{x_c},{x_S},{x_b},{x_t}} \right) + {f_{29}}\left( {{x_c},{x_S},{x_b},{x_t}} \right),\\[0.2cm]
{f_{32}}\left( {{x_c},{x_S},{x_b},{x_t}} \right) &= \frac{{5{x_b}{x_c}{x_S}{x_t}\left( {x_S^2\left( {x_t^2 + {x_t} + 2} \right) - {x_S}{x_t}\left( {3{x_t} + 5} \right) + 4x_t^2} \right)}}{{9{{\left( {{x_S} - 1} \right)}^3}\left( {{x_S} - 3{x_c}} \right){{\left( {{x_S} - {x_t}} \right)}^3}}},\\[0.2cm]
{f_{33}}\left( {{x_c},{x_S},{x_b},{x_t}} \right) &= \frac{{5{x_b}{x_c}x_t^2}}{{9{{\left( {{x_t} - 1} \right)}^3}\left( {{x_t} - 3{x_c}} \right){{\left( {{x_S} - {x_t}} \right)}^3}}},\\[0.2cm]
{f_{34}}\left( {{x_c},{x_S},{x_b},{x_t}} \right) &=  x_S^2\left( {x_t^2 - 2{x_t} + 2} \right) + {x_S}{x_t}\left( { - 3x_t^2 + 6{x_t} - 5} \right)
 + x_t^2\left( {3x_t^2 - 6{x_t} + 4} \right),\\[0.2cm]
{f_{35}}\left( {{x_c},{x_S},{x_b},{x_t}} \right) &= \frac{{5{x_b}{x_c}{x_S}x_t^2}}{{18{{\left( {{x_S} - 1} \right)}^2}{{\left( {{x_t} - 1} \right)}^2}{{\left( {{x_S} - {x_t}} \right)}^2}\left( {2{x_c}\left( {{x_S}\left( {{x_t} + 1} \right) + {x_t}} \right) - {x_S}{x_t}} \right)}},\\[0.2cm]
{f_{36}}\left( {{x_c},{x_S},{x_b},{x_t}} \right) & = x_S^3\left( {{x_t} - 3} \right) + x_S^2\left( { - 3x_t^2 + 5{x_t} + 4} \right),\\[0.2cm]
{f_{37}}\left( {{x_c},{x_S},{x_b},{x_t}} \right) &= {x_S}\left( {4x_t^3 - 13{x_t} + 3} \right) + {x_t}\left( { - 8x_t^2 + 15{x_t} - 5} \right),\\[0.2cm]
{f_{38}}\left( {{x_c},{x_S},{x_b},{x_t}} \right) &= \frac{{10{x_b}{x_S}x_t^2\left( {x_S^2\left( {x_t^2 + {x_t} + 1} \right) - 3{x_S}{x_t}\left( {{x_t} + 1} \right) + 3x_t^2} \right)}}{{9{{\left( {{x_S} - 1} \right)}^3}{{\left( {{x_S} - {x_t}} \right)}^4}}},\\[0.2cm]
{f_{39}}\left( {{x_c},{x_S},{x_b},{x_t}} \right) &= \frac{{10{x_b}x_t^2\left( {3{x_S}x_t^2 - 3x_S^2{x_t} + x_S^3 - \left( {x_t^2 - 3{x_t} + 3} \right)x_t^4} \right)}}{{9{{\left( {{x_t} - 1} \right)}^4}{{\left( {{x_S} - {x_t}} \right)}^4}}},\\[0.2cm]
{f_{40}}\left( {{x_c},{x_S},{x_b},{x_t}} \right) &= \frac{{5{x_b}x_t^2}}{{27{{\left( {{x_S} - 1} \right)}^2}{{\left( {{x_t} - 1} \right)}^3}{{\left( {{x_S} - {x_t}} \right)}^3}}},\\[0.2cm]
{f_{41}}\left( {{x_c},{x_S},{x_b},{x_t}} \right) &= x_S^4\left( {2x_t^2 - 7{x_t} + 11} \right) - x_S^3\left( {7x_t^3 - 19x_t^2 + 17{x_t} + 19} \right),\\[0.2cm]
{f_{42}}\left( {{x_c},{x_S},{x_b},{x_t}} \right) &= x_S^2\left( {11x_t^4 - 17x_t^3 - 15x_t^2 + 55{x_t} + 2} \right),\\[0.2cm]
{f_{43}}\left( {{x_c},{x_S},{x_b},{x_t}} \right) &=  - {x_S}{x_t}\left( {19x_t^3 - 55x_t^2 + 53{x_t} + 7} \right) + x_t^2\left( {2x_t^2 - 7{x_t} + 11} \right),
\end{align}

\end{appendix}


\begin{thebibliography}{100}

\bibitem{ATLAS:1}
  G.~Aad {\it et al.}  [ATLAS Collaboration],
  Phys.\ Lett.\ B {\bf 716} (2012) 1
  [arXiv:1207.7214 [hep-ex]].

\bibitem{CMS:1}
  S.~Chatrchyan {\it et al.}  [CMS Collaboration],
  Phys.\ Lett.\ B {\bf 716} (2012) 30
  [arXiv:1207.7235 [hep-ex]].

\bibitem{ATLAS:coupling}
  G.~Aad {\it et al.}  [ATLAS Collaboration],
  Phys.\ Lett.\ B {\bf 726} (2013) 88
   [Erratum-ibid.\ B {\bf 734} (2014) 406]
  [arXiv:1307.1427 [hep-ex]];
  Phys.\ Lett.\ B {\bf 726} (2013) 120
  [arXiv:1307.1432 [hep-ex]];
  The ATLAS collaboration,
  ATLAS-CONF-2014-009, ATLAS-COM-CONF-2014-013.

\bibitem{CMS:coupling}
  S.~Chatrchyan {\it et al.}  [CMS Collaboration],
  Nature Phys.\  {\bf 10} (2014) 557
  [arXiv:1401.6527 [hep-ex]];
  V.~Khachatryan {\it et al.}  [CMS Collaboration],
  arXiv:1411.3441 [hep-ex];
  arXiv:1412.8662 [hep-ex].

\bibitem{Aaltonen:2013kxa}
  T.~Aaltonen {\it et al.}  [CDF and D0 Collaborations],
  Phys.\ Rev.\ D {\bf 88} (2013) 052014
  [arXiv:1303.6346 [hep-ex]];
  B.~Tuchming [CDF and D0 Collaborations],
  arXiv:1405.5058 [hep-ex].

\bibitem{Lee:1973iz}
  T.~D.~Lee,
  Phys.\ Rev.\ D {\bf 8} (1973) 1226.

\bibitem{Branco:2011iw}
  G.~C.~Branco, P.~M.~Ferreira, L.~Lavoura, M.~N.~Rebelo, M.~Sher and J.~P.~Silva,
  Phys.\ Rept.\  {\bf 516} (2012) 1
  [arXiv:1106.0034 [hep-ph]];
  J.~F.~Gunion, H.~E.~Haber, G.~L.~Kane and S.~Dawson,
  Front.\ Phys.\  {\bf 80} (2000) 1.

\bibitem{GIM}
  S.~L.~Glashow, J.~Iliopoulos and L.~Maiani,
  Phys.\ Rev.\ D {\bf 2} (1970) 1285.

\bibitem{Glashow:1976nt}
  S.~L.~Glashow and S.~Weinberg,
  Phys.\ Rev.\ D {\bf 15} (1977) 1958.

\bibitem{MFV:1}
  A.~J.~Buras, P.~Gambino, M.~Gorbahn, S.~Jager and L.~Silvestrini,
  Phys.\ Lett.\ B {\bf 500} (2001) 161
  [hep-ph/0007085].

\bibitem{MFV:2}
  G.~D'Ambrosio, G.~F.~Giudice, G.~Isidori and A.~Strumia,
  Nucl.\ Phys.\ B {\bf 645} (2002) 155
  [hep-ph/0207036].

\bibitem{Buras:2010mh}
  A.~J.~Buras, M.~V.~Carlucci, S.~Gori and G.~Isidori,
  JHEP {\bf 1010}, 009 (2010)
  [arXiv:1005.5310 [hep-ph]];
  E.~Cervero and J.~M.~Gerard,
  Phys.\ Lett.\ B {\bf 712} (2012) 255
  [arXiv:1202.1973 [hep-ph]].

\bibitem{Chivukula:1987py}
  R.~S.~Chivukula and H.~Georgi,
  Phys.\ Lett.\ B {\bf 188} (1987) 99.

\bibitem{Hall:1990ac}
  L.~J.~Hall and L.~Randall,
  Phys.\ Rev.\ Lett.\  {\bf 65} (1990) 2939.

\bibitem{Cabibbo:1963yz}
  N.~Cabibbo,
  Phys.\ Rev.\ Lett.\  {\bf 10} (1963) 531.

\bibitem{Kobayashi:1973fv}
  M.~Kobayashi and T.~Maskawa,
  Prog.\ Theor.\ Phys.\  {\bf 49} (1973) 652.

\bibitem{Manohar:2006ga}
  A.~V.~Manohar and M.~B.~Wise,
  Phys.\ Rev.\ D {\bf 74} (2006) 035009
  [hep-ph/0606172].

\bibitem{Arnold:2009ay}
  J.~M.~Arnold, M.~Pospelov, M.~Trott and M.~B.~Wise,
  JHEP {\bf 1001} (2010) 073
  [arXiv:0911.2225 [hep-ph]].

\bibitem{Pich:2009sp}
  A.~Pich and P.~Tuzon,
  Phys.\ Rev.\ D {\bf 80} (2009) 091702
  [arXiv:0908.1554 [hep-ph]].

\bibitem{He:2013tla}
  X.~G.~He, H.~Phoon, Y.~Tang and G.~Valencia,
  JHEP {\bf 1305} (2013) 026
  [arXiv:1303.4848 [hep-ph]].

\bibitem{Li:2013vlx}
  X.~Q.~Li, Y.~D.~Yang and X.~B.~Yuan,
  Phys.\ Rev.\ D {\bf 89} (2014) 054024
  [arXiv:1311.2786 [hep-ph]].

\bibitem{Degrassi:2010ne}
  G.~Degrassi and P.~Slavich,
  Phys.\ Rev.\ D {\bf 81} (2010) 075001
  [arXiv:1002.1071 [hep-ph]].

\bibitem{Heo:2008sr}
  J.~H.~Heo and W.~Y.~Keung,
  Phys.\ Lett.\ B {\bf 661} (2008) 259
  [arXiv:0801.0231 [hep-ph]];
  S.~Fajfer and J.~O.~Eeg,
  Phys.\ Rev.\ D {\bf 89} (2014) 095030
  [arXiv:1401.2275 [hep-ph]].

\bibitem{Gresham:2007ri}
  M.~I.~Gresham and M.~B.~Wise,
  Phys.\ Rev.\ D {\bf 76} (2007) 075003
  [arXiv:0706.0909 [hep-ph]];
  B.~A.~Dobrescu, K.~Kong and R.~Mahbubani,
  Phys.\ Lett.\ B {\bf 670} (2008) 119
  [arXiv:0709.2378 [hep-ph]];
  M.~Gerbush, T.~J.~Khoo, D.~J.~Phalen, A.~Pierce and D.~Tucker-Smith,
  Phys.\ Rev.\ D {\bf 77} (2008) 095003
  [arXiv:0710.3133 [hep-ph]];
  C.~P.~Burgess, M.~Trott and S.~Zuberi,
  JHEP {\bf 0909} (2009) 082
  [arXiv:0907.2696 [hep-ph]];
  A.~Idilbi, C.~Kim and T.~Mehen,
  Phys.\ Rev.\ D {\bf 82} (2010) 075017
  [arXiv:1007.0865 [hep-ph]];
  X.~G.~He, G.~Valencia and H.~Yokoya,
  JHEP {\bf 1112} (2011) 030
  [arXiv:1110.2588 [hep-ph]];
  J.~M.~Arnold and B.~Fornal,
  Phys.\ Rev.\ D {\bf 85} (2012) 055020
  [arXiv:1112.0003 [hep-ph]];
  J.~Cao, P.~Wan, J.~M.~Yang and J.~Zhu,
  JHEP {\bf 1308} (2013) 009
  [arXiv:1303.2426 [hep-ph]];
  Z.~Heng, L.~Shang, Y.~Zhang and J.~Zhu,
  JHEP {\bf 1402} (2014) 083
  [arXiv:1312.4260 [hep-ph]];
  R.~Frederix and F.~Maltoni,
  JHEP {\bf 0901} (2009) 047
  [arXiv:0712.2355 [hep-ph]];
  A.~R.~Zerwekh, C.~O.~Dib and R.~Rosenfeld,
  Phys.\ Rev.\ D {\bf 77} (2008) 097703
  [arXiv:0802.4303 [hep-ph]];
  P.~Fileviez Perez, R.~Gavin, T.~McElmurry and F.~Petriello,
  Phys.\ Rev.\ D {\bf 78} (2008) 115017
  [arXiv:0809.2106 [hep-ph]];
  B.~Fornal and M.~Trott,
  JHEP {\bf 1006} (2010) 110
  [arXiv:1001.4287 [hep-ph]];
  T.~Han, I.~Lewis and Z.~Liu,
  JHEP {\bf 1012} (2010) 085
  [arXiv:1010.4309 [hep-ph]];
  P.~Fileviez Perez, T.~Han, S.~Spinner and M.~K.~Trenkel,
  JHEP {\bf 1101} (2011) 046
  [arXiv:1010.5802 [hep-ph]];
  Y.~Bai and B.~A.~Dobrescu,
  JHEP {\bf 1107} (2011) 100
  [arXiv:1012.5814 [hep-ph]].

\bibitem{Carpenter:2011yj}
  L.~M.~Carpenter and S.~Mantry,
  Phys.\ Lett.\ B {\bf 703} (2011) 479
  [arXiv:1104.5528 [hep-ph]];
  T.~Enkhbat, X.~G.~He, Y.~Mimura and H.~Yokoya,
  JHEP {\bf 1202} (2012) 058
  [arXiv:1105.2699 [hep-ph]].

\bibitem{Agashe:2014kda}
  K.~A.~Olive {\it et al.}  [Particle Data Group Collaboration],
  Chin.\ Phys.\ C {\bf 38} (2014) 090001.

\bibitem{Amhis:2014hma}
  Y.~Amhis {\it et al.}  [Heavy Flavor Averaging Group (HFAG) Collaboration],
  arXiv:1412.7515 [hep-ex].

\bibitem{Li:2014fea}
  X.~Q.~Li, J.~Lu and A.~Pich,
  JHEP {\bf 1406} (2014) 022  [arXiv:1404.5865 [hep-ph]].  

\bibitem{Chatrchyan:2013bka}
  S.~Chatrchyan {\it et al.}  [CMS Collaboration],
  Phys.\ Rev.\ Lett.\  {\bf 111} (2013) 101804
  [arXiv:1307.5025 [hep-ex]].

\bibitem{Aaij:2013aka}
  R.~Aaij {\it et al.}  [LHCb Collaboration],
  Phys.\ Rev.\ Lett.\  {\bf 111} (2013) 101805
  [arXiv:1307.5024 [hep-ex]].

\bibitem{CMS:2014xfa}
  V.~Khachatryan {\it et al.}  [CMS and LHCb Collaborations],
  arXiv:1411.4413 [hep-ex].

\bibitem{Bobeth:2013uxa}
  C.~Bobeth, M.~Gorbahn, T.~Hermann, M.~Misiak, E.~Stamou and M.~Steinhauser,
  Phys.\ Rev.\ Lett.\  {\bf 112} (2014) 101801
  [arXiv:1311.0903 [hep-ph]].

\bibitem{Bediaga:2012py}
  R.~Aaij {\it et al.}  [LHCb Collaboration],
  Eur.\ Phys.\ J.\ C {\bf 73} (2013) 2373
  [arXiv:1208.3355 [hep-ex]].

\bibitem{Aushev:2010bq}
  T.~Aushev, W.~Bartel, A.~Bondar, J.~Brodzicka, T.~E.~Browder, P.~Chang, Y.~Chao and K.~F.~Chen {\it et al.},
  arXiv:1002.5012 [hep-ex];
  B.~Meadows, M.~Blanke, A.~Stocchi, A.~Drutskoy, A.~Cervelli, M.~Giorgi, A.~Lusiani and A.~Perez {\it et al.},
  arXiv:1109.5028 [hep-ex].

\bibitem{Charles:2013aka}
  J.~Charles, S.~Descotes-Genon, Z.~Ligeti, S.~Monteil, M.~Papucci and K.~Trabelsi,
  Phys.\ Rev.\ D {\bf 89} (2014) 033016
  [arXiv:1309.2293 [hep-ph]].

\bibitem{Bai:2012ex}
  Y.~Bai, V.~Barger, L.~L.~Everett and G.~Shaughnessy,
  Phys.\ Rev.\ D {\bf 87} (2013) 115013
  [arXiv:1210.4922 [hep-ph]].

\bibitem{Altmannshofer:2012ar}
  W.~Altmannshofer, S.~Gori and G.~D.~Kribs,
  Phys.\ Rev.\ D {\bf 86} (2012) 115009
  [arXiv:1210.2465 [hep-ph]].

\bibitem{Christensen:2008py}
  N.~D.~Christensen and C.~Duhr,
  Comput.\ Phys.\ Commun.\  {\bf 180} (2009) 1614
  [arXiv:0806.4194 [hep-ph]];
  A.~Alloul, J.~D'Hondt, K.~De Causmaecker, B.~Fuks and M.~Rausch de Traubenberg,
  Eur.\ Phys.\ J.\ C {\bf 73} (2013) 2325
  [arXiv:1301.5932 [hep-ph]];
  A.~Alloul, N.~D.~Christensen, C.~Degrande, C.~Duhr and B.~Fuks,
  Comput.\ Phys.\ Commun.\  {\bf 185} (2014) 2250
  [arXiv:1310.1921 [hep-ph]].

\bibitem{Buras:2001ra}
  A.~J.~Buras, S.~Jager and J.~Urban,
  Nucl.\ Phys.\ B {\bf 605} (2001) 600
  [hep-ph/0102316].

\bibitem{Buras:2000if}
  A.~J.~Buras, M.~Misiak and J.~Urban,
  Nucl.\ Phys.\ B {\bf 586} (2000) 397
  [hep-ph/0005183].

\bibitem{Buras:1990fn}
  A.~J.~Buras, M.~Jamin and P.~H.~Weisz,
  Nucl.\ Phys.\ B {\bf 347} (1990) 491.

\bibitem{Urban:1997gw}
  J.~Urban, F.~Krauss, U.~Jentschura and G.~Soff,
  Nucl.\ Phys.\ B {\bf 523} (1998) 40
  [hep-ph/9710245].

\bibitem{Inami:1980fz}
  T.~Inami and C.~S.~Lim,
  Prog.\ Theor.\ Phys.\  {\bf 65} (1981) 297
   [Erratum-ibid.\  {\bf 65} (1981) 1772].

\bibitem{Urban:1997cm}
  J.~Urban, F.~Krauss and G.~Soff,
  J.\ Phys.\ G {\bf 23} (1997) L25.

\bibitem{Smirnov:1994tg}
  V.~A.~Smirnov,
  Mod.\ Phys.\ Lett.\ A {\bf 10} (1995) 1485
  [hep-th/9412063];
  Springer Tracts Mod.\ Phys.\  {\bf 177} (2002) 1.

\bibitem{Smirnov:2004ym}
  V.~A.~Smirnov,
  Springer Tracts Mod.\ Phys.\  {\bf 211} (2004) 1.

\bibitem{Passarino:1978jh}
  G.~Passarino and M.~J.~G.~Veltman,
  Nucl.\ Phys.\ B {\bf 160} (1979) 151.

\bibitem{Peskin:1995ev}
  M.~E.~Peskin and D.~V.~Schroeder,
  Reading, USA: Addison-Wesley (1995) 842 p

\bibitem{Carrasco:2013zta}
  N.~Carrasco {\it et al.}  [ETM Collaboration],
  JHEP {\bf 1403} (2014) 016
  [arXiv:1308.1851 [hep-lat]].

\bibitem{Buras:2013uqa}
  A.~J.~Buras, R.~Fleischer, J.~Girrbach and R.~Knegjens,
  JHEP {\bf 1307} (2013) 77
  [arXiv:1303.3820 [hep-ph]].

\bibitem{Bobeth:2013tba}
  C.~Bobeth, M.~Gorbahn and E.~Stamou,
  Phys.\ Rev.\ D {\bf 89} (2014) 034023
  [arXiv:1311.1348 [hep-ph]].

\bibitem{Hermann:2013kca}
  T.~Hermann, M.~Misiak and M.~Steinhauser,
  JHEP {\bf 1312} (2013) 097
  [arXiv:1311.1347 [hep-ph]].

\bibitem{Buchalla:1998ba}
  G.~Buchalla and A.~J.~Buras,
  Nucl.\ Phys.\ B {\bf 548} (1999) 309
  [hep-ph/9901288];
  Nucl.\ Phys.\ B {\bf 400} (1993) 225;
  Nucl.\ Phys.\ B {\bf 398} (1993) 285.

\bibitem{Misiak:1999yg}
  M.~Misiak and J.~Urban,
  Phys.\ Lett.\ B {\bf 451} (1999) 161
  [hep-ph/9901278].

\bibitem{DeBruyn:2012wk}
  K.~De Bruyn, R.~Fleischer, R.~Knegjens, P.~Koppenburg, M.~Merk, A.~Pellegrino and N.~Tuning,
  Phys.\ Rev.\ Lett.\  {\bf 109} (2012) 041801
  [arXiv:1204.1737 [hep-ph]].

\bibitem{ALEPH:2005ab}
  S.~Schael {\it et al.}  [ALEPH and DELPHI and L3 and OPAL and SLD and LEP Electroweak Working Group and SLD Electroweak Group and SLD Heavy Flavour Group Collaborations],
  Phys.\ Rept.\  {\bf 427} (2006) 257
  [hep-ex/0509008].

\bibitem{Baak:2014ora}
  M.~Baak {\it et al.}  [Gfitter Group Collaboration],
  Eur.\ Phys.\ J.\ C {\bf 74} (2014) 3046
  [arXiv:1407.3792 [hep-ph]];
  M.~Baak, M.~Goebel, J.~Haller, A.~Hoecker, D.~Kennedy, R.~Kogler, K.~Moenig and M.~Schott {\it et al.},
  Eur.\ Phys.\ J.\ C {\bf 72} (2012) 2205
  [arXiv:1209.2716 [hep-ph]].

\bibitem{Haber:1999zh}
  H.~E.~Haber and H.~E.~Logan,
  Phys.\ Rev.\ D {\bf 62} (2000) 015011
  [hep-ph/9909335].


\bibitem{Hurth:2010tk}
  T.~Hurth and M.~Nakao,
  Ann.\ Rev.\ Nucl.\ Part.\ Sci.\  {\bf 60} (2010) 645
  [arXiv:1005.1224 [hep-ph]];
  T.~Hurth,
  PoS CHARGED {\bf 2010} (2010) 020
  [arXiv:1104.5123 [hep-ph]].

\bibitem{Hermann:2012fc}
  T.~Hermann, M.~Misiak and M.~Steinhauser,
  JHEP {\bf 1211} (2012) 036
  [arXiv:1208.2788 [hep-ph]];
  F.~M.~Borzumati, C.~Greub,
  Phys.\ Rev.\ D {\bf 58} (1998) 074004
  [hep-ph/9802391];
  F.~M.~Borzumati and C.~Greub,
  Phys.\ Rev.\ D {\bf 59} (1999) 057501
  [hep-ph/9809438];
  M.~Ciuchini, G.~Degrassi, P.~Gambino, G.~F.~Giudice,
  Nucl.\ Phys.\ B {\bf 527} (1998) 21
  [hep-ph/9710335];
  P.~Ciafaloni, A.~Romanino, A.~Strumia,
  Nucl.\ Phys.\ B {\bf 524} (1998) 361
  [hep-ph/9710312].

\bibitem{Misiak:2015xwa}
  M.~Misiak, H.~M.~Asatrian, R.~Boughezal, M.~Czakon, T.~Ewerth, A.~Ferroglia, P.~Fiedler and P.~Gambino {\it et al.},
  arXiv:1503.01789 [hep-ph];
  M.~Misiak and M.~Steinhauser,
  Nucl.\ Phys.\ B {\bf 764} (2007) 62
  [hep-ph/0609241];
  M.~Misiak {\it et al.},
  Phys.\ Rev.\ Lett.\  {\bf 98} (2007) 022002
  [hep-ph/0609232].

\bibitem{Ball:2006eu}
  P.~Ball, G.~W.~Jones and R.~Zwicky,
  Phys.\ Rev.\ D {\bf 75} (2007) 054004
  [hep-ph/0612081];
  J.~Lyon and R.~Zwicky,
  Phys.\ Rev.\ D {\bf 88} (2013) 094004
  [arXiv:1305.4797 [hep-ph]].

\bibitem{Beneke:2001at}
  M.~Beneke, T.~Feldmann and D.~Seidel,
  Nucl.\ Phys.\ B {\bf 612} (2001) 25
  [hep-ph/0106067];
  Eur.\ Phys.\ J.\ C {\bf 41} (2005) 173
  [hep-ph/0412400].

\bibitem{Jung:2012vu}
  M.~Jung, X.~Q.~Li and A.~Pich,
  JHEP {\bf 1210} (2012) 063
  [arXiv:1208.1251 [hep-ph]].

\bibitem{Mahmoudi:2009zx}
  F.~Mahmoudi and O.~Stal,
  Phys.\ Rev.\ D {\bf 81} (2010) 035016
  [arXiv:0907.1791 [hep-ph]].

\bibitem{B2Vg-NP}
  W.~Altmannshofer and D.~M.~Straub,
  arXiv:1411.3161 [hep-ph];
  T.~Hurth and F.~Mahmoudi,
  Nucl.\ Phys.\ B {\bf 865} (2012) 461
  [arXiv:1207.0688 [hep-ph]];
  X.~Q.~Li, Y.~D.~Yang and X.~B.~Yuan,
  JHEP {\bf 1108} (2011) 075
  [arXiv:1105.0364 [hep-ph]];
  S.~Descotes-Genon, D.~Ghosh, J.~Matias and M.~Ramon,
  JHEP {\bf 1106} (2011) 099
  [arXiv:1104.3342 [hep-ph]];
  M.~R.~Ahmady and F.~Mahmoudi,
  Phys.\ Rev.\ D {\bf 75} (2007) 015007
  [hep-ph/0608212];
  M.~R.~Ahmady and F.~Chishtie,
  Int.\ J.\ Mod.\ Phys.\ A {\bf 20} (2005) 6229
  [hep-ph/0508105];
  Z.~-j.~Xiao and C.~Zhuang,
  Eur.\ Phys.\ J.\ C {\bf 33} (2004) 349
  [hep-ph/0310097].

\bibitem{WC1}
  G.~Buchalla, A.~J.~Buras and M.~E.~Lautenbacher,
  Rev.\ Mod.\ Phys.\  {\bf 68} (1996) 1125
  [hep-ph/9512380];
  K.~G.~Chetyrkin, M.~Misiak and M.~Munz,
  Phys.\ Lett.\ B {\bf 400} (1997) 206  [Erratum-ibid.\ B {\bf 425} (1998) 414]
  [hep-ph/9612313];
  A.~J.~Buras,
  hep-ph/9806471.

\bibitem{WC2}
  M.~Misiak and M.~Steinhauser,
  Nucl.\ Phys.\ B {\bf 683} (2004) 277
  [hep-ph/0401041];
  M.~Gorbahn and U.~Haisch,
  Nucl.\ Phys.\ B {\bf 713} (2005) 291
  [hep-ph/0411071];
  M.~Gorbahn, U.~Haisch and M.~Misiak,
  Phys.\ Rev.\ Lett.\  {\bf 95} (2005) 102004
  [hep-ph/0504194];
  M.~Czakon, U.~Haisch and M.~Misiak,
  JHEP {\bf 0703} (2007) 008
  [hep-ph/0612329].

\bibitem{Charles:2004jd}
  J.~Charles {\it et al.}  [CKMfitter Group Collaboration],
  Eur.\ Phys.\ J.\ C {\bf 41} (2005) 1
  [hep-ph/0406184], and online update at http://ckmfitter.in2p3.fr/.

\bibitem{Aoki:2013ldr}
  S.~Aoki, Y.~Aoki, C.~Bernard, T.~Blum, G.~Colangelo, M.~Della Morte, S.~D��rr and A.~X.~E.~Khadra {\it et al.},
  arXiv:1310.8555 [hep-lat], and online update at http://itpwiki.unibe.ch/flag.

\bibitem{Blum:2013mhx}
  T.~Blum, R.~S.~Van de Water, D.~Holmgren, R.~Brower, S.~Catterall, N.~Christ, A.~Kronfeld and J.~Kuti {\it et al.},
  arXiv:1310.6087 [hep-lat].

\bibitem{Aad:2011fq}
  G.~Aad {\it et al.}  [ATLAS Collaboration],
  Phys.\ Lett.\ B {\bf 708} (2012) 37
  [arXiv:1108.6311 [hep-ex]];
  JHEP {\bf 1301} (2013) 029
  [arXiv:1210.1718 [hep-ex]];
  Phys.\ Rev.\ D {\bf 91} (2015) 052007
  [arXiv:1407.1376 [hep-ex]].

\bibitem{CMS:2012yf}
  S.~Chatrchyan {\it et al.}  [CMS Collaboration],
  JHEP {\bf 1301} (2013) 013
  [arXiv:1210.2387 [hep-ex]];
  Phys.\ Rev.\ D {\bf 87} (2013) 114015
  [arXiv:1302.4794 [hep-ex]];
  V.~Khachatryan {\it et al.}  [CMS Collaboration],
  Phys.\ Rev.\ D {\bf 91} (2015) 052009
  [arXiv:1501.04198 [hep-ex]].

\bibitem{Aad:2011yh}
  G.~Aad {\it et al.}  [ATLAS Collaboration],
  Eur.\ Phys.\ J.\ C {\bf 71} (2011) 1828
  [arXiv:1110.2693 [hep-ex]];
  Eur.\ Phys.\ J.\ C {\bf 73} (2013) 2263
  [arXiv:1210.4826 [hep-ex]].

\bibitem{Chatrchyan:2013izb}
  S.~Chatrchyan {\it et al.}  [CMS Collaboration],
  Phys.\ Rev.\ Lett.\  {\bf 110} (2013) 141802
  [arXiv:1302.0531 [hep-ex]];
  V.~Khachatryan {\it et al.}  [CMS Collaboration],
  arXiv:1412.7706 [hep-ex].

\bibitem{Hocker:2001xe}
  A.~Hocker, H.~Lacker, S.~Laplace and F.~Le Diberder,
  Eur.\ Phys.\ J.\ C {\bf 21} (2001) 225
  [hep-ph/0104062].

\bibitem{Cheng:2014ova}
  X.~D.~Cheng, Y.~D.~Yang and X.~B.~Yuan,
  Eur.\ Phys.\ J.\ C {\bf 74} (2014) 3081
  [arXiv:1401.6657 [hep-ph]].

\bibitem{Hurth:2013ssa}
  T.~Hurth and F.~Mahmoudi,
  JHEP {\bf 1404} (2014) 097  [arXiv:1312.5267 [hep-ph]];
  T.~Hurth, F.~Mahmoudi and S.~Neshatpour,
  JHEP {\bf 1412} (2014) 053  [arXiv:1410.4545 [hep-ph]].

\end{thebibliography}
\end{document}